\newcommand{\punkt}[9]{\put(#3
   ,0){\setlength{\unitlength}{5cm}\begin{picture}(0,0)(0,-0.25)
   \put(0,#4){\circle*{   0.015}}
   \put(0,#4){\line(0,1){#5}}
   \put(0,#4){\line(0,-1){#5}}
   \end{picture}}}
\newcommand{\punkta}[9]{\put(#3
   ,0){\setlength{\unitlength}{5cm}\begin{picture}(0,0)(0,-0.25)
   \put(0,#4){\circle{   0.03}}
   \end{picture}}}

\newcommand{\T}[1]{\rule[1.5ex]{0cm}{#1cm}}

\documentstyle{laa}
\begin{document}

\thesaurus{03(11.17.3)}
\title{The Hamburg Quasar Monitoring Program (HQM) at Calar Alto\thanks{Based
    on observations collected at the
    German-Spanish Astronomical Centre, Calar Alto,
    operated by the Max-Planck-Institut f\"ur Astronomie (MPIA), Heidelberg,
    jointly with the Spanish National Commission for Astronomy}:\\
    I$\!$I. Lightcurves of weakly variable objects}
\author{K.-J.~Schramm
     \and
        U.~Borgeest
     \and
        D.~K\"uhl
     \and
        J.~v.~Linde
     \and
        M.D.~Linnert}
\institute{Hamburger Sternwarte, Gojenbergsweg 112,
           D-W\,2050 Hamburg 80, Germany }
\date{Received date; accepted date}
\maketitle

\begin{abstract}
HQM is an optical broad-band photometric monitoring program carried out
since Sept.~1988. We use a CCD camera equipped to the MPIA 1.2$\,$m
telescope. Fully automatic photometric reduction relative to stars in the
frames is done within a few minutes after each exposure, thus interesting
brightness changes can be followed in detail. The typical photometric error
is 1--2\,\% for a 17.5\,mag quasar. We here present lightcurves already
evaluated but not shown in Paper I. We also discuss existing
literature data.
\end{abstract}

\section{Introduction}

In this paper, being the second one in a series, we show some additional
lightcurves for quasars which were {\em not\/} known to be optically
violent variables prior to our program and for which measurements exist
sufficiently spread over a time-span $\ga$\,2\,yrs. Some interesting
properties of these quasars are given in Table~1 of Borgeest \& Schramm
(1993, hereafter Paper~I); Table~4 of Paper~I lists values of some
reasonable parameters through which the lightcurve shapes can be
quantified. We also give there the results of POSS photometry carried out
to obtain indications for variability on a longer timescale.

For HQM, we use a CCD camera which was equipped with different chips,
in 1988 with an RCA 15$ \mu $ chip ($640 \! \times \! 1024$, pixel size
0.315$''$) and later various, but similar, coated GEC 22$ \mu $ chips ($410
\! \times \! 580$, pixel size 0.462$''$). We measure the quasar fluxes
through a standard Johnson $R\/$ broad-band filter relative to stars
included in the frames. The data reduction is carried out automatically,
immediately after the observation, on a $\mu$VAX\,3200 workstation. The
software package ``HQM'' has been developed in Hamburg (for a short
description see Paper~I); it is much faster than standard image processing
software. A 0.01\,mag accuracy (in relative photometry) in the lightcurve
of a $\sim 17.5\,$mag quasar can be reached in this way also for
``non-photometric'' conditions with a typical exposure time of 500 sec.

Extensive monitoring programs which contain weakly variable quasars
have been carried out by several investigators:
Monitoring data obtained until 1973 are reviewed and critically discussed
by Grandi \& Tifft (\cite{GT74}, hereafter GT74).
Lloyd (\cite{Llo84}, hereafter L84) reports on lightcurves of 36 radio sources
from the Herstmonceux Optical Monitoring program for the period
1966-1980 (see also Tritton \& Selmes \cite{TS71}, hereafter TS71;
Selmes et al.\ \cite{STW75}, hereafter STW).
At the Rosemary Hill Observatory more than 200, mostly radio-selected
quasars were monitored since 1968, however not all objects
over the total period (Pica et al.~\cite{PPSL80}, hereafter PPSL;
Pica \& Smith \cite{PS83}, hereafter PS83; Smith et al.\ \cite{SNLC93},
hereafter SNLC, and Refs.\ therein).
Another program was carried out at the Asagio Observatory (e.g.\
Barbieri et al.~\cite{BRZ79}, hereafter BRZ) over the period 1967 to 1977.
Lightcurves of many bright quasars have been obtained from the Harvard
historical plate collection (e.g.\ Angione \cite{Ang73}, hereafter A73)
spanning periods of up to 100 years.
Netzer \& Sheffer (\cite{NS83}, hereafter NS83) compared the 1981 and
$\sim$\,1950 POSS magnitudes of 64 optically selected UM quasars.
Moore \& Stockman (\cite{MS84}, hereafter MS84) have collected a catalog
of the observational properties of 239 quasars, including variability data.
More recent publications on optical variability of quasars generally deal
with violently variable sources.

\begin{figure*}

\vspace*{0.4cm}

\begin{picture}(18 ,2.5 )(0,0)
\put(0,0){\setlength{\unitlength}{0.01059cm}%
\begin{picture}(1700, 236.072)(7350,0)
\put(7350,0){\framebox(1700, 236.072)[tl]{\begin{picture}(0,0)(0,0)
        \put(1700,-132){\makebox(0,0)[tr]{\normalsize{\bf 0003+158}\T{0.4}
                                 \hspace*{0.5cm}}}
    \end{picture}}}

\thicklines
\put(7350,0){\setlength{\unitlength}{5cm}\begin{picture}(0,0)(0,-0.25)
   \put(0,0){\setlength{\unitlength}{1cm}\begin{picture}(0,0)(0,0)
        \put(0,0){\line(1,0){0.3}}
        \end{picture}}
   \end{picture}}

\put(9050,0){\setlength{\unitlength}{5cm}\begin{picture}(0,0)(0,-0.25)
   \put(0,0){\setlength{\unitlength}{1cm}\begin{picture}(0,0)(0,0)
        \put(0,0){\line(-1,0){0.3}}
        \end{picture}}
   \end{picture}}

\thinlines
\put(7350,0){\setlength{\unitlength}{5cm}\begin{picture}(0,0)(0,-0.25)
   \multiput(0,0)(0,0.1){3}{\setlength{\unitlength}{1cm}%
\begin{picture}(0,0)(0,0)
        \put(0,0){\line(1,0){0.12}}
        \end{picture}}
   \end{picture}}

\put(7350,0){\setlength{\unitlength}{5cm}\begin{picture}(0,0)(0,-0.25)
   \multiput(0,0)(0,-0.1){3}{\setlength{\unitlength}{1cm}%
\begin{picture}(0,0)(0,0)
        \put(0,0){\line(1,0){0.12}}
        \end{picture}}
   \end{picture}}

\put(9050,0){\setlength{\unitlength}{5cm}\begin{picture}(0,0)(0,-0.25)
   \multiput(0,0)(0,0.1){3}{\setlength{\unitlength}{1cm}%
\begin{picture}(0,0)(0,0)
        \put(0,0){\line(-1,0){0.12}}
        \end{picture}}
   \end{picture}}

\put(9050,0){\setlength{\unitlength}{5cm}\begin{picture}(0,0)(0,-0.25)
   \multiput(0,0)(0,-0.1){3}{\setlength{\unitlength}{1cm}%
\begin{picture}(0,0)(0,0)
        \put(0,0){\line(-1,0){0.12}}
        \end{picture}}
   \end{picture}}

   \put(7527.5, 236.072){\setlength{\unitlength}{1cm}\begin{picture}(0,0)(0,0)
        \put(0,0){\line(0,-1){0.2}}
        \put(0,0.2){\makebox(0,0)[b]{\bf 1989}}
   \end{picture}}
   \put(7892.5, 236.072){\setlength{\unitlength}{1cm}\begin{picture}(0,0)(0,0)
        \put(0,0){\line(0,-1){0.2}}
        \put(0,0.2){\makebox(0,0)[b]{\bf 1990}}
   \end{picture}}
   \put(8257.5, 236.072){\setlength{\unitlength}{1cm}\begin{picture}(0,0)(0,0)
        \put(0,0){\line(0,-1){0.2}}
        \put(0,0.2){\makebox(0,0)[b]{\bf 1991}}
   \end{picture}}
   \put(8622.5, 236.072){\setlength{\unitlength}{1cm}\begin{picture}(0,0)(0,0)
        \put(0,0){\line(0,-1){0.2}}
        \put(0,0.2){\makebox(0,0)[b]{\bf 1992}}
   \end{picture}}
   \put(8987.5, 236.072){\setlength{\unitlength}{1cm}\begin{picture}(0,0)(0,0)
        \put(0,0){\line(0,-1){0.2}}
        \put(0,0.2){\makebox(0,0)[b]{\bf 1993}}
   \end{picture}}
    \multiput(7350,0)(50,0){33}%
        {\setlength{\unitlength}{1cm}\begin{picture}(0,0)(0,0)
        \put(0,0){\line(0,1){0.12}}
    \end{picture}}
    \put(7500,0){\setlength{\unitlength}{1cm}\begin{picture}(0,0)(0,0)
        \put(0,0){\line(0,1){0.2}}
    \end{picture}}
    \put(7750,0){\setlength{\unitlength}{1cm}\begin{picture}(0,0)(0,0)
        \put(0,0){\line(0,1){0.2}}
    \end{picture}}
    \put(8000,0){\setlength{\unitlength}{1cm}\begin{picture}(0,0)(0,0)
        \put(0,0){\line(0,1){0.2}}
    \end{picture}}
    \put(8250,0){\setlength{\unitlength}{1cm}\begin{picture}(0,0)(0,0)
        \put(0,0){\line(0,1){0.2}}
    \end{picture}}
    \put(8500,0){\setlength{\unitlength}{1cm}\begin{picture}(0,0)(0,0)
        \put(0,0){\line(0,1){0.2}}
    \end{picture}}
    \put(8750,0){\setlength{\unitlength}{1cm}\begin{picture}(0,0)(0,0)
        \put(0,0){\line(0,1){0.2}}
    \end{picture}}
    \put(9000,0){\setlength{\unitlength}{1cm}\begin{picture}(0,0)(0,0)
        \put(0,0){\line(0,1){0.2}}
    \end{picture}}

\punkta{07/09/88}{23:22}{7412.474}{  0.002}{0.043}{ 500}{1.61}{
640}{1.2CA}
\punkt{08/10/88}{21:36}{7443.401}{  0.048}{0.025}{ 500}{1.07}{
626}{1.2CA}
\punkta{13/08/89}{23:50}{7752.493}{  0.099}{0.014}{ 500}{1.35}{
1015}{1.2CA}
\punkta{02/11/89}{21:39}{7833.403}{  0.103}{0.034}{ 500}{1.51}{
571}{1.2CA}
\punkta{14/12/89}{19:11}{7875.300}{  0.059}{0.031}{ 500}{1.89}{
697}{1.2CA}
\punkta{20/12/89}{19:29}{7881.312}{  0.039}{0.033}{ 500}{1.70}{
694}{1.2CA}
\punkta{21/12/89}{18:22}{7882.266}{  0.078}{0.011}{ 500}{3.10}{
638}{1.2CA}
\punkt{27/07/91}{02:49}{8464.617}{ -0.088}{0.034}{ 150}{1.80}{
2627}{1.2CA}
\punkt{27/07/91}{02:54}{8464.622}{ -0.088}{0.032}{ 150}{1.69}{
2580}{1.2CA}
\punkt{22/08/91}{04:23}{8490.683}{ -0.114}{0.026}{ 500}{1.62}{
838}{1.2CA}
\punkta{23/08/91}{01:15}{8491.552}{ -0.113}{0.025}{ 500}{1.63}{
1398}{1.2CA}
\punkt{22/09/92}{23:35}{8888.483}{  0.024}{0.013}{ 100}{1.47}{
526}{1.2CA}
\punkt{22/09/92}{23:38}{8888.485}{  0.032}{0.015}{ 100}{1.53}{
543}{1.2CA}
\punkt{22/09/92}{23:38}{8888.485}{  0.032}{0.015}{ 100}{1.53}{
543}{1.2CA}

\end{picture}}

\end{picture}

\vspace*{-0.02cm}

\begin{picture}(18 ,3.5 )(0,0)
\put(0,0){\setlength{\unitlength}{0.01059cm}%
\begin{picture}(1700, 330.555)(7350,0)
\put(7350,0){\framebox(1700, 330.555)[tl]{\begin{picture}(0,0)(0,0)
        \put(1700,0){\makebox(0,0)[tr]{\bf{0007-$\!$-000}\T{0.4}
                                 \hspace*{0.5cm}}}
    \end{picture}}}

\thicklines
\put(7350,0){\setlength{\unitlength}{5cm}\begin{picture}(0,0)(0,-0.35)
   \put(0,0){\setlength{\unitlength}{1cm}\begin{picture}(0,0)(0,0)
        \put(0,0){\line(1,0){0.3}}
        \end{picture}}
   \end{picture}}

\put(9050,0){\setlength{\unitlength}{5cm}\begin{picture}(0,0)(0,-0.35)
   \put(0,0){\setlength{\unitlength}{1cm}\begin{picture}(0,0)(0,0)
        \put(0,0){\line(-1,0){0.3}}
        \end{picture}}
   \end{picture}}

\thinlines
\put(7350,0){\setlength{\unitlength}{5cm}\begin{picture}(0,0)(0,-0.35)
   \multiput(0,0)(0,0.1){4}{\setlength{\unitlength}{1cm}%
\begin{picture}(0,0)(0,0)
        \put(0,0){\line(1,0){0.12}}
        \end{picture}}
   \end{picture}}

\put(7350,0){\setlength{\unitlength}{5cm}\begin{picture}(0,0)(0,-0.35)
   \multiput(0,0)(0,-0.1){4}{\setlength{\unitlength}{1cm}%
\begin{picture}(0,0)(0,0)
        \put(0,0){\line(1,0){0.12}}
        \end{picture}}
   \end{picture}}

\put(9050,0){\setlength{\unitlength}{5cm}\begin{picture}(0,0)(0,-0.35)
   \multiput(0,0)(0,0.1){4}{\setlength{\unitlength}{1cm}%
\begin{picture}(0,0)(0,0)
        \put(0,0){\line(-1,0){0.12}}
        \end{picture}}
   \end{picture}}

\put(9050,0){\setlength{\unitlength}{5cm}\begin{picture}(0,0)(0,-0.35)
   \multiput(0,0)(0,-0.1){4}{\setlength{\unitlength}{1cm}%
\begin{picture}(0,0)(0,0)
        \put(0,0){\line(-1,0){0.12}}
        \end{picture}}
   \end{picture}}

   \put(7527.5, 330.555){\setlength{\unitlength}{1cm}\begin{picture}(0,0)(0,0)
        \put(0,0){\line(0,-1){0.2}}
   \end{picture}}
   \put(7892.5, 330.555){\setlength{\unitlength}{1cm}\begin{picture}(0,0)(0,0)
        \put(0,0){\line(0,-1){0.2}}
   \end{picture}}
   \put(8257.5, 330.555){\setlength{\unitlength}{1cm}\begin{picture}(0,0)(0,0)
        \put(0,0){\line(0,-1){0.2}}
   \end{picture}}
   \put(8622.5, 330.555){\setlength{\unitlength}{1cm}\begin{picture}(0,0)(0,0)
        \put(0,0){\line(0,-1){0.2}}
   \end{picture}}
   \put(8987.5, 330.555){\setlength{\unitlength}{1cm}\begin{picture}(0,0)(0,0)
        \put(0,0){\line(0,-1){0.2}}
   \end{picture}}
    \multiput(7350,0)(50,0){33}%
        {\setlength{\unitlength}{1cm}\begin{picture}(0,0)(0,0)
        \put(0,0){\line(0,1){0.12}}
    \end{picture}}
    \put(7500,0){\setlength{\unitlength}{1cm}\begin{picture}(0,0)(0,0)
        \put(0,0){\line(0,1){0.2}}
    \end{picture}}
    \put(7750,0){\setlength{\unitlength}{1cm}\begin{picture}(0,0)(0,0)
        \put(0,0){\line(0,1){0.2}}
    \end{picture}}
    \put(8000,0){\setlength{\unitlength}{1cm}\begin{picture}(0,0)(0,0)
        \put(0,0){\line(0,1){0.2}}
    \end{picture}}
    \put(8250,0){\setlength{\unitlength}{1cm}\begin{picture}(0,0)(0,0)
        \put(0,0){\line(0,1){0.2}}
    \end{picture}}
    \put(8500,0){\setlength{\unitlength}{1cm}\begin{picture}(0,0)(0,0)
        \put(0,0){\line(0,1){0.2}}
    \end{picture}}
    \put(8750,0){\setlength{\unitlength}{1cm}\begin{picture}(0,0)(0,0)
        \put(0,0){\line(0,1){0.2}}
    \end{picture}}
    \put(9000,0){\setlength{\unitlength}{1cm}\begin{picture}(0,0)(0,0)
        \put(0,0){\line(0,1){0.2}}
    \end{picture}}

\put(7000,0){\begin{picture}(0,0)(7000,-47.21435)
\punkt{04/09/88}{02:00}{7408.584}{  0.211}{0.024}{1000}{2.02}{
982}{2.2CA}
\punkt{14/08/89}{00:10}{7752.507}{  0.112}{0.016}{1000}{2.07}{
1718}{2.2CA}
\punkta{17/10/91}{21:50}{8547.410}{ -0.079}{0.050}{ 300}{2.50}{
1015}{2.2CA}
\punkt{23/09/92}{23:16}{8889.470}{ -0.150}{0.011}{ 100}{1.01}{
457}{2.2CA}
\punkt{23/09/92}{23:19}{8889.472}{ -0.147}{0.013}{ 100}{1.03}{
463}{2.2CA}
\punkt{23/09/92}{23:19}{8889.472}{ -0.147}{0.013}{ 100}{1.03}{
463}{2.2CA}
\end{picture}}

\end{picture}}

\end{picture}

\vspace*{-0.02cm}

\begin{picture}(18 ,2.5 )(0,0)
\put(0,0){\setlength{\unitlength}{0.01059cm}%
\begin{picture}(1700, 236.111)(7350,0)
\put(7350,0){\framebox(1700, 236.111)[tl]{\begin{picture}(0,0)(0,0)
        \put(1700,0){\makebox(0,0)[tr]{\bf{0013-$\!$-004}\T{0.4}
                                 \hspace*{0.5cm}}}
    \end{picture}}}

\thicklines
\put(7350,0){\setlength{\unitlength}{5cm}\begin{picture}(0,0)(0,-0.25)
   \put(0,0){\setlength{\unitlength}{1cm}\begin{picture}(0,0)(0,0)
        \put(0,0){\line(1,0){0.3}}
        \end{picture}}
   \end{picture}}

\put(9050,0){\setlength{\unitlength}{5cm}\begin{picture}(0,0)(0,-0.25)
   \put(0,0){\setlength{\unitlength}{1cm}\begin{picture}(0,0)(0,0)
        \put(0,0){\line(-1,0){0.3}}
        \end{picture}}
   \end{picture}}

\thinlines
\put(7350,0){\setlength{\unitlength}{5cm}\begin{picture}(0,0)(0,-0.25)
   \multiput(0,0)(0,0.1){3}{\setlength{\unitlength}{1cm}%
\begin{picture}(0,0)(0,0)
        \put(0,0){\line(1,0){0.12}}
        \end{picture}}
   \end{picture}}

\put(7350,0){\setlength{\unitlength}{5cm}\begin{picture}(0,0)(0,-0.25)
   \multiput(0,0)(0,-0.1){3}{\setlength{\unitlength}{1cm}%
\begin{picture}(0,0)(0,0)
        \put(0,0){\line(1,0){0.12}}
        \end{picture}}
   \end{picture}}

\put(9050,0){\setlength{\unitlength}{5cm}\begin{picture}(0,0)(0,-0.25)
   \multiput(0,0)(0,0.1){3}{\setlength{\unitlength}{1cm}%
\begin{picture}(0,0)(0,0)
        \put(0,0){\line(-1,0){0.12}}
        \end{picture}}
   \end{picture}}

\put(9050,0){\setlength{\unitlength}{5cm}\begin{picture}(0,0)(0,-0.25)
   \multiput(0,0)(0,-0.1){3}{\setlength{\unitlength}{1cm}%
\begin{picture}(0,0)(0,0)
        \put(0,0){\line(-1,0){0.12}}
        \end{picture}}
   \end{picture}}

   \put(7527.5, 236.111){\setlength{\unitlength}{1cm}\begin{picture}(0,0)(0,0)
        \put(0,0){\line(0,-1){0.2}}
   \end{picture}}
   \put(7892.5, 236.111){\setlength{\unitlength}{1cm}\begin{picture}(0,0)(0,0)
        \put(0,0){\line(0,-1){0.2}}
   \end{picture}}
   \put(8257.5, 236.111){\setlength{\unitlength}{1cm}\begin{picture}(0,0)(0,0)
        \put(0,0){\line(0,-1){0.2}}
   \end{picture}}
   \put(8622.5, 236.111){\setlength{\unitlength}{1cm}\begin{picture}(0,0)(0,0)
        \put(0,0){\line(0,-1){0.2}}
   \end{picture}}
   \put(8987.5, 236.111){\setlength{\unitlength}{1cm}\begin{picture}(0,0)(0,0)
        \put(0,0){\line(0,-1){0.2}}
   \end{picture}}
    \multiput(7350,0)(50,0){33}%
        {\setlength{\unitlength}{1cm}\begin{picture}(0,0)(0,0)
        \put(0,0){\line(0,1){0.12}}
    \end{picture}}
    \put(7500,0){\setlength{\unitlength}{1cm}\begin{picture}(0,0)(0,0)
        \put(0,0){\line(0,1){0.2}}
    \end{picture}}
    \put(7750,0){\setlength{\unitlength}{1cm}\begin{picture}(0,0)(0,0)
        \put(0,0){\line(0,1){0.2}}
    \end{picture}}
    \put(8000,0){\setlength{\unitlength}{1cm}\begin{picture}(0,0)(0,0)
        \put(0,0){\line(0,1){0.2}}
    \end{picture}}
    \put(8250,0){\setlength{\unitlength}{1cm}\begin{picture}(0,0)(0,0)
        \put(0,0){\line(0,1){0.2}}
    \end{picture}}
    \put(8500,0){\setlength{\unitlength}{1cm}\begin{picture}(0,0)(0,0)
        \put(0,0){\line(0,1){0.2}}
    \end{picture}}
    \put(8750,0){\setlength{\unitlength}{1cm}\begin{picture}(0,0)(0,0)
        \put(0,0){\line(0,1){0.2}}
    \end{picture}}
    \put(9000,0){\setlength{\unitlength}{1cm}\begin{picture}(0,0)(0,0)
        \put(0,0){\line(0,1){0.2}}
    \end{picture}}

\punkt{15/08/89}{01:36}{7753.567}{  0.011}{0.070}{ 500}{1.69}{
1133}{1.2CA}
\punkt{31/07/90}{03:15}{8103.636}{  0.030}{0.011}{ 500}{1.51}{
586}{1.2CA}
\punkt{21/10/90}{22:47}{8186.450}{  0.060}{0.047}{ 500}{3.53}{
881}{1.2CA}
\punkt{22/10/90}{22:47}{8187.450}{  0.067}{0.038}{ 500}{3.47}{
879}{1.2CA}
\punkt{04/08/91}{04:07}{8472.672}{ -0.010}{0.014}{ 500}{1.09}{
1026}{1.2CA}
\punkt{13/08/91}{02:38}{8481.610}{ -0.008}{0.024}{ 500}{1.68}{
518}{1.2CA}
\punkt{22/09/92}{00:56}{8887.539}{ -0.098}{0.041}{ 100}{2.47}{
419}{1.2CA}
\punkt{22/09/92}{00:56}{8887.539}{ -0.098}{0.041}{ 100}{2.47}{
419}{1.2CA}

\end{picture}}

\end{picture}

\vspace*{-0.02cm}

\begin{picture}(18 ,2.5 )(0,0)
\put(0,0){\setlength{\unitlength}{0.01059cm}%
\begin{picture}(1700, 236.111)(7350,0)
\put(7350,0){\framebox(1700, 236.111)[tl]{\begin{picture}(0,0)(0,0)
        \put(1700,0){\makebox(0,0)[tr]{\bf{0058+019}\T{0.4}
                                 \hspace*{0.5cm}}}
    \end{picture}}}

\thicklines
\put(7350,0){\setlength{\unitlength}{5cm}\begin{picture}(0,0)(0,-0.25)
   \put(0,0){\setlength{\unitlength}{1cm}\begin{picture}(0,0)(0,0)
        \put(0,0){\line(1,0){0.3}}
        \end{picture}}
   \end{picture}}

\put(9050,0){\setlength{\unitlength}{5cm}\begin{picture}(0,0)(0,-0.25)
   \put(0,0){\setlength{\unitlength}{1cm}\begin{picture}(0,0)(0,0)
        \put(0,0){\line(-1,0){0.3}}
        \end{picture}}
   \end{picture}}

\thinlines
\put(7350,0){\setlength{\unitlength}{5cm}\begin{picture}(0,0)(0,-0.25)
   \multiput(0,0)(0,0.1){3}{\setlength{\unitlength}{1cm}%
\begin{picture}(0,0)(0,0)
        \put(0,0){\line(1,0){0.12}}
        \end{picture}}
   \end{picture}}

\put(7350,0){\setlength{\unitlength}{5cm}\begin{picture}(0,0)(0,-0.25)
   \multiput(0,0)(0,-0.1){3}{\setlength{\unitlength}{1cm}%
\begin{picture}(0,0)(0,0)
        \put(0,0){\line(1,0){0.12}}
        \end{picture}}
   \end{picture}}

\put(9050,0){\setlength{\unitlength}{5cm}\begin{picture}(0,0)(0,-0.25)
   \multiput(0,0)(0,0.1){3}{\setlength{\unitlength}{1cm}%
\begin{picture}(0,0)(0,0)
        \put(0,0){\line(-1,0){0.12}}
        \end{picture}}
   \end{picture}}

\put(9050,0){\setlength{\unitlength}{5cm}\begin{picture}(0,0)(0,-0.25)
   \multiput(0,0)(0,-0.1){3}{\setlength{\unitlength}{1cm}%
\begin{picture}(0,0)(0,0)
        \put(0,0){\line(-1,0){0.12}}
        \end{picture}}
   \end{picture}}

   \put(7527.5, 236.111){\setlength{\unitlength}{1cm}\begin{picture}(0,0)(0,0)
        \put(0,0){\line(0,-1){0.2}}
   \end{picture}}
   \put(7892.5, 236.111){\setlength{\unitlength}{1cm}\begin{picture}(0,0)(0,0)
        \put(0,0){\line(0,-1){0.2}}
   \end{picture}}
   \put(8257.5, 236.111){\setlength{\unitlength}{1cm}\begin{picture}(0,0)(0,0)
        \put(0,0){\line(0,-1){0.2}}
   \end{picture}}
   \put(8622.5, 236.111){\setlength{\unitlength}{1cm}\begin{picture}(0,0)(0,0)
        \put(0,0){\line(0,-1){0.2}}
   \end{picture}}
   \put(8987.5, 236.111){\setlength{\unitlength}{1cm}\begin{picture}(0,0)(0,0)
        \put(0,0){\line(0,-1){0.2}}
   \end{picture}}
    \multiput(7350,0)(50,0){33}%
        {\setlength{\unitlength}{1cm}\begin{picture}(0,0)(0,0)
        \put(0,0){\line(0,1){0.12}}
    \end{picture}}
    \put(7500,0){\setlength{\unitlength}{1cm}\begin{picture}(0,0)(0,0)
        \put(0,0){\line(0,1){0.2}}
    \end{picture}}
    \put(7750,0){\setlength{\unitlength}{1cm}\begin{picture}(0,0)(0,0)
        \put(0,0){\line(0,1){0.2}}
    \end{picture}}
    \put(8000,0){\setlength{\unitlength}{1cm}\begin{picture}(0,0)(0,0)
        \put(0,0){\line(0,1){0.2}}
    \end{picture}}
    \put(8250,0){\setlength{\unitlength}{1cm}\begin{picture}(0,0)(0,0)
        \put(0,0){\line(0,1){0.2}}
    \end{picture}}
    \put(8500,0){\setlength{\unitlength}{1cm}\begin{picture}(0,0)(0,0)
        \put(0,0){\line(0,1){0.2}}
    \end{picture}}
    \put(8750,0){\setlength{\unitlength}{1cm}\begin{picture}(0,0)(0,0)
        \put(0,0){\line(0,1){0.2}}
    \end{picture}}
    \put(9000,0){\setlength{\unitlength}{1cm}\begin{picture}(0,0)(0,0)
        \put(0,0){\line(0,1){0.2}}
    \end{picture}}

\punkt{08/09/88}{01:15}{7412.552}{  0.012}{0.024}{ 500}{1.63}{
708}{1.2CA}
\punkt{09/09/88}{00:37}{7413.526}{  0.035}{0.042}{ 500}{1.91}{
641}{1.2CA}
\punkt{04/10/88}{23:33}{7439.482}{ -0.021}{0.024}{ 500}{2.39}{
548}{1.2CA}
\punkt{07/10/88}{22:53}{7442.454}{ -0.010}{0.013}{ 500}{1.94}{
570}{1.2CA}
\punkt{10/10/88}{21:55}{7445.414}{  0.034}{0.034}{ 500}{2.35}{
591}{1.2CA}
\punkt{15/10/88}{22:54}{7450.454}{ -0.047}{0.028}{ 500}{1.90}{
602}{1.2CA}
\punkt{14/08/89}{01:22}{7752.557}{  0.005}{0.013}{ 500}{1.84}{
826}{1.2CA}
\punkta{01/11/89}{21:36}{7832.400}{  0.065}{0.030}{ 500}{1.73}{
589}{1.2CA}
\punkt{31/07/90}{03:50}{8103.660}{  0.134}{0.015}{ 500}{1.35}{
614}{1.2CA}
\punkta{22/08/91}{03:10}{8490.632}{ -0.016}{0.038}{ 500}{1.67}{
728}{1.2CA}
\punkta{22/09/92}{00:50}{8887.535}{ -0.054}{0.029}{ 100}{1.44}{
391}{1.2CA}
\punkta{22/09/92}{00:53}{8887.537}{ -0.047}{0.013}{ 100}{1.79}{
391}{1.2CA}
\punkt{23/09/92}{23:36}{8889.483}{ -0.055}{0.013}{ 100}{1.11}{
392}{1.2CA}
\punkt{23/09/92}{23:39}{8889.486}{ -0.054}{0.014}{ 100}{1.13}{
401}{1.2CA}
\punkt{23/09/92}{23:39}{8889.486}{ -0.054}{0.014}{ 100}{1.13}{
401}{1.2CA}

\end{picture}}

\end{picture}

\vspace*{-0.02cm}

\begin{picture}(18 ,2.5 )(0,0)
\put(0,0){\setlength{\unitlength}{0.01059cm}%
\begin{picture}(1700, 236.072)(7350,0)
\put(7350,0){\framebox(1700, 236.072)[tl]{\begin{picture}(0,0)(0,0)
        \put(1700,-132){\makebox(0,0)[tr]{\normalsize{\bf 0104+318}\T{0.4}
                                 \hspace*{0.5cm}}}
    \end{picture}}}

\thicklines
\put(7350,0){\setlength{\unitlength}{5cm}\begin{picture}(0,0)(0,-0.25)
   \put(0,0){\setlength{\unitlength}{1cm}\begin{picture}(0,0)(0,0)
        \put(0,0){\line(1,0){0.3}}
        \end{picture}}
   \end{picture}}

\put(9050,0){\setlength{\unitlength}{5cm}\begin{picture}(0,0)(0,-0.25)
   \put(0,0){\setlength{\unitlength}{1cm}\begin{picture}(0,0)(0,0)
        \put(0,0){\line(-1,0){0.3}}
        \end{picture}}
   \end{picture}}

\thinlines
\put(7350,0){\setlength{\unitlength}{5cm}\begin{picture}(0,0)(0,-0.25)
   \multiput(0,0)(0,0.1){3}{\setlength{\unitlength}{1cm}%
\begin{picture}(0,0)(0,0)
        \put(0,0){\line(1,0){0.12}}
        \end{picture}}
   \end{picture}}

\put(7350,0){\setlength{\unitlength}{5cm}\begin{picture}(0,0)(0,-0.25)
   \multiput(0,0)(0,-0.1){3}{\setlength{\unitlength}{1cm}%
\begin{picture}(0,0)(0,0)
        \put(0,0){\line(1,0){0.12}}
        \end{picture}}
   \end{picture}}

\put(9050,0){\setlength{\unitlength}{5cm}\begin{picture}(0,0)(0,-0.25)
   \multiput(0,0)(0,0.1){3}{\setlength{\unitlength}{1cm}%
\begin{picture}(0,0)(0,0)
        \put(0,0){\line(-1,0){0.12}}
        \end{picture}}
   \end{picture}}

\put(9050,0){\setlength{\unitlength}{5cm}\begin{picture}(0,0)(0,-0.25)
   \multiput(0,0)(0,-0.1){3}{\setlength{\unitlength}{1cm}%
\begin{picture}(0,0)(0,0)
        \put(0,0){\line(-1,0){0.12}}
        \end{picture}}
   \end{picture}}

   \put(7527.5, 236.072){\setlength{\unitlength}{1cm}\begin{picture}(0,0)(0,0)
        \put(0,0){\line(0,-1){0.2}}
   \end{picture}}
   \put(7892.5, 236.072){\setlength{\unitlength}{1cm}\begin{picture}(0,0)(0,0)
        \put(0,0){\line(0,-1){0.2}}
   \end{picture}}
   \put(8257.5, 236.072){\setlength{\unitlength}{1cm}\begin{picture}(0,0)(0,0)
        \put(0,0){\line(0,-1){0.2}}
   \end{picture}}
   \put(8622.5, 236.072){\setlength{\unitlength}{1cm}\begin{picture}(0,0)(0,0)
        \put(0,0){\line(0,-1){0.2}}
   \end{picture}}
   \put(8987.5, 236.072){\setlength{\unitlength}{1cm}\begin{picture}(0,0)(0,0)
        \put(0,0){\line(0,-1){0.2}}
   \end{picture}}
    \multiput(7350,0)(50,0){33}%
        {\setlength{\unitlength}{1cm}\begin{picture}(0,0)(0,0)
        \put(0,0){\line(0,1){0.12}}
    \end{picture}}
    \put(7500,0){\setlength{\unitlength}{1cm}\begin{picture}(0,0)(0,0)
        \put(0,0){\line(0,1){0.2}}
    \end{picture}}
    \put(7750,0){\setlength{\unitlength}{1cm}\begin{picture}(0,0)(0,0)
        \put(0,0){\line(0,1){0.2}}
    \end{picture}}
    \put(8000,0){\setlength{\unitlength}{1cm}\begin{picture}(0,0)(0,0)
        \put(0,0){\line(0,1){0.2}}
    \end{picture}}
    \put(8250,0){\setlength{\unitlength}{1cm}\begin{picture}(0,0)(0,0)
        \put(0,0){\line(0,1){0.2}}
    \end{picture}}
    \put(8500,0){\setlength{\unitlength}{1cm}\begin{picture}(0,0)(0,0)
        \put(0,0){\line(0,1){0.2}}
    \end{picture}}
    \put(8750,0){\setlength{\unitlength}{1cm}\begin{picture}(0,0)(0,0)
        \put(0,0){\line(0,1){0.2}}
    \end{picture}}
    \put(9000,0){\setlength{\unitlength}{1cm}\begin{picture}(0,0)(0,0)
        \put(0,0){\line(0,1){0.2}}
    \end{picture}}

\punkt{17/08/89}{14:05}{7756.087}{  0.097}{0.057}{****}{1.78}{
600}{1.2CA}
\punkt{01/11/89}{21:51}{7832.411}{  0.067}{0.054}{ 500}{1.75}{
541}{1.2CA}
\punkt{27/09/90}{23:28}{8162.478}{  0.168}{0.070}{ 500}{1.64}{
1013}{1.2CA}
\punkt{12/08/91}{02:57}{8480.623}{ -0.078}{0.069}{ 500}{1.64}{
496}{1.2CA}
\punkt{20/09/91}{01:00}{8519.542}{ -0.091}{0.048}{ 100}{0.87}{
441}{1.2CA}
\punkt{23/09/92}{01:55}{8888.580}{  0.103}{0.028}{ 150}{1.23}{
529}{1.2CA}
\punkt{23/09/92}{01:59}{8888.583}{  0.102}{0.028}{ 100}{1.19}{
475}{1.2CA}

\end{picture}}

\end{picture}

\vspace*{-0.02cm}

\begin{picture}(18 ,4.5 )(0,0)
\put(0,0){\setlength{\unitlength}{0.01059cm}%
\begin{picture}(1700, 424.999)(7350,0)
\put(7350,0){\framebox(1700, 424.999)[tl]{\begin{picture}(0,0)(0,0)
        \put(1700,0){\makebox(0,0)[tr]{\bf{0151+045}\T{0.4}
                                 \hspace*{0.5cm}}}
    \end{picture}}}

\thicklines
\put(7350,0){\setlength{\unitlength}{5cm}\begin{picture}(0,0)(0,-0.45)
   \put(0,0){\setlength{\unitlength}{1cm}\begin{picture}(0,0)(0,0)
        \put(0,0){\line(1,0){0.3}}
        \end{picture}}
   \end{picture}}

\put(9050,0){\setlength{\unitlength}{5cm}\begin{picture}(0,0)(0,-0.45)
   \put(0,0){\setlength{\unitlength}{1cm}\begin{picture}(0,0)(0,0)
        \put(0,0){\line(-1,0){0.3}}
        \end{picture}}
   \end{picture}}

\thinlines
\put(7350,0){\setlength{\unitlength}{5cm}\begin{picture}(0,0)(0,-0.45)
   \multiput(0,0)(0,0.1){5}{\setlength{\unitlength}{1cm}%
\begin{picture}(0,0)(0,0)
        \put(0,0){\line(1,0){0.12}}
        \end{picture}}
   \end{picture}}

\put(7350,0){\setlength{\unitlength}{5cm}\begin{picture}(0,0)(0,-0.45)
   \multiput(0,0)(0,-0.1){5}{\setlength{\unitlength}{1cm}%
\begin{picture}(0,0)(0,0)
        \put(0,0){\line(1,0){0.12}}
        \end{picture}}
   \end{picture}}

\put(9050,0){\setlength{\unitlength}{5cm}\begin{picture}(0,0)(0,-0.45)
   \multiput(0,0)(0,0.1){5}{\setlength{\unitlength}{1cm}%
\begin{picture}(0,0)(0,0)
        \put(0,0){\line(-1,0){0.12}}
        \end{picture}}
   \end{picture}}

\put(9050,0){\setlength{\unitlength}{5cm}\begin{picture}(0,0)(0,-0.45)
   \multiput(0,0)(0,-0.1){5}{\setlength{\unitlength}{1cm}%
\begin{picture}(0,0)(0,0)
        \put(0,0){\line(-1,0){0.12}}
        \end{picture}}
   \end{picture}}

   \put(7527.5, 424.999){\setlength{\unitlength}{1cm}\begin{picture}(0,0)(0,0)
        \put(0,0){\line(0,-1){0.2}}
   \end{picture}}
   \put(7892.5, 424.999){\setlength{\unitlength}{1cm}\begin{picture}(0,0)(0,0)
        \put(0,0){\line(0,-1){0.2}}
   \end{picture}}
   \put(8257.5, 424.999){\setlength{\unitlength}{1cm}\begin{picture}(0,0)(0,0)
       \put(0,0){\line(0,-1){0.2}}
   \end{picture}}
   \put(8622.5, 424.999){\setlength{\unitlength}{1cm}\begin{picture}(0,0)(0,0)
        \put(0,0){\line(0,-1){0.2}}
   \end{picture}}
   \put(8987.5, 424.999){\setlength{\unitlength}{1cm}\begin{picture}(0,0)(0,0)
        \put(0,0){\line(0,-1){0.2}}
   \end{picture}}
    \multiput(7350,0)(50,0){33}%
        {\setlength{\unitlength}{1cm}\begin{picture}(0,0)(0,0)
        \put(0,0){\line(0,1){0.12}}
    \end{picture}}
    \put(7500,0){\setlength{\unitlength}{1cm}\begin{picture}(0,0)(0,0)
        \put(0,0){\line(0,1){0.2}}
    \end{picture}}
    \put(7750,0){\setlength{\unitlength}{1cm}\begin{picture}(0,0)(0,0)
        \put(0,0){\line(0,1){0.2}}
    \end{picture}}
    \put(8000,0){\setlength{\unitlength}{1cm}\begin{picture}(0,0)(0,0)
        \put(0,0){\line(0,1){0.2}}
    \end{picture}}
    \put(8250,0){\setlength{\unitlength}{1cm}\begin{picture}(0,0)(0,0)
        \put(0,0){\line(0,1){0.2}}
    \end{picture}}
    \put(8500,0){\setlength{\unitlength}{1cm}\begin{picture}(0,0)(0,0)
        \put(0,0){\line(0,1){0.2}}
   \end{picture}}
    \put(8750,0){\setlength{\unitlength}{1cm}\begin{picture}(0,0)(0,0)
        \put(0,0){\line(0,1){0.2}}
    \end{picture}}
    \put(9000,0){\setlength{\unitlength}{1cm}\begin{picture}(0,0)(0,0)
        \put(0,0){\line(0,1){0.2}}
    \end{picture}}

\put(7000,0){\begin{picture}(0,0)(7000,-94.4287)
\punkt{09/09/88}{01:00}{7413.542}{ -0.098}{0.044}{ 500}{2.01}{
630}{1.2CA}
\punkt{04/10/88}{00:53}{7438.537}{ -0.226}{0.056}{ 500}{2.29}{
768}{1.2CA}
\punkt{07/10/88}{23:46}{7442.490}{ -0.107}{0.017}{ 500}{1.64}{
561}{1.2CA}
\punkt{16/10/88}{10:56}{7450.956}{ -0.031}{0.077}{****}{1.92}{
455}{1.2CA}
\punkt{17/10/88}{00:17}{7451.512}{ -0.142}{0.027}{ 500}{2.05}{
570}{1.2CA}
\punkt{16/08/89}{02:38}{7754.610}{ -0.289}{0.015}{ 500}{1.55}{
1436}{1.2CA}
\punkt{03/09/89}{03:16}{7772.637}{ -0.361}{0.016}{ 500}{2.04}{
1114}{1.2CA}
\punkt{02/11/89}{23:08}{7833.464}{ -0.258}{0.073}{ 500}{2.34}{
541}{1.2CA}
\punkt{15/12/89}{21:19}{7876.389}{ -0.188}{0.062}{ 500}{2.37}{
747}{1.2CA}
\punkt{25/09/90}{02:28}{8159.603}{  0.076}{0.014}{ 500}{1.94}{
851}{1.2CA}
\punkt{26/09/90}{02:28}{8160.603}{  0.064}{0.010}{ 500}{1.93}{
854}{1.2CA}
\punkt{26/09/90}{23:44}{8161.489}{ -0.044}{0.019}{ 500}{1.30}{
722}{1.2CA}
\punkt{27/09/90}{11:41}{8161.987}{  0.022}{0.008}{ 500}{1.34}{
800}{1.2CA}
\punkt{27/09/90}{23:41}{8162.487}{  0.015}{0.007}{ 500}{1.34}{
800}{1.2CA}
\punkt{17/10/90}{02:21}{8181.598}{  0.057}{0.012}{ 500}{2.30}{
821}{1.2CA}
\punkt{18/10/90}{02:21}{8182.598}{  0.065}{0.011}{ 500}{2.28}{
821}{1.2CA}
\punkt{20/10/90}{00:12}{8184.509}{  0.073}{0.011}{ 500}{1.48}{
836}{1.2CA}
\punkt{20/10/90}{01:26}{8184.560}{  0.087}{0.017}{ 500}{2.34}{
875}{1.2CA}
\punkt{24/08/91}{02:31}{8492.605}{  0.187}{0.039}{ 500}{1.55}{
1701}{1.2CA}
\punkt{19/09/91}{03:54}{8518.663}{  0.208}{0.029}{ 500}{1.63}{
504}{1.2CA}
\punkt{19/10/91}{02:15}{8548.594}{  0.255}{0.022}{ 100}{2.13}{
326}{1.2CA}
\punkt{29/10/91}{01:09}{8558.549}{  0.168}{0.030}{ 100}{1.15}{
380}{1.2CA}
\punkt{21/09/92}{02:38}{8886.610}{  0.062}{0.008}{ 100}{0.81}{
511}{1.2CA}
\punkt{21/09/92}{02:41}{8886.612}{  0.067}{0.010}{ 100}{0.87}{
516}{1.2CA}
\punkt{21/09/92}{02:41}{8886.612}{  0.067}{0.010}{ 100}{0.87}{
516}{1.2CA}

\end{picture}}
\end{picture}}

\end{picture}

\vspace*{-0.02cm}

\begin{picture}(18 ,2.5 )(0,0)
\put(0,0){\setlength{\unitlength}{0.01059cm}%
\begin{picture}(1700, 236.111)(7350,0)
\put(7350,0){\framebox(1700, 236.111)[tl]{\begin{picture}(0,0)(0,0)
        \put(1700,0){\makebox(0,0)[tr]{\bf{0731+653}\T{0.4}
                                 \hspace*{0.5cm}}}
      \put(1700,-
236.111){\setlength{\unitlength}{1cm}\begin{picture}(0,0)(0,0)
            \put(0,-1){\makebox(0,0)[br]{\bf J.D.\,2,440,000\,+}}
        \end{picture}}
    \end{picture}}}

\thicklines
\put(7350,0){\setlength{\unitlength}{5cm}\begin{picture}(0,0)(0,-0.25)
   \put(0,0){\setlength{\unitlength}{1cm}\begin{picture}(0,0)(0,0)
        \put(0,0){\line(1,0){0.3}}
        \end{picture}}
   \end{picture}}

\put(9050,0){\setlength{\unitlength}{5cm}\begin{picture}(0,0)(0,-0.25)
   \put(0,0){\setlength{\unitlength}{1cm}\begin{picture}(0,0)(0,0)
        \put(0,0){\line(-1,0){0.3}}
        \end{picture}}
   \end{picture}}

\thinlines
\put(7350,0){\setlength{\unitlength}{5cm}\begin{picture}(0,0)(0,-0.25)
   \multiput(0,0)(0,0.1){3}{\setlength{\unitlength}{1cm}%
\begin{picture}(0,0)(0,0)
        \put(0,0){\line(1,0){0.12}}
        \end{picture}}
   \end{picture}}

\put(7350,0){\setlength{\unitlength}{5cm}\begin{picture}(0,0)(0,-0.25)
   \multiput(0,0)(0,-0.1){3}{\setlength{\unitlength}{1cm}%
\begin{picture}(0,0)(0,0)
        \put(0,0){\line(1,0){0.12}}
        \end{picture}}
   \end{picture}}

\put(9050,0){\setlength{\unitlength}{5cm}\begin{picture}(0,0)(0,-0.25)
   \multiput(0,0)(0,0.1){3}{\setlength{\unitlength}{1cm}%
\begin{picture}(0,0)(0,0)
        \put(0,0){\line(-1,0){0.12}}
        \end{picture}}
   \end{picture}}

\put(9050,0){\setlength{\unitlength}{5cm}\begin{picture}(0,0)(0,-0.25)
   \multiput(0,0)(0,-0.1){3}{\setlength{\unitlength}{1cm}%
\begin{picture}(0,0)(0,0)
        \put(0,0){\line(-1,0){0.12}}
        \end{picture}}
   \end{picture}}

   \put(7527.5, 236.111){\setlength{\unitlength}{1cm}\begin{picture}(0,0)(0,0)
        \put(0,0){\line(0,-1){0.2}}
   \end{picture}}
   \put(7892.5, 236.111){\setlength{\unitlength}{1cm}\begin{picture}(0,0)(0,0)
        \put(0,0){\line(0,-1){0.2}}
   \end{picture}}
   \put(8257.5, 236.111){\setlength{\unitlength}{1cm}\begin{picture}(0,0)(0,0)
        \put(0,0){\line(0,-1){0.2}}
   \end{picture}}
   \put(8622.5, 236.111){\setlength{\unitlength}{1cm}\begin{picture}(0,0)(0,0)
        \put(0,0){\line(0,-1){0.2}}
   \end{picture}}
   \put(8987.5, 236.111){\setlength{\unitlength}{1cm}\begin{picture}(0,0)(0,0)
        \put(0,0){\line(0,-1){0.2}}
   \end{picture}}
    \multiput(7350,0)(50,0){33}%
        {\setlength{\unitlength}{1cm}\begin{picture}(0,0)(0,0)
        \put(0,0){\line(0,1){0.12}}
    \end{picture}}
    \put(7500,0){\setlength{\unitlength}{1cm}\begin{picture}(0,0)(0,0)
        \put(0,0){\line(0,1){0.2}}
        \put(0,-0.2){\makebox(0,0)[t]{\bf 7500}}
    \end{picture}}
    \put(7750,0){\setlength{\unitlength}{1cm}\begin{picture}(0,0)(0,0)
        \put(0,0){\line(0,1){0.2}}
        \put(0,-0.2){\makebox(0,0)[t]{\bf 7750}}
    \end{picture}}
    \put(8000,0){\setlength{\unitlength}{1cm}\begin{picture}(0,0)(0,0)
        \put(0,0){\line(0,1){0.2}}
        \put(0,-0.2){\makebox(0,0)[t]{\bf 8000}}
    \end{picture}}
    \put(8250,0){\setlength{\unitlength}{1cm}\begin{picture}(0,0)(0,0)
        \put(0,0){\line(0,1){0.2}}
        \put(0,-0.2){\makebox(0,0)[t]{\bf 8250}}
    \end{picture}}
    \put(8500,0){\setlength{\unitlength}{1cm}\begin{picture}(0,0)(0,0)
        \put(0,0){\line(0,1){0.2}}
        \put(0,-0.2){\makebox(0,0)[t]{\bf 8500}}
    \end{picture}}
    \put(8750,0){\setlength{\unitlength}{1cm}\begin{picture}(0,0)(0,0)
        \put(0,0){\line(0,1){0.2}}
        \put(0,-0.2){\makebox(0,0)[t]{\bf 8750}}
    \end{picture}}
    \put(9000,0){\setlength{\unitlength}{1cm}\begin{picture}(0,0)(0,0)
        \put(0,0){\line(0,1){0.2}}
        \put(0,-0.2){\makebox(0,0)[t]{\bf 9000}}
    \end{picture}}

\punkt{01/11/89}{01:52}{7831.578}{ -0.087}{0.023}{ 500}{1.78}{
534}{1.2CA}
\punkt{02/11/89}{01:55}{7832.580}{ -0.035}{0.043}{ 500}{1.34}{
549}{1.2CA}
\punkta{03/11/89}{02:52}{7833.620}{ -0.070}{0.098}{ 500}{2.24}{
549}{1.2CA}
\punkt{20/12/89}{01:03}{7880.544}{ -0.052}{0.028}{ 250}{1.77}{
542}{1.2CA}
\punkt{20/12/89}{01:10}{7880.549}{ -0.036}{0.027}{ 250}{1.65}{
571}{1.2CA}
\punkta{26/01/90}{21:49}{7918.409}{ -0.003}{0.019}{ 500}{3.35}{
2055}{1.2CA}
\punkt{27/02/90}{20:15}{7950.344}{ -0.025}{0.017}{ 500}{1.29}{
678}{1.2CA}
\punkta{28/09/90}{04:15}{8162.678}{  0.115}{0.050}{ 150}{1.66}{
596}{1.2CA}
\punkta{18/10/90}{04:59}{8182.708}{ -0.140}{0.050}{ 436}{2.46}{
783}{1.2CA}
\punkta{22/12/90}{11:43}{8247.989}{ -0.086}{0.050}{ 500}{3.79}{
622}{1.2CA}
\punkt{09/02/91}{21:57}{8297.415}{ -0.041}{0.023}{ 300}{1.72}{
450}{1.2CA}
\punkta{11/02/91}{23:14}{8299.468}{ -0.058}{0.069}{ 500}{3.47}{
544}{1.2CA}
\punkta{14/02/91}{23:46}{8302.491}{ -0.056}{0.067}{ 500}{2.76}{
559}{1.2CA}
\punkt{25/02/91}{01:53}{8312.579}{  0.003}{0.006}{ 500}{2.22}{
1243}{1.2CA}
\punkt{07/02/92}{22:01}{8660.418}{  0.104}{0.012}{ 500}{1.45}{
496}{1.2CA}
\punkt{08/02/92}{23:04}{8661.462}{  0.102}{0.011}{ 500}{1.37}{
487}{1.2CA}
\punkt{10/02/92}{21:17}{8663.387}{  0.128}{0.009}{ 500}{1.01}{
685}{1.2CA}
\punkta{14/02/92}{20:55}{8667.372}{  0.165}{0.073}{ 500}{1.52}{
2858}{1.2CA}

\end{picture}}

\end{picture}

\vspace*{1cm}

\caption{HQM-lightcurves in the $R\/$-band. Dashes on the vertical axes
represent 0.1\,mag steps. Plotted are variations $\Delta R$=$R_0$$-$$R$;
the reference magnitude $R_0$ is indicated by thick dashes. Measurements
obtained under bad atmospheric conditions or those with only one reference
star are shown by open circles; reliable error bars can in these cases not
be given}
\end{figure*}
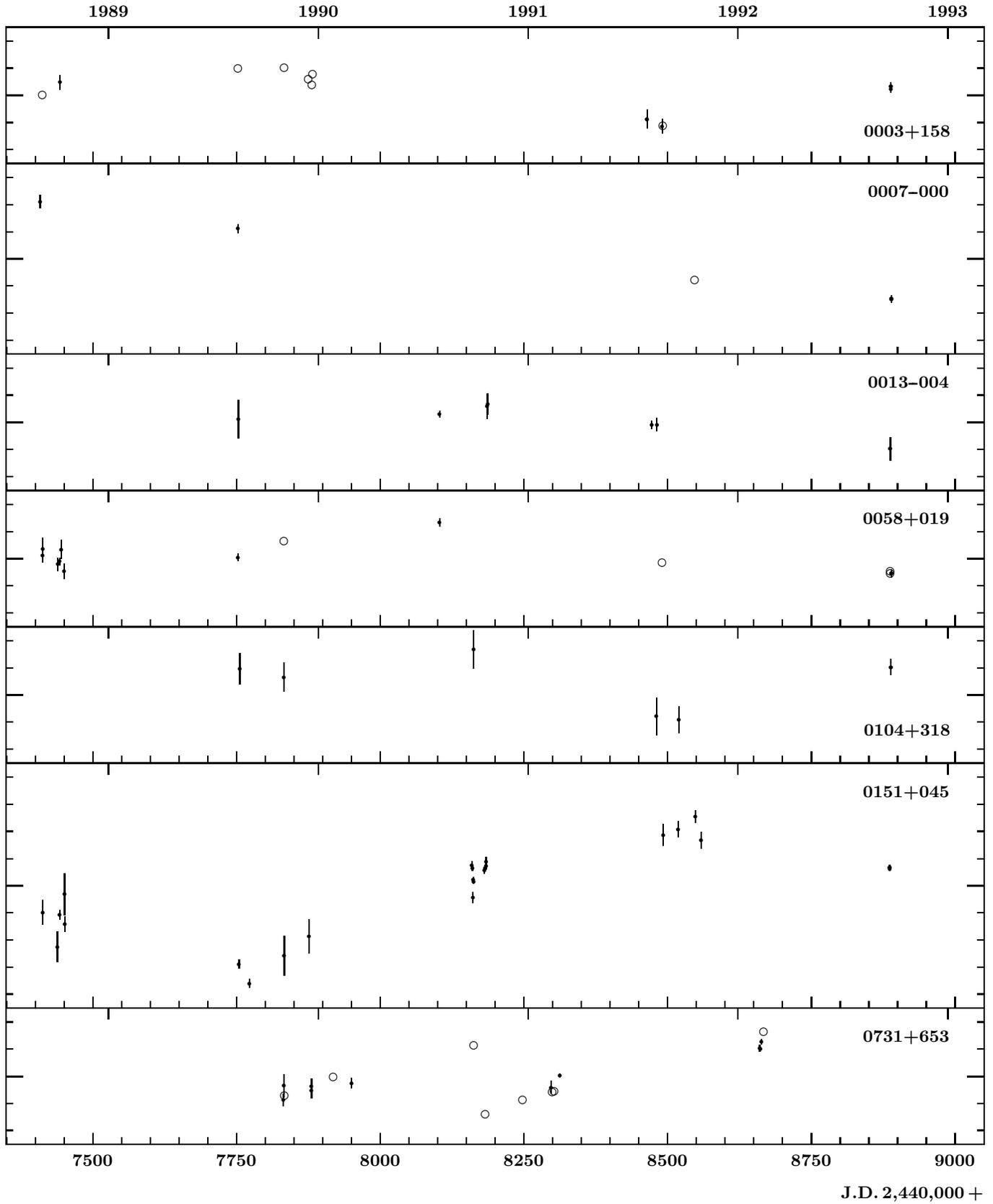

\begin{figure*}

\vspace*{0.4cm}

\begin{picture}(18 ,3.5 )(0,0)
\put(0,0){\setlength{\unitlength}{0.01059cm}%
\begin{picture}(1700, 330.555)(7350,0)
\put(7350,0){\framebox(1700, 330.555)[tl]{\begin{picture}(0,0)(0,0)
        \put(1700,0){\makebox(0,0)[tr]{\bf{0745+557}\T{0.4}
                                 \hspace*{0.5cm}}}
    \end{picture}}}

\thicklines
\put(7350,0){\setlength{\unitlength}{5cm}\begin{picture}(0,0)(0,-0.35)
   \put(0,0){\setlength{\unitlength}{1cm}\begin{picture}(0,0)(0,0)
        \put(0,0){\line(1,0){0.3}}
        \end{picture}}
   \end{picture}}

\put(9050,0){\setlength{\unitlength}{5cm}\begin{picture}(0,0)(0,-0.35)
   \put(0,0){\setlength{\unitlength}{1cm}\begin{picture}(0,0)(0,0)
        \put(0,0){\line(-1,0){0.3}}
        \end{picture}}
   \end{picture}}

\thinlines
\put(7350,0){\setlength{\unitlength}{5cm}\begin{picture}(0,0)(0,-0.35)
   \multiput(0,0)(0,0.1){4}{\setlength{\unitlength}{1cm}%
\begin{picture}(0,0)(0,0)
        \put(0,0){\line(1,0){0.12}}
        \end{picture}}
   \end{picture}}

\put(7350,0){\setlength{\unitlength}{5cm}\begin{picture}(0,0)(0,-0.35)
   \multiput(0,0)(0,-0.1){4}{\setlength{\unitlength}{1cm}%
\begin{picture}(0,0)(0,0)
        \put(0,0){\line(1,0){0.12}}
        \end{picture}}
   \end{picture}}

\put(9050,0){\setlength{\unitlength}{5cm}\begin{picture}(0,0)(0,-0.35)
   \multiput(0,0)(0,0.1){4}{\setlength{\unitlength}{1cm}%
\begin{picture}(0,0)(0,0)
        \put(0,0){\line(-1,0){0.12}}
        \end{picture}}
   \end{picture}}

\put(9050,0){\setlength{\unitlength}{5cm}\begin{picture}(0,0)(0,-0.35)
   \multiput(0,0)(0,-0.1){4}{\setlength{\unitlength}{1cm}%
\begin{picture}(0,0)(0,0)
        \put(0,0){\line(-1,0){0.12}}
        \end{picture}}
   \end{picture}}

   \put(7527.5, 330.555){\setlength{\unitlength}{1cm}\begin{picture}(0,0)(0,0)
        \put(0,0){\line(0,-1){0.2}}
        \put(0,0.2){\makebox(0,0)[b]{\bf 1989}}
   \end{picture}}
   \put(7892.5, 330.555){\setlength{\unitlength}{1cm}\begin{picture}(0,0)(0,0)
        \put(0,0){\line(0,-1){0.2}}
        \put(0,0.2){\makebox(0,0)[b]{\bf 1990}}
   \end{picture}}
   \put(8257.5, 330.555){\setlength{\unitlength}{1cm}\begin{picture}(0,0)(0,0)
        \put(0,0){\line(0,-1){0.2}}
        \put(0,0.2){\makebox(0,0)[b]{\bf 1991}}
   \end{picture}}
   \put(8622.5, 330.555){\setlength{\unitlength}{1cm}\begin{picture}(0,0)(0,0)
        \put(0,0){\line(0,-1){0.2}}
        \put(0,0.2){\makebox(0,0)[b]{\bf 1992}}
   \end{picture}}
   \put(8987.5, 330.555){\setlength{\unitlength}{1cm}\begin{picture}(0,0)(0,0)
        \put(0,0){\line(0,-1){0.2}}
        \put(0,0.2){\makebox(0,0)[b]{\bf 1993}}
   \end{picture}}
    \multiput(7350,0)(50,0){33}%
        {\setlength{\unitlength}{1cm}\begin{picture}(0,0)(0,0)
        \put(0,0){\line(0,1){0.12}}
    \end{picture}}
    \put(7500,0){\setlength{\unitlength}{1cm}\begin{picture}(0,0)(0,0)
        \put(0,0){\line(0,1){0.2}}
    \end{picture}}
    \put(7750,0){\setlength{\unitlength}{1cm}\begin{picture}(0,0)(0,0)
        \put(0,0){\line(0,1){0.2}}
    \end{picture}}
    \put(8000,0){\setlength{\unitlength}{1cm}\begin{picture}(0,0)(0,0)
        \put(0,0){\line(0,1){0.2}}
    \end{picture}}
    \put(8250,0){\setlength{\unitlength}{1cm}\begin{picture}(0,0)(0,0)
        \put(0,0){\line(0,1){0.2}}
    \end{picture}}
    \put(8500,0){\setlength{\unitlength}{1cm}\begin{picture}(0,0)(0,0)
        \put(0,0){\line(0,1){0.2}}
    \end{picture}}
    \put(8750,0){\setlength{\unitlength}{1cm}\begin{picture}(0,0)(0,0)
        \put(0,0){\line(0,1){0.2}}
    \end{picture}}
    \put(9000,0){\setlength{\unitlength}{1cm}\begin{picture}(0,0)(0,0)
        \put(0,0){\line(0,1){0.2}}
    \end{picture}}

\put(7000,0){\begin{picture}(0,0)(7000,-47.21435)
\punkt{03/11/89}{03:07}{7833.630}{  0.130}{0.072}{ 500}{2.18}{
536}{1.2CA}
\punkt{20/12/89}{00:45}{7880.532}{  0.090}{0.029}{ 250}{1.74}{
539}{1.2CA}
\punkt{21/12/89}{02:12}{7881.592}{  0.073}{0.025}{ 500}{1.78}{
664}{1.2CA}
\punkta{06/01/90}{23:31}{7898.480}{  0.047}{0.078}{ 500}{1.92}{
991}{1.2CA}
\punkta{26/02/90}{22:08}{7949.423}{ -0.018}{0.023}{ 500}{1.80}{
623}{1.2CA}
\punkt{22/02/91}{22:12}{8310.425}{  0.077}{0.035}{ 500}{1.76}{
1003}{1.2CA}
\punkta{25/05/91}{20:38}{8402.360}{  0.074}{0.411}{ 300}{3.41}{
1183}{1.2CA}
\punkt{09/02/92}{22:38}{8662.443}{ -0.155}{0.017}{ 500}{1.21}{
494}{1.2CA}
\punkt{11/02/92}{21:11}{8664.383}{ -0.171}{0.014}{ 500}{1.12}{
828}{1.2CA}
\end{picture}}

\end{picture}}

\end{picture}

\vspace*{-0.02cm}

\begin{picture}(18 ,2.5 )(0,0)
\put(0,0){\setlength{\unitlength}{0.01059cm}%
\begin{picture}(1700, 236.111)(7350,0)
\put(7350,0){\framebox(1700, 236.111)[tl]{\begin{picture}(0,0)(0,0)
        \put(1700,0){\makebox(0,0)[tr]{\bf{0805+046}\T{0.4}
                                 \hspace*{0.5cm}}}
    \end{picture}}}

\thicklines
\put(7350,0){\setlength{\unitlength}{5cm}\begin{picture}(0,0)(0,-0.25)
   \put(0,0){\setlength{\unitlength}{1cm}\begin{picture}(0,0)(0,0)
        \put(0,0){\line(1,0){0.3}}
        \end{picture}}
   \end{picture}}

\put(9050,0){\setlength{\unitlength}{5cm}\begin{picture}(0,0)(0,-0.25)
   \put(0,0){\setlength{\unitlength}{1cm}\begin{picture}(0,0)(0,0)
        \put(0,0){\line(-1,0){0.3}}
        \end{picture}}
   \end{picture}}

\thinlines
\put(7350,0){\setlength{\unitlength}{5cm}\begin{picture}(0,0)(0,-0.25)
   \multiput(0,0)(0,0.1){3}{\setlength{\unitlength}{1cm}%
\begin{picture}(0,0)(0,0)
        \put(0,0){\line(1,0){0.12}}
        \end{picture}}
   \end{picture}}

\put(7350,0){\setlength{\unitlength}{5cm}\begin{picture}(0,0)(0,-0.25)
   \multiput(0,0)(0,-0.1){3}{\setlength{\unitlength}{1cm}%
\begin{picture}(0,0)(0,0)
        \put(0,0){\line(1,0){0.12}}
        \end{picture}}
   \end{picture}}

\put(9050,0){\setlength{\unitlength}{5cm}\begin{picture}(0,0)(0,-0.25)
   \multiput(0,0)(0,0.1){3}{\setlength{\unitlength}{1cm}%
\begin{picture}(0,0)(0,0)
        \put(0,0){\line(-1,0){0.12}}
        \end{picture}}
   \end{picture}}

\put(9050,0){\setlength{\unitlength}{5cm}\begin{picture}(0,0)(0,-0.25)
   \multiput(0,0)(0,-0.1){3}{\setlength{\unitlength}{1cm}%
\begin{picture}(0,0)(0,0)
        \put(0,0){\line(-1,0){0.12}}
        \end{picture}}
   \end{picture}}

   \put(7527.5, 236.111){\setlength{\unitlength}{1cm}\begin{picture}(0,0)(0,0)
        \put(0,0){\line(0,-1){0.2}}
   \end{picture}}
   \put(7892.5, 236.111){\setlength{\unitlength}{1cm}\begin{picture}(0,0)(0,0)
        \put(0,0){\line(0,-1){0.2}}
   \end{picture}}
   \put(8257.5, 236.111){\setlength{\unitlength}{1cm}\begin{picture}(0,0)(0,0)
        \put(0,0){\line(0,-1){0.2}}
   \end{picture}}
   \put(8622.5, 236.111){\setlength{\unitlength}{1cm}\begin{picture}(0,0)(0,0)
        \put(0,0){\line(0,-1){0.2}}
   \end{picture}}
   \put(8987.5, 236.111){\setlength{\unitlength}{1cm}\begin{picture}(0,0)(0,0)
        \put(0,0){\line(0,-1){0.2}}
   \end{picture}}
    \multiput(7350,0)(50,0){33}%
        {\setlength{\unitlength}{1cm}\begin{picture}(0,0)(0,0)
        \put(0,0){\line(0,1){0.12}}
    \end{picture}}
    \put(7500,0){\setlength{\unitlength}{1cm}\begin{picture}(0,0)(0,0)
        \put(0,0){\line(0,1){0.2}}
    \end{picture}}
    \put(7750,0){\setlength{\unitlength}{1cm}\begin{picture}(0,0)(0,0)
        \put(0,0){\line(0,1){0.2}}
    \end{picture}}
    \put(8000,0){\setlength{\unitlength}{1cm}\begin{picture}(0,0)(0,0)
        \put(0,0){\line(0,1){0.2}}
    \end{picture}}
    \put(8250,0){\setlength{\unitlength}{1cm}\begin{picture}(0,0)(0,0)
        \put(0,0){\line(0,1){0.2}}
    \end{picture}}
    \put(8500,0){\setlength{\unitlength}{1cm}\begin{picture}(0,0)(0,0)
        \put(0,0){\line(0,1){0.2}}
    \end{picture}}
    \put(8750,0){\setlength{\unitlength}{1cm}\begin{picture}(0,0)(0,0)
        \put(0,0){\line(0,1){0.2}}
    \end{picture}}
    \put(9000,0){\setlength{\unitlength}{1cm}\begin{picture}(0,0)(0,0)
        \put(0,0){\line(0,1){0.2}}
    \end{picture}}

\punkt{02/11/89}{05:21}{7832.723}{ -0.127}{0.060}{ 500}{1.66}{
592}{1.2CA}
\punkt{26/02/90}{22:35}{7949.441}{ -0.168}{0.037}{ 500}{2.02}{
640}{1.2CA}
\punkt{23/02/91}{21:31}{8311.397}{  0.068}{0.020}{ 500}{1.16}{
1167}{1.2CA}
\punkt{13/02/92}{21:09}{8666.381}{  0.063}{0.064}{ 500}{2.41}{
2030}{1.2CA}
\punkt{24/03/92}{23:44}{8706.489}{  0.061}{0.039}{ 300}{1.56}{
1325}{1.2CA}
\punkt{24/03/92}{23:44}{8706.489}{  0.061}{0.039}{ 300}{1.56}{
1325}{1.2CA}

\end{picture}}

\end{picture}

\vspace*{-0.02cm}

\begin{picture}(18 ,2.5 )(0,0)
\put(0,0){\setlength{\unitlength}{0.01059cm}%
\begin{picture}(1700, 236.111)(7350,0)
\put(7350,0){\framebox(1700, 236.111)[tl]{\begin{picture}(0,0)(0,0)
        \put(1700,0){\makebox(0,0)[tr]{\bf{0809+483}\T{0.4}
                                 \hspace*{0.5cm}}}
    \end{picture}}}

\thicklines
\put(7350,0){\setlength{\unitlength}{5cm}\begin{picture}(0,0)(0,-0.25)
   \put(0,0){\setlength{\unitlength}{1cm}\begin{picture}(0,0)(0,0)
        \put(0,0){\line(1,0){0.3}}
        \end{picture}}
   \end{picture}}

\put(9050,0){\setlength{\unitlength}{5cm}\begin{picture}(0,0)(0,-0.25)
   \put(0,0){\setlength{\unitlength}{1cm}\begin{picture}(0,0)(0,0)
        \put(0,0){\line(-1,0){0.3}}
        \end{picture}}
   \end{picture}}

\thinlines
\put(7350,0){\setlength{\unitlength}{5cm}\begin{picture}(0,0)(0,-0.25)
   \multiput(0,0)(0,0.1){3}{\setlength{\unitlength}{1cm}%
\begin{picture}(0,0)(0,0)
        \put(0,0){\line(1,0){0.12}}
        \end{picture}}
   \end{picture}}

\put(7350,0){\setlength{\unitlength}{5cm}\begin{picture}(0,0)(0,-0.25)
   \multiput(0,0)(0,-0.1){3}{\setlength{\unitlength}{1cm}%
\begin{picture}(0,0)(0,0)
        \put(0,0){\line(1,0){0.12}}
        \end{picture}}
   \end{picture}}

\put(9050,0){\setlength{\unitlength}{5cm}\begin{picture}(0,0)(0,-0.25)
   \multiput(0,0)(0,0.1){3}{\setlength{\unitlength}{1cm}%
\begin{picture}(0,0)(0,0)
        \put(0,0){\line(-1,0){0.12}}
        \end{picture}}
   \end{picture}}

\put(9050,0){\setlength{\unitlength}{5cm}\begin{picture}(0,0)(0,-0.25)
   \multiput(0,0)(0,-0.1){3}{\setlength{\unitlength}{1cm}%
\begin{picture}(0,0)(0,0)
        \put(0,0){\line(-1,0){0.12}}
        \end{picture}}
   \end{picture}}

   \put(7527.5, 236.111){\setlength{\unitlength}{1cm}\begin{picture}(0,0)(0,0)
        \put(0,0){\line(0,-1){0.2}}
   \end{picture}}
   \put(7892.5, 236.111){\setlength{\unitlength}{1cm}\begin{picture}(0,0)(0,0)
        \put(0,0){\line(0,-1){0.2}}
   \end{picture}}
   \put(8257.5, 236.111){\setlength{\unitlength}{1cm}\begin{picture}(0,0)(0,0)
        \put(0,0){\line(0,-1){0.2}}
   \end{picture}}
   \put(8622.5, 236.111){\setlength{\unitlength}{1cm}\begin{picture}(0,0)(0,0)
        \put(0,0){\line(0,-1){0.2}}
   \end{picture}}
   \put(8987.5, 236.111){\setlength{\unitlength}{1cm}\begin{picture}(0,0)(0,0)
        \put(0,0){\line(0,-1){0.2}}
   \end{picture}}
    \multiput(7350,0)(50,0){33}%
        {\setlength{\unitlength}{1cm}\begin{picture}(0,0)(0,0)
        \put(0,0){\line(0,1){0.12}}
    \end{picture}}
    \put(7500,0){\setlength{\unitlength}{1cm}\begin{picture}(0,0)(0,0)
        \put(0,0){\line(0,1){0.2}}
    \end{picture}}
    \put(7750,0){\setlength{\unitlength}{1cm}\begin{picture}(0,0)(0,0)
        \put(0,0){\line(0,1){0.2}}
    \end{picture}}
    \put(8000,0){\setlength{\unitlength}{1cm}\begin{picture}(0,0)(0,0)
        \put(0,0){\line(0,1){0.2}}
    \end{picture}}
    \put(8250,0){\setlength{\unitlength}{1cm}\begin{picture}(0,0)(0,0)
        \put(0,0){\line(0,1){0.2}}
    \end{picture}}
    \put(8500,0){\setlength{\unitlength}{1cm}\begin{picture}(0,0)(0,0)
        \put(0,0){\line(0,1){0.2}}
    \end{picture}}
    \put(8750,0){\setlength{\unitlength}{1cm}\begin{picture}(0,0)(0,0)
        \put(0,0){\line(0,1){0.2}}
    \end{picture}}
    \put(9000,0){\setlength{\unitlength}{1cm}\begin{picture}(0,0)(0,0)
        \put(0,0){\line(0,1){0.2}}
    \end{picture}}

\punkt{02/11/89}{02:30}{7832.604}{ -0.071}{0.052}{ 500}{1.68}{
531}{1.2CA}
\punkta{21/12/89}{01:09}{7881.549}{ -0.092}{0.028}{ 500}{2.54}{
609}{1.2CA}
\punkt{21/12/89}{02:24}{7881.601}{ -0.108}{0.027}{ 500}{2.01}{
705}{1.2CA}
\punkt{26/02/90}{23:18}{7949.471}{ -0.106}{0.078}{ 500}{2.37}{
634}{1.2CA}
\punkt{22/12/90}{12:28}{8248.020}{ -0.102}{0.050}{ 500}{2.24}{
905}{1.2CA}
\punkt{02/02/91}{22:56}{8290.456}{ -0.061}{0.047}{ 500}{3.13}{
902}{1.2CA}
\punkt{11/02/91}{22:51}{8299.452}{ -0.022}{0.051}{ 500}{3.83}{
601}{1.2CA}
\punkt{22/02/91}{23:03}{8310.460}{ -0.086}{0.028}{ 500}{1.30}{
1899}{1.2CA}
\punkt{17/10/91}{04:58}{8546.707}{  0.014}{0.023}{ 500}{2.21}{
748}{1.2CA}
\punkt{02/02/92}{01:24}{8654.559}{  0.087}{0.012}{ 500}{1.61}{
482}{1.2CA}
\punkt{03/02/92}{01:14}{8655.552}{  0.094}{0.040}{ 500}{1.65}{
565}{1.2CA}
\punkt{06/02/92}{23:48}{8659.492}{  0.102}{0.013}{ 500}{1.74}{
464}{1.2CA}
\punkt{10/02/92}{00:14}{8662.510}{  0.068}{0.015}{ 500}{1.03}{
511}{1.2CA}
\punkt{11/02/92}{19:55}{8664.330}{  0.078}{0.008}{ 500}{1.10}{
834}{1.2CA}
\punkt{14/02/92}{22:43}{8667.447}{  0.127}{0.021}{ 500}{1.25}{
3155}{1.2CA}
\punkt{14/02/92}{22:43}{8667.447}{  0.127}{0.021}{ 500}{1.25}{
3155}{1.2CA}

\end{picture}}

\end{picture}

\vspace*{-0.02cm}

\begin{picture}(18 ,3.5 )(0,0)
\put(0,0){\setlength{\unitlength}{0.01059cm}%
\begin{picture}(1700, 330.555)(7350,0)
\put(7350,0){\framebox(1700, 330.555)[tl]{\begin{picture}(0,0)(0,0)
        \put(1700,0){\makebox(0,0)[tr]{\bf{0903+175}\T{0.4}
                                 \hspace*{0.5cm}}}
    \end{picture}}}

\thicklines
\put(7350,0){\setlength{\unitlength}{5cm}\begin{picture}(0,0)(0,-0.35)
   \put(0,0){\setlength{\unitlength}{1cm}\begin{picture}(0,0)(0,0)
        \put(0,0){\line(1,0){0.3}}
        \end{picture}}
   \end{picture}}

\put(9050,0){\setlength{\unitlength}{5cm}\begin{picture}(0,0)(0,-0.35)
   \put(0,0){\setlength{\unitlength}{1cm}\begin{picture}(0,0)(0,0)
        \put(0,0){\line(-1,0){0.3}}
        \end{picture}}
   \end{picture}}

\thinlines
\put(7350,0){\setlength{\unitlength}{5cm}\begin{picture}(0,0)(0,-0.35)
   \multiput(0,0)(0,0.1){4}{\setlength{\unitlength}{1cm}%
\begin{picture}(0,0)(0,0)
        \put(0,0){\line(1,0){0.12}}
        \end{picture}}
   \end{picture}}

\put(7350,0){\setlength{\unitlength}{5cm}\begin{picture}(0,0)(0,-0.35)
   \multiput(0,0)(0,-0.1){4}{\setlength{\unitlength}{1cm}%
\begin{picture}(0,0)(0,0)
        \put(0,0){\line(1,0){0.12}}
        \end{picture}}
   \end{picture}}

\put(9050,0){\setlength{\unitlength}{5cm}\begin{picture}(0,0)(0,-0.35)
   \multiput(0,0)(0,0.1){4}{\setlength{\unitlength}{1cm}%
\begin{picture}(0,0)(0,0)
        \put(0,0){\line(-1,0){0.12}}
        \end{picture}}
   \end{picture}}

\put(9050,0){\setlength{\unitlength}{5cm}\begin{picture}(0,0)(0,-0.35)
   \multiput(0,0)(0,-0.1){4}{\setlength{\unitlength}{1cm}%
\begin{picture}(0,0)(0,0)
        \put(0,0){\line(-1,0){0.12}}
        \end{picture}}
   \end{picture}}

   \put(7527.5, 330.555){\setlength{\unitlength}{1cm}\begin{picture}(0,0)(0,0)
        \put(0,0){\line(0,-1){0.2}}
   \end{picture}}
   \put(7892.5, 330.555){\setlength{\unitlength}{1cm}\begin{picture}(0,0)(0,0)
        \put(0,0){\line(0,-1){0.2}}
   \end{picture}}
   \put(8257.5, 330.555){\setlength{\unitlength}{1cm}\begin{picture}(0,0)(0,0)
        \put(0,0){\line(0,-1){0.2}}
   \end{picture}}
   \put(8622.5, 330.555){\setlength{\unitlength}{1cm}\begin{picture}(0,0)(0,0)
        \put(0,0){\line(0,-1){0.2}}
   \end{picture}}
   \put(8987.5, 330.555){\setlength{\unitlength}{1cm}\begin{picture}(0,0)(0,0)
        \put(0,0){\line(0,-1){0.2}}
   \end{picture}}
    \multiput(7350,0)(50,0){33}%
        {\setlength{\unitlength}{1cm}\begin{picture}(0,0)(0,0)
        \put(0,0){\line(0,1){0.12}}
    \end{picture}}
    \put(7500,0){\setlength{\unitlength}{1cm}\begin{picture}(0,0)(0,0)
        \put(0,0){\line(0,1){0.2}}
    \end{picture}}
    \put(7750,0){\setlength{\unitlength}{1cm}\begin{picture}(0,0)(0,0)
        \put(0,0){\line(0,1){0.2}}
   \end{picture}}
    \put(8000,0){\setlength{\unitlength}{1cm}\begin{picture}(0,0)(0,0)
        \put(0,0){\line(0,1){0.2}}
    \end{picture}}
    \put(8250,0){\setlength{\unitlength}{1cm}\begin{picture}(0,0)(0,0)
        \put(0,0){\line(0,1){0.2}}
    \end{picture}}
    \put(8500,0){\setlength{\unitlength}{1cm}\begin{picture}(0,0)(0,0)
        \put(0,0){\line(0,1){0.2}}
   \end{picture}}
    \put(8750,0){\setlength{\unitlength}{1cm}\begin{picture}(0,0)(0,0)
        \put(0,0){\line(0,1){0.2}}
    \end{picture}}
    \put(9000,0){\setlength{\unitlength}{1cm}\begin{picture}(0,0)(0,0)
        \put(0,0){\line(0,1){0.2}}
    \end{picture}}

\put(7000,0){\begin{picture}(0,0)(7000,-47.2144)
\punkt{01/11/89}{04:55}{7831.705}{ -0.022}{0.016}{ 500}{1.27}{
556}{1.2CA}
\punkta{06/01/90}{02:35}{7897.608}{ -0.060}{0.055}{ 500}{2.06}{
628}{1.2CA}
\punkt{27/02/90}{22:38}{7950.443}{ -0.049}{0.023}{ 500}{1.71}{
701}{1.2CA}
\punkta{22/12/90}{14:39}{8248.111}{ -0.031}{0.111}{ 500}{2.36}{
812}{1.2CA}
\punkt{03/02/91}{00:49}{8290.534}{  0.056}{0.058}{ 500}{2.72}{
1162}{1.2CA}
\punkt{22/02/91}{23:51}{8310.494}{ -0.006}{0.019}{ 500}{1.18}{
1005}{1.2CA}
\punkt{25/02/91}{03:46}{8312.657}{  0.010}{0.014}{ 500}{2.46}{
2124}{1.2CA}
\punkt{18/03/91}{01:39}{8333.569}{  0.027}{0.026}{  60}{1.49}{
876}{1.2CA}
\punkt{23/05/91}{20:26}{8400.352}{  0.014}{0.046}{ 500}{1.28}{
1749}{1.2CA}
\punkt{02/02/92}{02:30}{8654.605}{ -0.065}{0.058}{ 500}{1.72}{
524}{1.2CA}
\punkt{03/02/92}{01:54}{8655.580}{ -0.043}{0.064}{ 500}{1.54}{
574}{1.2CA}
\punkta{07/02/92}{00:42}{8659.529}{  0.040}{0.012}{ 100}{0.93}{
300}{1.2CA}
\punkta{08/02/92}{01:14}{8660.552}{ -0.019}{0.124}{ 100}{1.21}{
303}{1.2CA}
\punkt{11/02/92}{01:46}{8663.574}{ -0.003}{0.012}{ 500}{1.06}{
508}{1.2CA}
\punkta{14/02/92}{21:34}{8667.399}{ -0.032}{0.024}{ 500}{1.33}{
3677}{1.2CA}
\punkta{14/02/92}{23:49}{8667.493}{ -0.003}{0.052}{ 500}{1.14}{
2629}{1.2CA}
\punkt{09/03/92}{02:48}{8690.617}{ -0.111}{0.031}{ 300}{1.45}{
287}{1.2CA}
\punkta{20/03/92}{02:05}{8701.587}{ -0.050}{0.078}{ 300}{1.70}{
801}{1.2CA}
\punkta{23/03/92}{01:39}{8704.569}{ -0.075}{0.044}{ 300}{1.55}{
339}{1.2CA}
\punkta{23/03/92}{01:39}{8704.569}{ -0.075}{0.044}{ 300}{1.55}{
339}{1.2CA}
\end{picture}}

\end{picture}}

\end{picture}

\vspace*{-0.02cm}

\begin{picture}(18 ,2.5 )(0,0)
\put(0,0){\setlength{\unitlength}{0.01059cm}%
\begin{picture}(1700, 236.111)(7350,0)
\put(7350,0){\framebox(1700, 236.111)[tl]{\begin{picture}(0,0)(0,0)
        \put(1700,0){\makebox(0,0)[tr]{\bf{1011+250}\T{0.4}
                                 \hspace*{0.5cm}}}
    \end{picture}}}

\thicklines
\put(7350,0){\setlength{\unitlength}{5cm}\begin{picture}(0,0)(0,-0.25)
   \put(0,0){\setlength{\unitlength}{1cm}\begin{picture}(0,0)(0,0)
        \put(0,0){\line(1,0){0.3}}
        \end{picture}}
   \end{picture}}

\put(9050,0){\setlength{\unitlength}{5cm}\begin{picture}(0,0)(0,-0.25)
   \put(0,0){\setlength{\unitlength}{1cm}\begin{picture}(0,0)(0,0)
        \put(0,0){\line(-1,0){0.3}}
        \end{picture}}
   \end{picture}}

\thinlines
\put(7350,0){\setlength{\unitlength}{5cm}\begin{picture}(0,0)(0,-0.25)
   \multiput(0,0)(0,0.1){3}{\setlength{\unitlength}{1cm}%
\begin{picture}(0,0)(0,0)
        \put(0,0){\line(1,0){0.12}}
        \end{picture}}
   \end{picture}}

\put(7350,0){\setlength{\unitlength}{5cm}\begin{picture}(0,0)(0,-0.25)
   \multiput(0,0)(0,-0.1){3}{\setlength{\unitlength}{1cm}%
\begin{picture}(0,0)(0,0)
        \put(0,0){\line(1,0){0.12}}
        \end{picture}}
   \end{picture}}

\put(9050,0){\setlength{\unitlength}{5cm}\begin{picture}(0,0)(0,-0.25)
   \multiput(0,0)(0,0.1){3}{\setlength{\unitlength}{1cm}%
\begin{picture}(0,0)(0,0)
        \put(0,0){\line(-1,0){0.12}}
        \end{picture}}
   \end{picture}}

\put(9050,0){\setlength{\unitlength}{5cm}\begin{picture}(0,0)(0,-0.25)
   \multiput(0,0)(0,-0.1){3}{\setlength{\unitlength}{1cm}%
\begin{picture}(0,0)(0,0)
        \put(0,0){\line(-1,0){0.12}}
        \end{picture}}
   \end{picture}}

   \put(7527.5, 236.111){\setlength{\unitlength}{1cm}\begin{picture}(0,0)(0,0)
        \put(0,0){\line(0,-1){0.2}}
   \end{picture}}
   \put(7892.5, 236.111){\setlength{\unitlength}{1cm}\begin{picture}(0,0)(0,0)
        \put(0,0){\line(0,-1){0.2}}
   \end{picture}}
   \put(8257.5, 236.111){\setlength{\unitlength}{1cm}\begin{picture}(0,0)(0,0)
        \put(0,0){\line(0,-1){0.2}}
   \end{picture}}
   \put(8622.5, 236.111){\setlength{\unitlength}{1cm}\begin{picture}(0,0)(0,0)
        \put(0,0){\line(0,-1){0.2}}
   \end{picture}}
   \put(8987.5, 236.111){\setlength{\unitlength}{1cm}\begin{picture}(0,0)(0,0)
        \put(0,0){\line(0,-1){0.2}}
   \end{picture}}
    \multiput(7350,0)(50,0){33}%
        {\setlength{\unitlength}{1cm}\begin{picture}(0,0)(0,0)
        \put(0,0){\line(0,1){0.12}}
    \end{picture}}
    \put(7500,0){\setlength{\unitlength}{1cm}\begin{picture}(0,0)(0,0)
        \put(0,0){\line(0,1){0.2}}
    \end{picture}}
    \put(7750,0){\setlength{\unitlength}{1cm}\begin{picture}(0,0)(0,0)
        \put(0,0){\line(0,1){0.2}}
    \end{picture}}
    \put(8000,0){\setlength{\unitlength}{1cm}\begin{picture}(0,0)(0,0)
        \put(0,0){\line(0,1){0.2}}
    \end{picture}}
    \put(8250,0){\setlength{\unitlength}{1cm}\begin{picture}(0,0)(0,0)
        \put(0,0){\line(0,1){0.2}}
    \end{picture}}
    \put(8500,0){\setlength{\unitlength}{1cm}\begin{picture}(0,0)(0,0)
        \put(0,0){\line(0,1){0.2}}
    \end{picture}}
    \put(8750,0){\setlength{\unitlength}{1cm}\begin{picture}(0,0)(0,0)
        \put(0,0){\line(0,1){0.2}}
    \end{picture}}
    \put(9000,0){\setlength{\unitlength}{1cm}\begin{picture}(0,0)(0,0)
        \put(0,0){\line(0,1){0.2}}
    \end{picture}}

\punkta{02/11/89}{05:07}{7832.714}{ -0.011}{0.018}{ 500}{1.97}{
600}{1.2CA}
\punkta{14/12/89}{04:44}{7874.698}{ -0.008}{0.009}{ 500}{1.68}{
1246}{1.2CA}
\punkta{15/12/89}{03:24}{7875.642}{ -0.011}{0.038}{ 500}{2.57}{
1313}{1.2CA}
\punkta{28/02/90}{00:22}{7950.515}{ -0.014}{0.038}{ 500}{1.59}{
698}{1.2CA}
\punkta{24/05/90}{22:57}{8036.456}{ -0.010}{0.024}{ 400}{2.00}{
2266}{1.2CA}
\punkt{10/02/91}{01:00}{8297.542}{ -0.010}{0.032}{ 200}{2.06}{
405}{1.2CA}
\punkt{}{}{8310.540}{0.022}{0.010}{}{}{}{}
\punkt{}{}{8401.407}{0.000}{0.03}{}{}{}{}
\punkta{11/02/92}{03:32}{8663.648}{ -0.006}{0.027}{ 500}{1.10}{
529}{1.2CA}
\punkt{14/02/92}{00:05}{8666.504}{ -0.006}{0.012}{ 100}{1.24}{
539}{1.2CA}
\punkt{15/02/92}{01:41}{8667.570}{ -0.015}{0.010}{ 300}{1.45}{
1242}{1.2CA}
\punkta{}{}{8692.655}{0.050}{0.03}{}{}{}{}
%

\end{picture}}

\end{picture}

\vspace*{-0.02cm}

\begin{picture}(18 ,3.5 )(0,0)
\put(0,0){\setlength{\unitlength}{0.01059cm}%
\begin{picture}(1700, 330.555)(7350,0)
\put(7350,0){\framebox(1700, 330.555)[tl]{\begin{picture}(0,0)(0,0)
        \put(1700,0){\makebox(0,0)[tr]{\bf{1109+357}\T{0.4}
                                 \hspace*{0.5cm}}}
    \end{picture}}}

\thicklines
\put(7350,0){\setlength{\unitlength}{5cm}\begin{picture}(0,0)(0,-0.35)
   \put(0,0){\setlength{\unitlength}{1cm}\begin{picture}(0,0)(0,0)
        \put(0,0){\line(1,0){0.3}}
        \end{picture}}
   \end{picture}}

\put(9050,0){\setlength{\unitlength}{5cm}\begin{picture}(0,0)(0,-0.35)
   \put(0,0){\setlength{\unitlength}{1cm}\begin{picture}(0,0)(0,0)
        \put(0,0){\line(-1,0){0.3}}
        \end{picture}}
   \end{picture}}

\thinlines
\put(7350,0){\setlength{\unitlength}{5cm}\begin{picture}(0,0)(0,-0.35)
   \multiput(0,0)(0,0.1){4}{\setlength{\unitlength}{1cm}%
\begin{picture}(0,0)(0,0)
        \put(0,0){\line(1,0){0.12}}
        \end{picture}}
   \end{picture}}

\put(7350,0){\setlength{\unitlength}{5cm}\begin{picture}(0,0)(0,-0.35)
   \multiput(0,0)(0,-0.1){4}{\setlength{\unitlength}{1cm}%
\begin{picture}(0,0)(0,0)
        \put(0,0){\line(1,0){0.12}}
        \end{picture}}
   \end{picture}}

\put(9050,0){\setlength{\unitlength}{5cm}\begin{picture}(0,0)(0,-0.35)
   \multiput(0,0)(0,0.1){4}{\setlength{\unitlength}{1cm}%
\begin{picture}(0,0)(0,0)
        \put(0,0){\line(-1,0){0.12}}
        \end{picture}}
   \end{picture}}

\put(9050,0){\setlength{\unitlength}{5cm}\begin{picture}(0,0)(0,-0.35)
   \multiput(0,0)(0,-0.1){4}{\setlength{\unitlength}{1cm}%
\begin{picture}(0,0)(0,0)
        \put(0,0){\line(-1,0){0.12}}
        \end{picture}}
   \end{picture}}

   \put(7527.5, 330.555){\setlength{\unitlength}{1cm}\begin{picture}(0,0)(0,0)
        \put(0,0){\line(0,-1){0.2}}
   \end{picture}}
   \put(7892.5, 330.555){\setlength{\unitlength}{1cm}\begin{picture}(0,0)(0,0)
        \put(0,0){\line(0,-1){0.2}}
   \end{picture}}
   \put(8257.5, 330.555){\setlength{\unitlength}{1cm}\begin{picture}(0,0)(0,0)
        \put(0,0){\line(0,-1){0.2}}
   \end{picture}}
   \put(8622.5, 330.555){\setlength{\unitlength}{1cm}\begin{picture}(0,0)(0,0)
        \put(0,0){\line(0,-1){0.2}}
   \end{picture}}
   \put(8987.5, 330.555){\setlength{\unitlength}{1cm}\begin{picture}(0,0)(0,0)
        \put(0,0){\line(0,-1){0.2}}
   \end{picture}}
    \multiput(7350,0)(50,0){33}%
        {\setlength{\unitlength}{1cm}\begin{picture}(0,0)(0,0)
        \put(0,0){\line(0,1){0.12}}
    \end{picture}}
    \put(7500,0){\setlength{\unitlength}{1cm}\begin{picture}(0,0)(0,0)
        \put(0,0){\line(0,1){0.2}}
    \end{picture}}
    \put(7750,0){\setlength{\unitlength}{1cm}\begin{picture}(0,0)(0,0)
        \put(0,0){\line(0,1){0.2}}
    \end{picture}}
    \put(8000,0){\setlength{\unitlength}{1cm}\begin{picture}(0,0)(0,0)
        \put(0,0){\line(0,1){0.2}}
    \end{picture}}
    \put(8250,0){\setlength{\unitlength}{1cm}\begin{picture}(0,0)(0,0)
        \put(0,0){\line(0,1){0.2}}
    \end{picture}}
    \put(8500,0){\setlength{\unitlength}{1cm}\begin{picture}(0,0)(0,0)
        \put(0,0){\line(0,1){0.2}}
    \end{picture}}
    \put(8750,0){\setlength{\unitlength}{1cm}\begin{picture}(0,0)(0,0)
        \put(0,0){\line(0,1){0.2}}
    \end{picture}}
    \put(9000,0){\setlength{\unitlength}{1cm}\begin{picture}(0,0)(0,0)
        \put(0,0){\line(0,1){0.2}}
    \end{picture}}

\put(7000,0){\begin{picture}(0,0)(7000,-47.214)
\punkt{14/12/89}{05:33}{7874.732}{  0.062}{0.073}{ 500}{3.23}{
1317}{1.2CA}
\punkt{20/12/89}{06:05}{7880.754}{  0.078}{0.077}{ 500}{1.96}{
913}{1.2CA}
\punkta{27/01/90}{00:54}{7918.538}{  0.037}{0.033}{ 500}{2.92}{
2396}{1.2CA}
\punkt{28/02/90}{00:40}{7950.528}{  0.066}{0.038}{ 500}{1.78}{
630}{1.2CA}
\punkta{10/02/91}{02:49}{8297.617}{ -0.243}{0.175}{ 200}{2.57}{
394}{1.2CA}
\punkta{25/05/91}{21:43}{8402.405}{ -0.072}{0.139}{ 500}{2.76}{
1491}{1.2CA}
\punkt{12/02/92}{03:28}{8664.645}{ -0.077}{0.022}{ 500}{1.27}{
548}{1.2CA}

\end{picture}}

\end{picture}}

\end{picture}

\vspace*{-0.02cm}

\begin{picture}(18 ,2.5 )(0,0)
\put(0,0){\setlength{\unitlength}{0.01059cm}%
\begin{picture}(1700, 236.111)(7350,0)
\put(7350,0){\framebox(1700, 236.111)[tl]{\begin{picture}(0,0)(0,0)
        \put(1700,0){\makebox(0,0)[tr]{\bf{1209+107}\T{0.4}
                                 \hspace*{0.5cm}}}
        \put(1700,-
236.111){\setlength{\unitlength}{1cm}\begin{picture}(0,0)(0,0)
            \put(0,-1){\makebox(0,0)[br]{\bf J.D.\,2,440,000\,+}}
        \end{picture}}
    \end{picture}}}

\thicklines
\put(7350,0){\setlength{\unitlength}{5cm}\begin{picture}(0,0)(0,-0.25)
   \put(0,0){\setlength{\unitlength}{1cm}\begin{picture}(0,0)(0,0)
        \put(0,0){\line(1,0){0.3}}
        \end{picture}}
   \end{picture}}

\put(9050,0){\setlength{\unitlength}{5cm}\begin{picture}(0,0)(0,-0.25)
   \put(0,0){\setlength{\unitlength}{1cm}\begin{picture}(0,0)(0,0)
        \put(0,0){\line(-1,0){0.3}}
        \end{picture}}
   \end{picture}}

\thinlines
\put(7350,0){\setlength{\unitlength}{5cm}\begin{picture}(0,0)(0,-0.25)
   \multiput(0,0)(0,0.1){3}{\setlength{\unitlength}{1cm}%
\begin{picture}(0,0)(0,0)
        \put(0,0){\line(1,0){0.12}}
        \end{picture}}
   \end{picture}}

\put(7350,0){\setlength{\unitlength}{5cm}\begin{picture}(0,0)(0,-0.25)
   \multiput(0,0)(0,-0.1){3}{\setlength{\unitlength}{1cm}%
\begin{picture}(0,0)(0,0)
        \put(0,0){\line(1,0){0.12}}
        \end{picture}}
   \end{picture}}

\put(9050,0){\setlength{\unitlength}{5cm}\begin{picture}(0,0)(0,-0.25)
   \multiput(0,0)(0,0.1){3}{\setlength{\unitlength}{1cm}%
\begin{picture}(0,0)(0,0)
        \put(0,0){\line(-1,0){0.12}}
        \end{picture}}
   \end{picture}}

\put(9050,0){\setlength{\unitlength}{5cm}\begin{picture}(0,0)(0,-0.25)
   \multiput(0,0)(0,-0.1){3}{\setlength{\unitlength}{1cm}%
\begin{picture}(0,0)(0,0)
        \put(0,0){\line(-1,0){0.12}}
        \end{picture}}
   \end{picture}}

   \put(7527.5, 236.111){\setlength{\unitlength}{1cm}\begin{picture}(0,0)(0,0)
        \put(0,0){\line(0,-1){0.2}}
   \end{picture}}
   \put(7892.5, 236.111){\setlength{\unitlength}{1cm}\begin{picture}(0,0)(0,0)
        \put(0,0){\line(0,-1){0.2}}
   \end{picture}}
   \put(8257.5, 236.111){\setlength{\unitlength}{1cm}\begin{picture}(0,0)(0,0)
        \put(0,0){\line(0,-1){0.2}}
   \end{picture}}
   \put(8622.5, 236.111){\setlength{\unitlength}{1cm}\begin{picture}(0,0)(0,0)
        \put(0,0){\line(0,-1){0.2}}
   \end{picture}}
   \put(8987.5, 236.111){\setlength{\unitlength}{1cm}\begin{picture}(0,0)(0,0)
        \put(0,0){\line(0,-1){0.2}}
   \end{picture}}
    \multiput(7350,0)(50,0){33}%
        {\setlength{\unitlength}{1cm}\begin{picture}(0,0)(0,0)
        \put(0,0){\line(0,1){0.12}}
    \end{picture}}
    \put(7500,0){\setlength{\unitlength}{1cm}\begin{picture}(0,0)(0,0)
        \put(0,0){\line(0,1){0.2}}
        \put(0,-0.2){\makebox(0,0)[t]{\bf 7500}}
    \end{picture}}
    \put(7750,0){\setlength{\unitlength}{1cm}\begin{picture}(0,0)(0,0)
        \put(0,0){\line(0,1){0.2}}
        \put(0,-0.2){\makebox(0,0)[t]{\bf 7750}}
    \end{picture}}
    \put(8000,0){\setlength{\unitlength}{1cm}\begin{picture}(0,0)(0,0)
        \put(0,0){\line(0,1){0.2}}
        \put(0,-0.2){\makebox(0,0)[t]{\bf 8000}}
    \end{picture}}
    \put(8250,0){\setlength{\unitlength}{1cm}\begin{picture}(0,0)(0,0)
        \put(0,0){\line(0,1){0.2}}
        \put(0,-0.2){\makebox(0,0)[t]{\bf 8250}}
    \end{picture}}
    \put(8500,0){\setlength{\unitlength}{1cm}\begin{picture}(0,0)(0,0)
        \put(0,0){\line(0,1){0.2}}
        \put(0,-0.2){\makebox(0,0)[t]{\bf 8500}}
    \end{picture}}
    \put(8750,0){\setlength{\unitlength}{1cm}\begin{picture}(0,0)(0,0)
        \put(0,0){\line(0,1){0.2}}
        \put(0,-0.2){\makebox(0,0)[t]{\bf 8750}}
    \end{picture}}
    \put(9000,0){\setlength{\unitlength}{1cm}\begin{picture}(0,0)(0,0)
        \put(0,0){\line(0,1){0.2}}
        \put(0,-0.2){\makebox(0,0)[t]{\bf 9000}}
    \end{picture}}

\punkt{09/02/90}{00:15}{7931.511}{  0.162}{0.040}{ 500}{1.77}{
1523}{1.2CA}
\punkt{28/02/90}{01:47}{7950.574}{  0.084}{0.027}{ 500}{1.60}{
612}{1.2CA}
\punkt{22/12/90}{16:43}{8248.197}{  0.110}{0.058}{ 500}{2.51}{
630}{1.2CA}
\punkta{16/03/91}{02:18}{8331.596}{  0.137}{0.050}{ 100}{2.88}{
314}{1.2CA}
\punkta{26/05/91}{22:14}{8403.427}{  0.094}{0.050}{ 500}{3.27}{
3328}{1.2CA}
\punkt{02/02/92}{04:48}{8654.700}{ -0.046}{0.018}{ 500}{1.91}{
525}{1.2CA}
\punkt{03/02/92}{04:30}{8655.688}{ -0.075}{0.046}{ 500}{1.70}{
513}{1.2CA}
\punkt{07/02/92}{03:47}{8659.658}{ -0.028}{0.023}{ 500}{2.30}{
485}{1.2CA}
\punkt{09/02/92}{04:17}{8661.679}{ -0.058}{0.011}{ 500}{1.04}{
474}{1.2CA}
\punkt{14/02/92}{01:59}{8666.583}{  0.005}{0.034}{ 500}{2.39}{
803}{1.2CA}
\punkt{15/02/92}{03:17}{8667.637}{ -0.073}{0.013}{ 500}{1.41}{
947}{1.2CA}
\punkt{16/02/92}{03:13}{8668.634}{ -0.058}{0.024}{ 500}{1.46}{
1646}{1.2CA}
\punkt{16/02/92}{03:13}{8668.634}{ -0.058}{0.024}{ 500}{1.46}{
1646}{1.2CA}

    \put(7350,0){\setlength{\unitlength}{1cm}\begin{picture}(0,0)(0,0)
        \put(0,-1.3){\makebox(0,0)[tl]{\footnotesize{\bf Fig.~1.} (continued)}}
    \end{picture}}

\end{picture}}

\end{picture}

\vspace*{1.3cm}
\end{figure*}

\begin{figure*}

\vspace*{0.4cm}

\begin{picture}(18 ,2.5 )(0,0)
\put(0,0){\setlength{\unitlength}{0.01059cm}%
\begin{picture}(1700, 236.111)(7350,0)
\put(7350,0){\framebox(1700, 236.111)[tl]{\begin{picture}(0,0)(0,0)
        \put(1700,0){\makebox(0,0)[tr]{\bf{1222+228}\T{0.4}
                                 \hspace*{0.5cm}}}
    \end{picture}}}

\thicklines
\put(7350,0){\setlength{\unitlength}{5cm}\begin{picture}(0,0)(0,-0.25)
   \put(0,0){\setlength{\unitlength}{1cm}\begin{picture}(0,0)(0,0)
        \put(0,0){\line(1,0){0.3}}
        \end{picture}}
   \end{picture}}

\put(9050,0){\setlength{\unitlength}{5cm}\begin{picture}(0,0)(0,-0.25)
   \put(0,0){\setlength{\unitlength}{1cm}\begin{picture}(0,0)(0,0)
        \put(0,0){\line(-1,0){0.3}}
        \end{picture}}
   \end{picture}}

\thinlines
\put(7350,0){\setlength{\unitlength}{5cm}\begin{picture}(0,0)(0,-0.25)
   \multiput(0,0)(0,0.1){3}{\setlength{\unitlength}{1cm}%
\begin{picture}(0,0)(0,0)
        \put(0,0){\line(1,0){0.12}}
        \end{picture}}
   \end{picture}}

\put(7350,0){\setlength{\unitlength}{5cm}\begin{picture}(0,0)(0,-0.25)
   \multiput(0,0)(0,-0.1){3}{\setlength{\unitlength}{1cm}%
\begin{picture}(0,0)(0,0)
        \put(0,0){\line(1,0){0.12}}
        \end{picture}}
   \end{picture}}

\put(9050,0){\setlength{\unitlength}{5cm}\begin{picture}(0,0)(0,-0.25)
   \multiput(0,0)(0,0.1){3}{\setlength{\unitlength}{1cm}%
\begin{picture}(0,0)(0,0)
        \put(0,0){\line(-1,0){0.12}}
        \end{picture}}
   \end{picture}}

\put(9050,0){\setlength{\unitlength}{5cm}\begin{picture}(0,0)(0,-0.25)
   \multiput(0,0)(0,-0.1){3}{\setlength{\unitlength}{1cm}%
\begin{picture}(0,0)(0,0)
        \put(0,0){\line(-1,0){0.12}}
        \end{picture}}
   \end{picture}}

   \put(7527.5, 236.111){\setlength{\unitlength}{1cm}\begin{picture}(0,0)(0,0)
        \put(0,0){\line(0,-1){0.2}}
        \put(0,0.2){\makebox(0,0)[b]{\bf 1989}}
   \end{picture}}
   \put(7892.5, 236.111){\setlength{\unitlength}{1cm}\begin{picture}(0,0)(0,0)
        \put(0,0){\line(0,-1){0.2}}
        \put(0,0.2){\makebox(0,0)[b]{\bf 1990}}
   \end{picture}}
   \put(8257.5, 236.111){\setlength{\unitlength}{1cm}\begin{picture}(0,0)(0,0)
        \put(0,0){\line(0,-1){0.2}}
        \put(0,0.2){\makebox(0,0)[b]{\bf 1991}}
   \end{picture}}
   \put(8622.5, 236.111){\setlength{\unitlength}{1cm}\begin{picture}(0,0)(0,0)
        \put(0,0){\line(0,-1){0.2}}
        \put(0,0.2){\makebox(0,0)[b]{\bf 1992}}
   \end{picture}}
   \put(8987.5, 236.111){\setlength{\unitlength}{1cm}\begin{picture}(0,0)(0,0)
        \put(0,0){\line(0,-1){0.2}}
        \put(0,0.2){\makebox(0,0)[b]{\bf 1993}}
   \end{picture}}
    \multiput(7350,0)(50,0){33}%
        {\setlength{\unitlength}{1cm}\begin{picture}(0,0)(0,0)
        \put(0,0){\line(0,1){0.12}}
    \end{picture}}
    \put(7500,0){\setlength{\unitlength}{1cm}\begin{picture}(0,0)(0,0)
        \put(0,0){\line(0,1){0.2}}
    \end{picture}}
    \put(7750,0){\setlength{\unitlength}{1cm}\begin{picture}(0,0)(0,0)
        \put(0,0){\line(0,1){0.2}}
    \end{picture}}
    \put(8000,0){\setlength{\unitlength}{1cm}\begin{picture}(0,0)(0,0)
       \put(0,0){\line(0,1){0.2}}
    \end{picture}}
    \put(8250,0){\setlength{\unitlength}{1cm}\begin{picture}(0,0)(0,0)
        \put(0,0){\line(0,1){0.2}}
    \end{picture}}
    \put(8500,0){\setlength{\unitlength}{1cm}\begin{picture}(0,0)(0,0)
        \put(0,0){\line(0,1){0.2}}
    \end{picture}}
    \put(8750,0){\setlength{\unitlength}{1cm}\begin{picture}(0,0)(0,0)
        \put(0,0){\line(0,1){0.2}}
    \end{picture}}
    \put(9000,0){\setlength{\unitlength}{1cm}\begin{picture}(0,0)(0,0)
        \put(0,0){\line(0,1){0.2}}
    \end{picture}}

%
\punkt{23/06/89}{21:05}{7701.379}{  0.013}{0.013}{1000}{2.02}{
3187}{1.2CA}
\punkt{24/06/89}{21:22}{7702.391}{ -0.006}{0.007}{1000}{1.48}{
1314}{1.2CA}
\punkt{26/06/89}{21:18}{7704.388}{ -0.014}{0.013}{1000}{2.09}{
1299}{1.2CA}
\punkt{15/12/89}{05:11}{7875.717}{  0.027}{0.021}{ 500}{2.69}{
992}{1.2CA}
\punkt{23/05/90}{21:22}{8035.391}{  0.007}{0.017}{ 504}{3.27}{
16043}{1.2CA}
\punkt{23/05/90}{21:31}{8035.397}{  0.021}{0.021}{ 400}{3.36}{
1609}{1.2CA}
\punkt{23/05/90}{21:40}{8035.403}{  0.021}{0.024}{ 400}{3.43}{
1632}{1.2CA}
\punkt{15/02/91}{04:17}{8302.679}{ -0.031}{0.013}{ 500}{2.82}{
560}{1.2CA}
\punkt{25/05/91}{22:12}{8402.426}{ -0.045}{0.051}{ 150}{2.37}{
733}{1.2CA}

\end{picture}}

\end{picture}

\vspace*{-0.02cm}

\begin{picture}(18 ,2.5 )(0,0)
\put(0,0){\setlength{\unitlength}{0.01059cm}%
\begin{picture}(1700, 236.111)(7350,0)
\put(7350,0){\framebox(1700, 236.111)[tl]{\begin{picture}(0,0)(0,0)
        \put(1700,0){\makebox(0,0)[tr]{\bf{1332+552}\T{0.4}
                                 \hspace*{0.5cm}}}
    \end{picture}}}

\thicklines
\put(7350,0){\setlength{\unitlength}{5cm}\begin{picture}(0,0)(0,-0.25)
   \put(0,0){\setlength{\unitlength}{1cm}\begin{picture}(0,0)(0,0)
        \put(0,0){\line(1,0){0.3}}
        \end{picture}}
   \end{picture}}

\put(9050,0){\setlength{\unitlength}{5cm}\begin{picture}(0,0)(0,-0.25)
   \put(0,0){\setlength{\unitlength}{1cm}\begin{picture}(0,0)(0,0)
        \put(0,0){\line(-1,0){0.3}}
        \end{picture}}
   \end{picture}}

\thinlines
\put(7350,0){\setlength{\unitlength}{5cm}\begin{picture}(0,0)(0,-0.25)
   \multiput(0,0)(0,0.1){3}{\setlength{\unitlength}{1cm}%
\begin{picture}(0,0)(0,0)
        \put(0,0){\line(1,0){0.12}}
        \end{picture}}
   \end{picture}}

\put(7350,0){\setlength{\unitlength}{5cm}\begin{picture}(0,0)(0,-0.25)
   \multiput(0,0)(0,-0.1){3}{\setlength{\unitlength}{1cm}%
\begin{picture}(0,0)(0,0)
        \put(0,0){\line(1,0){0.12}}
        \end{picture}}
   \end{picture}}

\put(9050,0){\setlength{\unitlength}{5cm}\begin{picture}(0,0)(0,-0.25)
   \multiput(0,0)(0,0.1){3}{\setlength{\unitlength}{1cm}%
\begin{picture}(0,0)(0,0)
        \put(0,0){\line(-1,0){0.12}}
        \end{picture}}
   \end{picture}}

\put(9050,0){\setlength{\unitlength}{5cm}\begin{picture}(0,0)(0,-0.25)
   \multiput(0,0)(0,-0.1){3}{\setlength{\unitlength}{1cm}%
\begin{picture}(0,0)(0,0)
        \put(0,0){\line(-1,0){0.12}}
        \end{picture}}
   \end{picture}}

   \put(7527.5, 236.111){\setlength{\unitlength}{1cm}\begin{picture}(0,0)(0,0)
        \put(0,0){\line(0,-1){0.2}}
   \end{picture}}
   \put(7892.5, 236.111){\setlength{\unitlength}{1cm}\begin{picture}(0,0)(0,0)
        \put(0,0){\line(0,-1){0.2}}
   \end{picture}}
   \put(8257.5, 236.111){\setlength{\unitlength}{1cm}\begin{picture}(0,0)(0,0)
        \put(0,0){\line(0,-1){0.2}}
   \end{picture}}
   \put(8622.5, 236.111){\setlength{\unitlength}{1cm}\begin{picture}(0,0)(0,0)
        \put(0,0){\line(0,-1){0.2}}
   \end{picture}}
   \put(8987.5, 236.111){\setlength{\unitlength}{1cm}\begin{picture}(0,0)(0,0)
        \put(0,0){\line(0,-1){0.2}}
   \end{picture}}

    \multiput(7350,0)(50,0){33}%
        {\setlength{\unitlength}{1cm}\begin{picture}(0,0)(0,0)
        \put(0,0){\line(0,1){0.12}}
    \end{picture}}
    \put(7500,0){\setlength{\unitlength}{1cm}\begin{picture}(0,0)(0,0)
        \put(0,0){\line(0,1){0.2}}
    \end{picture}}
    \put(7750,0){\setlength{\unitlength}{1cm}\begin{picture}(0,0)(0,0)
        \put(0,0){\line(0,1){0.2}}
    \end{picture}}
    \put(8000,0){\setlength{\unitlength}{1cm}\begin{picture}(0,0)(0,0)
        \put(0,0){\line(0,1){0.2}}
    \end{picture}}
    \put(8250,0){\setlength{\unitlength}{1cm}\begin{picture}(0,0)(0,0)
        \put(0,0){\line(0,1){0.2}}
    \end{picture}}
    \put(8500,0){\setlength{\unitlength}{1cm}\begin{picture}(0,0)(0,0)
        \put(0,0){\line(0,1){0.2}}
    \end{picture}}
    \put(8750,0){\setlength{\unitlength}{1cm}\begin{picture}(0,0)(0,0)
        \put(0,0){\line(0,1){0.2}}
    \end{picture}}
    \put(9000,0){\setlength{\unitlength}{1cm}\begin{picture}(0,0)(0,0)
        \put(0,0){\line(0,1){0.2}}
    \end{picture}}

\punkt{15/08/89}{20:27}{7754.352}{  0.008}{0.016}{1000}{2.26}{
2355}{1.2CA}
\punkt{23/05/90}{23:04}{8035.462}{  0.008}{0.019}{ 400}{2.82}{
1498}{1.2CA}
\punkt{01/08/90}{21:14}{8105.385}{  0.043}{0.004}{ 500}{2.41}{
1004}{1.2CA}
\punkta{03/02/91}{03:13}{8290.634}{ -0.033}{0.091}{ 300}{3.08}{
667}{1.2CA}
\punkt{24/02/91}{01:24}{8311.559}{ -0.006}{0.025}{ 500}{1.52}{
964}{1.2CA}
\punkta{24/05/91}{22:23}{8401.433}{ -0.009}{0.096}{ 500}{2.14}{
1249}{1.2CA}

\end{picture}}

\end{picture}

\vspace*{-0.02cm}

\begin{picture}(18 ,2.5 )(0,0)
\put(0,0){\setlength{\unitlength}{0.01059cm}%
\begin{picture}(1700, 236.111)(7350,0)
\put(7350,0){\framebox(1700, 236.111)[tl]{\begin{picture}(0,0)(0,0)
        \put(1700,0){\makebox(0,0)[tr]{\bf{1421+330}\T{0.4}
                                 \hspace*{0.5cm}}}
        \put(1700,-
236.111){\setlength{\unitlength}{1cm}\begin{picture}(0,0)(0,0)
        \end{picture}}
    \end{picture}}}

\thicklines
\put(7350,0){\setlength{\unitlength}{5cm}\begin{picture}(0,0)(0,-0.25)
   \put(0,0){\setlength{\unitlength}{1cm}\begin{picture}(0,0)(0,0)
        \put(0,0){\line(1,0){0.3}}
        \end{picture}}
   \end{picture}}

\put(9050,0){\setlength{\unitlength}{5cm}\begin{picture}(0,0)(0,-0.25)
   \put(0,0){\setlength{\unitlength}{1cm}\begin{picture}(0,0)(0,0)
        \put(0,0){\line(-1,0){0.3}}
        \end{picture}}
   \end{picture}}

\thinlines
\put(7350,0){\setlength{\unitlength}{5cm}\begin{picture}(0,0)(0,-0.25)
   \multiput(0,0)(0,0.1){3}{\setlength{\unitlength}{1cm}%
\begin{picture}(0,0)(0,0)
        \put(0,0){\line(1,0){0.12}}
        \end{picture}}
   \end{picture}}

\put(7350,0){\setlength{\unitlength}{5cm}\begin{picture}(0,0)(0,-0.25)
   \multiput(0,0)(0,-0.1){3}{\setlength{\unitlength}{1cm}%
\begin{picture}(0,0)(0,0)
        \put(0,0){\line(1,0){0.12}}
        \end{picture}}
   \end{picture}}

\put(9050,0){\setlength{\unitlength}{5cm}\begin{picture}(0,0)(0,-0.25)
   \multiput(0,0)(0,0.1){3}{\setlength{\unitlength}{1cm}%
\begin{picture}(0,0)(0,0)
        \put(0,0){\line(-1,0){0.12}}
        \end{picture}}
   \end{picture}}

\put(9050,0){\setlength{\unitlength}{5cm}\begin{picture}(0,0)(0,-0.25)
   \multiput(0,0)(0,-0.1){3}{\setlength{\unitlength}{1cm}%
\begin{picture}(0,0)(0,0)
        \put(0,0){\line(-1,0){0.12}}
        \end{picture}}
   \end{picture}}

   \put(7527.5, 236.111){\setlength{\unitlength}{1cm}\begin{picture}(0,0)(0,0)
        \put(0,0){\line(0,-1){0.2}}
   \end{picture}}
   \put(7892.5, 236.111){\setlength{\unitlength}{1cm}\begin{picture}(0,0)(0,0)
        \put(0,0){\line(0,-1){0.2}}
   \end{picture}}
   \put(8257.5, 236.111){\setlength{\unitlength}{1cm}\begin{picture}(0,0)(0,0)
        \put(0,0){\line(0,-1){0.2}}
   \end{picture}}
   \put(8622.5, 236.111){\setlength{\unitlength}{1cm}\begin{picture}(0,0)(0,0)
        \put(0,0){\line(0,-1){0.2}}
   \end{picture}}
   \put(8987.5, 236.111){\setlength{\unitlength}{1cm}\begin{picture}(0,0)(0,0)
        \put(0,0){\line(0,-1){0.2}}
   \end{picture}}
    \multiput(7350,0)(50,0){33}%
        {\setlength{\unitlength}{1cm}\begin{picture}(0,0)(0,0)
        \put(0,0){\line(0,1){0.12}}
    \end{picture}}
    \put(7500,0){\setlength{\unitlength}{1cm}\begin{picture}(0,0)(0,0)
        \put(0,0){\line(0,1){0.2}}
    \end{picture}}
    \put(7750,0){\setlength{\unitlength}{1cm}\begin{picture}(0,0)(0,0)
        \put(0,0){\line(0,1){0.2}}
    \end{picture}}
    \put(8000,0){\setlength{\unitlength}{1cm}\begin{picture}(0,0)(0,0)
        \put(0,0){\line(0,1){0.2}}
    \end{picture}}
    \put(8250,0){\setlength{\unitlength}{1cm}\begin{picture}(0,0)(0,0)
        \put(0,0){\line(0,1){0.2}}
    \end{picture}}
    \put(8500,0){\setlength{\unitlength}{1cm}\begin{picture}(0,0)(0,0)
        \put(0,0){\line(0,1){0.2}}
    \end{picture}}
    \put(8750,0){\setlength{\unitlength}{1cm}\begin{picture}(0,0)(0,0)
        \put(0,0){\line(0,1){0.2}}
    \end{picture}}
    \put(9000,0){\setlength{\unitlength}{1cm}\begin{picture}(0,0)(0,0)
        \put(0,0){\line(0,1){0.2}}
    \end{picture}}

\punkt{10/06/89}{22:33}{7688.440}{ -0.026}{0.012}{ 500}{1.62}{
1491}{1.2CA}
\punkt{23/07/89}{22:05}{7731.421}{ -0.019}{0.009}{ 500}{2.29}{
1178}{1.2CA}
\punkt{13/08/89}{21:06}{7752.379}{ -0.036}{0.007}{ 500}{1.90}{
1340}{1.2CA}
\punkt{01/08/90}{21:31}{8105.397}{  0.038}{0.002}{ 500}{1.93}{
1068}{1.2CA}
\punkt{24/05/91}{23:15}{8401.469}{  0.009}{0.034}{ 300}{2.20}{
1082}{1.2CA}
\punkt{28/05/91}{23:16}{8405.470}{ -0.001}{0.029}{ 500}{1.42}{
2822}{1.2CA}
\punkt{07/08/91}{21:21}{8476.390}{  0.013}{0.014}{ 500}{2.00}{
663}{1.2CA}
\punkta{11/02/92}{06:08}{8663.756}{ -0.013}{0.031}{ 500}{1.18}{
1362}{1.2CA}
\punkt{14/02/92}{05:59}{8666.750}{ -0.018}{0.007}{ 500}{1.28}{
687}{1.2CA}
\punkt{14/02/92}{05:59}{8666.750}{ -0.018}{0.007}{ 500}{1.28}{
687}{1.2CA}

\end{picture}}

\end{picture}

\vspace*{-0.02cm}

\begin{picture}(18 ,2.5 )(0,0)
\put(0,0){\setlength{\unitlength}{0.01059cm}%
\begin{picture}(1700, 236.111)(7350,0)
\put(7350,0){\framebox(1700, 236.111)[tl]{\begin{picture}(0,0)(0,0)
        \put(1700,0){\makebox(0,0)[tr]{\bf{1435+638}\T{0.4}
                                 \hspace*{0.5cm}}}
    \end{picture}}}

\thicklines
\put(7350,0){\setlength{\unitlength}{5cm}\begin{picture}(0,0)(0,-0.25)
   \put(0,0){\setlength{\unitlength}{1cm}\begin{picture}(0,0)(0,0)
        \put(0,0){\line(1,0){0.3}}
        \end{picture}}
   \end{picture}}

\put(9050,0){\setlength{\unitlength}{5cm}\begin{picture}(0,0)(0,-0.25)
   \put(0,0){\setlength{\unitlength}{1cm}\begin{picture}(0,0)(0,0)
        \put(0,0){\line(-1,0){0.3}}
        \end{picture}}
   \end{picture}}

\thinlines
\put(7350,0){\setlength{\unitlength}{5cm}\begin{picture}(0,0)(0,-0.25)
   \multiput(0,0)(0,0.1){3}{\setlength{\unitlength}{1cm}%
\begin{picture}(0,0)(0,0)
        \put(0,0){\line(1,0){0.12}}
        \end{picture}}
   \end{picture}}

\put(7350,0){\setlength{\unitlength}{5cm}\begin{picture}(0,0)(0,-0.25)
   \multiput(0,0)(0,-0.1){3}{\setlength{\unitlength}{1cm}%
\begin{picture}(0,0)(0,0)
        \put(0,0){\line(1,0){0.12}}
        \end{picture}}
   \end{picture}}

\put(9050,0){\setlength{\unitlength}{5cm}\begin{picture}(0,0)(0,-0.25)
   \multiput(0,0)(0,0.1){3}{\setlength{\unitlength}{1cm}%
\begin{picture}(0,0)(0,0)
        \put(0,0){\line(-1,0){0.12}}
        \end{picture}}
   \end{picture}}

\put(9050,0){\setlength{\unitlength}{5cm}\begin{picture}(0,0)(0,-0.25)
   \multiput(0,0)(0,-0.1){3}{\setlength{\unitlength}{1cm}%
\begin{picture}(0,0)(0,0)
        \put(0,0){\line(-1,0){0.12}}
        \end{picture}}
   \end{picture}}

   \put(7527.5, 236.111){\setlength{\unitlength}{1cm}\begin{picture}(0,0)(0,0)
        \put(0,0){\line(0,-1){0.2}}
   \end{picture}}
   \put(7892.5, 236.111){\setlength{\unitlength}{1cm}\begin{picture}(0,0)(0,0)
        \put(0,0){\line(0,-1){0.2}}
   \end{picture}}
   \put(8257.5, 236.111){\setlength{\unitlength}{1cm}\begin{picture}(0,0)(0,0)
        \put(0,0){\line(0,-1){0.2}}
   \end{picture}}
   \put(8622.5, 236.111){\setlength{\unitlength}{1cm}\begin{picture}(0,0)(0,0)
        \put(0,0){\line(0,-1){0.2}}
   \end{picture}}
   \put(8987.5, 236.111){\setlength{\unitlength}{1cm}\begin{picture}(0,0)(0,0)
        \put(0,0){\line(0,-1){0.2}}
   \end{picture}}
    \multiput(7350,0)(50,0){33}%
        {\setlength{\unitlength}{1cm}\begin{picture}(0,0)(0,0)
        \put(0,0){\line(0,1){0.12}}
    \end{picture}}
    \put(7500,0){\setlength{\unitlength}{1cm}\begin{picture}(0,0)(0,0)
        \put(0,0){\line(0,1){0.2}}
    \end{picture}}
    \put(7750,0){\setlength{\unitlength}{1cm}\begin{picture}(0,0)(0,0)
        \put(0,0){\line(0,1){0.2}}
    \end{picture}}
    \put(8000,0){\setlength{\unitlength}{1cm}\begin{picture}(0,0)(0,0)
        \put(0,0){\line(0,1){0.2}}
    \end{picture}}
    \put(8250,0){\setlength{\unitlength}{1cm}\begin{picture}(0,0)(0,0)
        \put(0,0){\line(0,1){0.2}}
    \end{picture}}
    \put(8500,0){\setlength{\unitlength}{1cm}\begin{picture}(0,0)(0,0)
        \put(0,0){\line(0,1){0.2}}
    \end{picture}}
    \put(8750,0){\setlength{\unitlength}{1cm}\begin{picture}(0,0)(0,0)
        \put(0,0){\line(0,1){0.2}}
    \end{picture}}
    \put(9000,0){\setlength{\unitlength}{1cm}\begin{picture}(0,0)(0,0)
        \put(0,0){\line(0,1){0.2}}
    \end{picture}}

\punkt{05/09/88}{19:38}{7410.318}{ -0.113}{0.013}{ 500}{1.58}{
1037}{1.2CA}
\punkt{10/06/89}{22:57}{7688.456}{ -0.047}{0.011}{ 500}{1.62}{
1092}{1.2CA}
\punkt{25/06/89}{21:31}{7703.397}{ -0.033}{0.012}{1000}{1.63}{
1270}{1.2CA}
\punkt{23/07/89}{22:19}{7731.431}{ -0.035}{0.009}{ 500}{2.19}{
1170}{1.2CA}
\punkt{16/08/89}{20:09}{7755.340}{ -0.030}{0.008}{ 500}{1.54}{
1724}{1.2CA}
\punkt{02/09/89}{07:41}{7771.821}{  0.001}{0.009}{****}{3.37}{
1413}{1.2CA}
\punkt{}{}{8402.457}{0.037}{0.04}{}{}{}{}
\punkta{06/02/92}{04:23}{8658.683}{  0.031}{0.023}{1000}{1.65}{
1163}{1.2CA}
\punkt{16/02/92}{04:39}{8668.694}{  0.040}{0.011}{ 500}{1.50}{
1038}{1.2CA}
\punkt{16/02/92}{04:39}{8668.694}{  0.040}{0.011}{ 500}{1.50}{
1038}{1.2CA}

\end{picture}}

\end{picture}

\vspace*{-0.02cm}

\begin{picture}(18 ,2.5 )(0,0)
\put(0,0){\setlength{\unitlength}{0.01059cm}%
\begin{picture}(1700, 236.111)(7350,0)
\put(7350,0){\framebox(1700, 236.111)[tl]{\begin{picture}(0,0)(0,0)
        \put(1700,0){\makebox(0,0)[tr]{\bf{1520+413}\T{0.4}
                                 \hspace*{0.5cm}}}
    \end{picture}}}

\thicklines
\put(7350,0){\setlength{\unitlength}{5cm}\begin{picture}(0,0)(0,-0.25)
   \put(0,0){\setlength{\unitlength}{1cm}\begin{picture}(0,0)(0,0)
        \put(0,0){\line(1,0){0.3}}
        \end{picture}}
   \end{picture}}

\put(9050,0){\setlength{\unitlength}{5cm}\begin{picture}(0,0)(0,-0.25)
   \put(0,0){\setlength{\unitlength}{1cm}\begin{picture}(0,0)(0,0)
        \put(0,0){\line(-1,0){0.3}}
        \end{picture}}
   \end{picture}}

\thinlines
\put(7350,0){\setlength{\unitlength}{5cm}\begin{picture}(0,0)(0,-0.25)
   \multiput(0,0)(0,0.1){3}{\setlength{\unitlength}{1cm}%
\begin{picture}(0,0)(0,0)
        \put(0,0){\line(1,0){0.12}}
        \end{picture}}
   \end{picture}}

\put(7350,0){\setlength{\unitlength}{5cm}\begin{picture}(0,0)(0,-0.25)
   \multiput(0,0)(0,-0.1){3}{\setlength{\unitlength}{1cm}%
\begin{picture}(0,0)(0,0)
        \put(0,0){\line(1,0){0.12}}
        \end{picture}}
   \end{picture}}

\put(9050,0){\setlength{\unitlength}{5cm}\begin{picture}(0,0)(0,-0.25)
   \multiput(0,0)(0,0.1){3}{\setlength{\unitlength}{1cm}%
\begin{picture}(0,0)(0,0)
        \put(0,0){\line(-1,0){0.12}}
        \end{picture}}
   \end{picture}}

\put(9050,0){\setlength{\unitlength}{5cm}\begin{picture}(0,0)(0,-0.25)
   \multiput(0,0)(0,-0.1){3}{\setlength{\unitlength}{1cm}%
\begin{picture}(0,0)(0,0)
        \put(0,0){\line(-1,0){0.12}}
        \end{picture}}
   \end{picture}}

   \put(7527.5, 236.111){\setlength{\unitlength}{1cm}\begin{picture}(0,0)(0,0)
        \put(0,0){\line(0,-1){0.2}}
   \end{picture}}
   \put(7892.5, 236.111){\setlength{\unitlength}{1cm}\begin{picture}(0,0)(0,0)
        \put(0,0){\line(0,-1){0.2}}
   \end{picture}}
   \put(8257.5, 236.111){\setlength{\unitlength}{1cm}\begin{picture}(0,0)(0,0)
        \put(0,0){\line(0,-1){0.2}}
   \end{picture}}
   \put(8622.5, 236.111){\setlength{\unitlength}{1cm}\begin{picture}(0,0)(0,0)
        \put(0,0){\line(0,-1){0.2}}
   \end{picture}}
   \put(8987.5, 236.111){\setlength{\unitlength}{1cm}\begin{picture}(0,0)(0,0)
        \put(0,0){\line(0,-1){0.2}}
   \end{picture}}
    \multiput(7350,0)(50,0){33}%
        {\setlength{\unitlength}{1cm}\begin{picture}(0,0)(0,0)
        \put(0,0){\line(0,1){0.12}}
    \end{picture}}
    \put(7500,0){\setlength{\unitlength}{1cm}\begin{picture}(0,0)(0,0)
        \put(0,0){\line(0,1){0.2}}
    \end{picture}}
    \put(7750,0){\setlength{\unitlength}{1cm}\begin{picture}(0,0)(0,0)
        \put(0,0){\line(0,1){0.2}}
   \end{picture}}
    \put(8000,0){\setlength{\unitlength}{1cm}\begin{picture}(0,0)(0,0)
        \put(0,0){\line(0,1){0.2}}
    \end{picture}}
    \put(8250,0){\setlength{\unitlength}{1cm}\begin{picture}(0,0)(0,0)
        \put(0,0){\line(0,1){0.2}}
    \end{picture}}
    \put(8500,0){\setlength{\unitlength}{1cm}\begin{picture}(0,0)(0,0)
        \put(0,0){\line(0,1){0.2}}
    \end{picture}}
    \put(8750,0){\setlength{\unitlength}{1cm}\begin{picture}(0,0)(0,0)
        \put(0,0){\line(0,1){0.2}}
    \end{picture}}
    \put(9000,0){\setlength{\unitlength}{1cm}\begin{picture}(0,0)(0,0)
        \put(0,0){\line(0,1){0.2}}
    \end{picture}}

\punkt{10/06/89}{23:22}{7688.474}{  0.029}{0.026}{1000}{1.53}{
1643}{1.2CA}
\punkt{27/06/89}{00:10}{7704.508}{  0.031}{0.034}{1000}{1.43}{
1463}{1.2CA}
\punkt{13/08/89}{21:35}{7752.400}{  0.013}{0.031}{1000}{1.55}{
2119}{1.2CA}
\punkt{15/08/89}{20:51}{7754.369}{  0.003}{0.040}{ 500}{1.18}{
1476}{1.2CA}
\punkt{16/08/89}{21:00}{7755.375}{  0.008}{0.037}{1000}{1.44}{
3160}{1.2CA}
\punkta{24/05/90}{00:53}{8035.537}{ -0.025}{0.054}{ 400}{2.76}{
1328}{1.2CA}
\punkt{}{}{8036.625}{0.045}{0.03}{}{}{}{}
\punkt{20/03/91}{05:04}{8335.711}{ -0.080}{0.026}{ 500}{1.18}{
520}{1.2CA}
\punkta{25/05/91}{23:29}{8402.479}{ -0.064}{0.050}{ 150}{2.17}{
594}{1.2CA}
\punkt{12/08/91}{22:01}{8481.418}{  0.003}{0.032}{ 500}{1.58}{
547}{1.2CA}
\punkt{12/08/91}{22:01}{8481.418}{  0.003}{0.032}{ 500}{1.58}{
547}{1.2CA}

\end{picture}}

\end{picture}

\vspace*{-0.02cm}

\begin{picture}(18 ,3.5 )(0,0)
\put(0,0){\setlength{\unitlength}{0.01059cm}%
\begin{picture}(1700, 330.555)(7350,0)
\put(7350,0){\framebox(1700, 330.555)[tl]{\begin{picture}(0,0)(0,0)
        \put(1700,0){\makebox(0,0)[tr]{\bf{1604+290}\T{0.4}
                                 \hspace*{0.5cm}}}
        \put(1700,-
330.555){\setlength{\unitlength}{1cm}\begin{picture}(0,0)(0,0)
        \end{picture}}
    \end{picture}}}

\thicklines
\put(7350,0){\setlength{\unitlength}{5cm}\begin{picture}(0,0)(0,-0.35)
   \put(0,0){\setlength{\unitlength}{1cm}\begin{picture}(0,0)(0,0)
        \put(0,0){\line(1,0){0.3}}
        \end{picture}}
   \end{picture}}

\put(9050,0){\setlength{\unitlength}{5cm}\begin{picture}(0,0)(0,-0.35)
   \put(0,0){\setlength{\unitlength}{1cm}\begin{picture}(0,0)(0,0)
        \put(0,0){\line(-1,0){0.3}}
        \end{picture}}
   \end{picture}}

\thinlines
\put(7350,0){\setlength{\unitlength}{5cm}\begin{picture}(0,0)(0,-0.35)
   \multiput(0,0)(0,0.1){4}{\setlength{\unitlength}{1cm}%
\begin{picture}(0,0)(0,0)
        \put(0,0){\line(1,0){0.12}}
        \end{picture}}
   \end{picture}}

\put(7350,0){\setlength{\unitlength}{5cm}\begin{picture}(0,0)(0,-0.35)
   \multiput(0,0)(0,-0.1){4}{\setlength{\unitlength}{1cm}%
\begin{picture}(0,0)(0,0)
        \put(0,0){\line(1,0){0.12}}
        \end{picture}}
   \end{picture}}

\put(9050,0){\setlength{\unitlength}{5cm}\begin{picture}(0,0)(0,-0.35)
   \multiput(0,0)(0,0.1){4}{\setlength{\unitlength}{1cm}%
\begin{picture}(0,0)(0,0)
        \put(0,0){\line(-1,0){0.12}}
        \end{picture}}
   \end{picture}}

\put(9050,0){\setlength{\unitlength}{5cm}\begin{picture}(0,0)(0,-0.35)
   \multiput(0,0)(0,-0.1){4}{\setlength{\unitlength}{1cm}%
\begin{picture}(0,0)(0,0)
        \put(0,0){\line(-1,0){0.12}}
        \end{picture}}
   \end{picture}}

   \put(7527.5, 330.555){\setlength{\unitlength}{1cm}\begin{picture}(0,0)(0,0)
        \put(0,0){\line(0,-1){0.2}}
   \end{picture}}
   \put(7892.5, 330.555){\setlength{\unitlength}{1cm}\begin{picture}(0,0)(0,0)
        \put(0,0){\line(0,-1){0.2}}
   \end{picture}}
   \put(8257.5, 330.555){\setlength{\unitlength}{1cm}\begin{picture}(0,0)(0,0)
        \put(0,0){\line(0,-1){0.2}}
   \end{picture}}
   \put(8622.5, 330.555){\setlength{\unitlength}{1cm}\begin{picture}(0,0)(0,0)
        \put(0,0){\line(0,-1){0.2}}
   \end{picture}}
   \put(8987.5, 330.555){\setlength{\unitlength}{1cm}\begin{picture}(0,0)(0,0)
        \put(0,0){\line(0,-1){0.2}}
   \end{picture}}
    \multiput(7350,0)(50,0){33}%
        {\setlength{\unitlength}{1cm}\begin{picture}(0,0)(0,0)
        \put(0,0){\line(0,1){0.12}}
    \end{picture}}
    \put(7500,0){\setlength{\unitlength}{1cm}\begin{picture}(0,0)(0,0)
        \put(0,0){\line(0,1){0.2}}
    \end{picture}}
    \put(7750,0){\setlength{\unitlength}{1cm}\begin{picture}(0,0)(0,0)
        \put(0,0){\line(0,1){0.2}}
    \end{picture}}
    \put(8000,0){\setlength{\unitlength}{1cm}\begin{picture}(0,0)(0,0)
        \put(0,0){\line(0,1){0.2}}
    \end{picture}}
    \put(8250,0){\setlength{\unitlength}{1cm}\begin{picture}(0,0)(0,0)
        \put(0,0){\line(0,1){0.2}}
    \end{picture}}
    \put(8500,0){\setlength{\unitlength}{1cm}\begin{picture}(0,0)(0,0)
        \put(0,0){\line(0,1){0.2}}
    \end{picture}}
    \put(8750,0){\setlength{\unitlength}{1cm}\begin{picture}(0,0)(0,0)
        \put(0,0){\line(0,1){0.2}}
    \end{picture}}
    \put(9000,0){\setlength{\unitlength}{1cm}\begin{picture}(0,0)(0,0)
        \put(0,0){\line(0,1){0.2}}
    \end{picture}}

\put(7000,0){\begin{picture}(0,0)(7000,-47.214)
\punkt{11/06/89}{23:43}{7689.489}{ -0.147}{0.038}{ 500}{1.68}{
1146}{3.5CA}
\punkt{14/06/89}{00:11}{7691.508}{ -0.154}{0.088}{ 500}{1.54}{
1380}{3.5CA}
\punkt{28/06/89}{23:35}{7706.483}{ -0.247}{0.097}{1000}{2.56}{
1320}{3.5CA}
\punkt{23/07/89}{22:54}{7731.454}{ -0.186}{0.023}{ 500}{2.02}{
1109}{3.5CA}
\punkt{11/08/89}{21:13}{7750.385}{ -0.046}{0.021}{ 500}{2.07}{
1188}{3.5CA}
\punkt{24/05/90}{01:19}{8035.555}{  0.019}{0.023}{ 400}{3.58}{
1320}{3.5CA}
\punkt{25/05/90}{01:41}{8036.570}{  0.025}{0.021}{ 400}{2.09}{
1317}{3.5CA}
\punkt{25/05/90}{02:06}{8036.588}{  0.028}{0.022}{ 400}{1.63}{
1424}{3.5CA}
\punkt{31/07/90}{22:13}{8104.426}{  0.053}{0.024}{ 500}{1.79}{
1066}{3.5CA}
\punkta{24/05/91}{23:31}{8401.480}{  0.084}{0.140}{ 500}{2.20}{
1271}{3.5CA}
\punkt{25/05/91}{23:44}{8402.489}{  0.112}{0.05}{ 500}{1.97}{    1432}{3.5CA}
\punkt{24/07/91}{21:38}{8462.402}{  0.071}{0.040}{ 500}{1.45}{
1800}{3.5CA}
\punkt{25/07/91}{22:03}{8463.419}{  0.128}{0.042}{ 500}{1.51}{
1639}{3.5CA}
\punkt{30/07/91}{21:10}{8468.383}{  0.031}{0.050}{ 500}{2.47}{
765}{3.5CA}
\punkt{31/07/91}{21:42}{8469.405}{  0.078}{0.011}{ 500}{1.22}{
548}{3.5CA}
\punkta{09/08/91}{21:43}{8478.405}{  0.086}{0.104}{1000}{5.34}{
852}{3.5CA}
\punkta{10/08/91}{21:16}{8479.387}{ -0.002}{0.075}{ 500}{3.40}{
561}{3.5CA}
\punkt{20/10/91}{19:05}{8550.296}{ -0.010}{0.044}{ 500}{1.55}{
1502}{3.5CA}
\punkt{16/02/92}{05:41}{8668.737}{ -0.026}{0.016}{ 500}{1.34}{
582}{3.5CA}
\punkt{16/02/92}{05:41}{8668.737}{ -0.026}{0.016}{ 500}{1.34}{
582}{3.5CA}

\end{picture}}

\end{picture}}

\end{picture}

\vspace*{-0.02cm}

\begin{picture}(18 ,2.5 )(0,0)
\put(0,0){\setlength{\unitlength}{0.01059cm}%
\begin{picture}(1700, 236.111)(7350,0)
\put(7350,0){\framebox(1700, 236.111)[tl]{\begin{picture}(0,0)(0,0)
        \put(1700,0){\makebox(0,0)[tr]{\bf{1630+377}\T{0.4}
                                 \hspace*{0.5cm}}}
    \end{picture}}}

\thicklines
\put(7350,0){\setlength{\unitlength}{5cm}\begin{picture}(0,0)(0,-0.25)
   \put(0,0){\setlength{\unitlength}{1cm}\begin{picture}(0,0)(0,0)
        \put(0,0){\line(1,0){0.3}}
        \end{picture}}
   \end{picture}}

\put(9050,0){\setlength{\unitlength}{5cm}\begin{picture}(0,0)(0,-0.25)
   \put(0,0){\setlength{\unitlength}{1cm}\begin{picture}(0,0)(0,0)
        \put(0,0){\line(-1,0){0.3}}
        \end{picture}}
   \end{picture}}

\thinlines
\put(7350,0){\setlength{\unitlength}{5cm}\begin{picture}(0,0)(0,-0.25)
   \multiput(0,0)(0,0.1){3}{\setlength{\unitlength}{1cm}%
\begin{picture}(0,0)(0,0)
        \put(0,0){\line(1,0){0.12}}
        \end{picture}}
   \end{picture}}

\put(7350,0){\setlength{\unitlength}{5cm}\begin{picture}(0,0)(0,-0.25)
   \multiput(0,0)(0,-0.1){3}{\setlength{\unitlength}{1cm}%
\begin{picture}(0,0)(0,0)
        \put(0,0){\line(1,0){0.12}}
        \end{picture}}
   \end{picture}}

\put(9050,0){\setlength{\unitlength}{5cm}\begin{picture}(0,0)(0,-0.25)
   \multiput(0,0)(0,0.1){3}{\setlength{\unitlength}{1cm}%
\begin{picture}(0,0)(0,0)
        \put(0,0){\line(-1,0){0.12}}
        \end{picture}}
   \end{picture}}

\put(9050,0){\setlength{\unitlength}{5cm}\begin{picture}(0,0)(0,-0.25)
   \multiput(0,0)(0,-0.1){3}{\setlength{\unitlength}{1cm}%
\begin{picture}(0,0)(0,0)
        \put(0,0){\line(-1,0){0.12}}
        \end{picture}}
   \end{picture}}

   \put(7527.5, 236.111){\setlength{\unitlength}{1cm}\begin{picture}(0,0)(0,0)
        \put(0,0){\line(0,-1){0.2}}
   \end{picture}}
   \put(7892.5, 236.111){\setlength{\unitlength}{1cm}\begin{picture}(0,0)(0,0)
        \put(0,0){\line(0,-1){0.2}}
   \end{picture}}
   \put(8257.5, 236.111){\setlength{\unitlength}{1cm}\begin{picture}(0,0)(0,0)
        \put(0,0){\line(0,-1){0.2}}
   \end{picture}}
   \put(8622.5, 236.111){\setlength{\unitlength}{1cm}\begin{picture}(0,0)(0,0)
        \put(0,0){\line(0,-1){0.2}}
   \end{picture}}
   \put(8987.5, 236.111){\setlength{\unitlength}{1cm}\begin{picture}(0,0)(0,0)
        \put(0,0){\line(0,-1){0.2}}
   \end{picture}}
    \multiput(7350,0)(50,0){33}%
        {\setlength{\unitlength}{1cm}\begin{picture}(0,0)(0,0)
        \put(0,0){\line(0,1){0.12}}
    \end{picture}}
    \put(7500,0){\setlength{\unitlength}{1cm}\begin{picture}(0,0)(0,0)
        \put(0,0){\line(0,1){0.2}}
    \end{picture}}
    \put(7750,0){\setlength{\unitlength}{1cm}\begin{picture}(0,0)(0,0)
        \put(0,0){\line(0,1){0.2}}
    \end{picture}}
    \put(8000,0){\setlength{\unitlength}{1cm}\begin{picture}(0,0)(0,0)
        \put(0,0){\line(0,1){0.2}}
    \end{picture}}
    \put(8250,0){\setlength{\unitlength}{1cm}\begin{picture}(0,0)(0,0)
        \put(0,0){\line(0,1){0.2}}
    \end{picture}}
    \put(8500,0){\setlength{\unitlength}{1cm}\begin{picture}(0,0)(0,0)
        \put(0,0){\line(0,1){0.2}}
    \end{picture}}
    \put(8750,0){\setlength{\unitlength}{1cm}\begin{picture}(0,0)(0,0)
        \put(0,0){\line(0,1){0.2}}
    \end{picture}}
    \put(9000,0){\setlength{\unitlength}{1cm}\begin{picture}(0,0)(0,0)
        \put(0,0){\line(0,1){0.2}}
    \end{picture}}

\punkt{}{}{7690.475}{-0.022}{0.02}{}{}{}{}
\punkt{23/07/89}{23:10}{7731.465}{ -0.054}{0.026}{ 500}{2.03}{
983}{1.2CA}
\punkt{15/08/89}{21:08}{7754.381}{ -0.060}{0.015}{ 500}{1.52}{
1457}{1.2CA}
\punkt{23/05/90}{01:55}{8034.580}{ -0.027}{0.035}{ 400}{2.02}{
1427}{1.2CA}
\punkt{25/05/90}{03:43}{8036.655}{ -0.043}{0.030}{ 200}{1.59}{
2050}{1.2CA}
\punkt{13/08/90}{00:05}{8116.504}{  0.013}{0.054}{ 221}{1.47}{
1971}{1.2CA}
\punkt{03/02/91}{05:13}{8290.718}{ -0.019}{0.014}{ 500}{2.27}{
953}{1.2CA}
\punkt{25/02/91}{05:15}{8312.719}{ -0.029}{0.019}{ 300}{3.58}{
1068}{1.2CA}
\punkt{23/05/91}{00:01}{8399.501}{ -0.006}{0.030}{ 200}{1.50}{
418}{1.2CA}
\punkt{24/05/91}{01:07}{8400.547}{  0.018}{0.019}{ 300}{1.36}{
544}{1.2CA}
\punkt{29/05/91}{00:26}{8405.518}{ -0.013}{0.029}{ 500}{1.38}{
2708}{1.2CA}
\punkt{}{}{8464.395}{0.019}{0.01}{}{}{}{}
\punkt{10/08/91}{21:37}{8479.401}{  0.009}{0.018}{ 500}{3.20}{
502}{1.2CA}
\punkt{20/10/91}{19:21}{8550.306}{  0.006}{0.012}{ 500}{1.63}{
1403}{1.2CA}
\punkt{28/10/91}{19:39}{8558.319}{  0.030}{0.015}{ 500}{1.26}{
1536}{1.2CA}
\punkt{28/10/91}{19:39}{8558.319}{  0.030}{0.015}{ 500}{1.26}{
1536}{1.2CA}

\end{picture}}

\end{picture}

\vspace*{-0.02cm}

\begin{picture}(18 ,2.5 )(0,0)
\put(0,0){\setlength{\unitlength}{0.01059cm}%
\begin{picture}(1700, 236.111)(7350,0)
\put(7350,0){\framebox(1700, 236.111)[tl]{\begin{picture}(0,0)(0,0)
        \put(1700,0){\makebox(0,0)[tr]{\bf{1633+267}\T{0.4}
                                 \hspace*{0.5cm}}}
   \put(1700,- 236.111){\setlength{\unitlength}{1cm}\begin{picture}(0,0)(0,0)
            \put(0,-1){\makebox(0,0)[br]{\bf J.D.\,2,440,000\,+}}
        \end{picture}}
    \end{picture}}}

\thicklines
\put(7350,0){\setlength{\unitlength}{5cm}\begin{picture}(0,0)(0,-0.25)
   \put(0,0){\setlength{\unitlength}{1cm}\begin{picture}(0,0)(0,0)
        \put(0,0){\line(1,0){0.3}}
        \end{picture}}
   \end{picture}}

\put(9050,0){\setlength{\unitlength}{5cm}\begin{picture}(0,0)(0,-0.25)
   \put(0,0){\setlength{\unitlength}{1cm}\begin{picture}(0,0)(0,0)
        \put(0,0){\line(-1,0){0.3}}
        \end{picture}}
   \end{picture}}

\thinlines
\put(7350,0){\setlength{\unitlength}{5cm}\begin{picture}(0,0)(0,-0.25)
   \multiput(0,0)(0,0.1){3}{\setlength{\unitlength}{1cm}%
\begin{picture}(0,0)(0,0)
        \put(0,0){\line(1,0){0.12}}
        \end{picture}}
   \end{picture}}

\put(7350,0){\setlength{\unitlength}{5cm}\begin{picture}(0,0)(0,-0.25)
   \multiput(0,0)(0,-0.1){3}{\setlength{\unitlength}{1cm}%
\begin{picture}(0,0)(0,0)
        \put(0,0){\line(1,0){0.12}}
        \end{picture}}
   \end{picture}}

\put(9050,0){\setlength{\unitlength}{5cm}\begin{picture}(0,0)(0,-0.25)
   \multiput(0,0)(0,0.1){3}{\setlength{\unitlength}{1cm}%
\begin{picture}(0,0)(0,0)
        \put(0,0){\line(-1,0){0.12}}
        \end{picture}}
   \end{picture}}

\put(9050,0){\setlength{\unitlength}{5cm}\begin{picture}(0,0)(0,-0.25)
   \multiput(0,0)(0,-0.1){3}{\setlength{\unitlength}{1cm}%
\begin{picture}(0,0)(0,0)
        \put(0,0){\line(-1,0){0.12}}
        \end{picture}}
   \end{picture}}

   \put(7527.5, 236.111){\setlength{\unitlength}{1cm}\begin{picture}(0,0)(0,0)
        \put(0,0){\line(0,-1){0.2}}
   \end{picture}}
   \put(7892.5, 236.111){\setlength{\unitlength}{1cm}\begin{picture}(0,0)(0,0)
        \put(0,0){\line(0,-1){0.2}}
   \end{picture}}
   \put(8257.5, 236.111){\setlength{\unitlength}{1cm}\begin{picture}(0,0)(0,0)
        \put(0,0){\line(0,-1){0.2}}
   \end{picture}}
   \put(8622.5, 236.111){\setlength{\unitlength}{1cm}\begin{picture}(0,0)(0,0)
        \put(0,0){\line(0,-1){0.2}}
   \end{picture}}
   \put(8987.5, 236.111){\setlength{\unitlength}{1cm}\begin{picture}(0,0)(0,0)
        \put(0,0){\line(0,-1){0.2}}
   \end{picture}}
    \multiput(7350,0)(50,0){33}%
        {\setlength{\unitlength}{1cm}\begin{picture}(0,0)(0,0)
        \put(0,0){\line(0,1){0.12}}
    \end{picture}}
    \put(7500,0){\setlength{\unitlength}{1cm}\begin{picture}(0,0)(0,0)
        \put(0,0){\line(0,1){0.2}}
        \put(0,-0.2){\makebox(0,0)[t]{\bf 7500}}
   \end{picture}}
    \put(7750,0){\setlength{\unitlength}{1cm}\begin{picture}(0,0)(0,0)
        \put(0,0){\line(0,1){0.2}}
        \put(0,-0.2){\makebox(0,0)[t]{\bf 7750}}
    \end{picture}}
   \put(8000,0){\setlength{\unitlength}{1cm}\begin{picture}(0,0)(0,0)
        \put(0,0){\line(0,1){0.2}}
        \put(0,-0.2){\makebox(0,0)[t]{\bf 8000}}
    \end{picture}}
    \put(8250,0){\setlength{\unitlength}{1cm}\begin{picture}(0,0)(0,0)
       \put(0,0){\line(0,1){0.2}}
        \put(0,-0.2){\makebox(0,0)[t]{\bf 8250}}
    \end{picture}}
    \put(8500,0){\setlength{\unitlength}{1cm}\begin{picture}(0,0)(0,0)
        \put(0,0){\line(0,1){0.2}}
        \put(0,-0.2){\makebox(0,0)[t]{\bf 8500}}
    \end{picture}}
    \put(8750,0){\setlength{\unitlength}{1cm}\begin{picture}(0,0)(0,0)
        \put(0,0){\line(0,1){0.2}}
        \put(0,-0.2){\makebox(0,0)[t]{\bf 8750}}
    \end{picture}}
   \put(9000,0){\setlength{\unitlength}{1cm}\begin{picture}(0,0)(0,0)
        \put(0,0){\line(0,1){0.2}}
        \put(0,-0.2){\makebox(0,0)[t]{\bf 9000}}
    \end{picture}}

\punkt{12/06/89}{00:20}{7689.514}{  0.068}{0.021}{ 500}{1.66}{
999}{1.2CA}
\punkt{12/06/89}{23:28}{7690.478}{  0.041}{0.016}{1000}{1.56}{
1660}{1.2CA}
\punkt{25/06/89}{02:19}{7702.597}{  0.083}{0.027}{ 500}{1.81}{
1041}{1.2CA}
\punkt{23/07/89}{23:26}{7731.477}{  0.020}{0.021}{ 500}{1.84}{
979}{1.2CA}
\punkt{15/08/89}{21:30}{7754.396}{  0.072}{0.020}{ 500}{1.82}{
1441}{1.2CA}
\punkta{04/09/89}{07:48}{7773.825}{  0.042}{0.018}{****}{3.54}{
712}{1.2CA}
\punkt{23/05/90}{02:15}{8034.594}{  0.035}{0.055}{ 400}{1.99}{
1519}{1.2CA}
\punkt{13/08/90}{00:11}{8116.508}{  0.018}{0.026}{  90}{1.66}{
930}{1.2CA}
\punkt{14/08/90}{20:12}{8118.342}{  0.060}{0.044}{ 100}{1.63}{
751}{1.2CA}
\punkta{26/05/91}{00:06}{8402.504}{ -0.095}{0.05}{ 300}{2.24}{
1053}{1.2CA}
\punkta{26/07/91}{21:46}{8464.408}{ -0.131}{0.072}{ 300}{1.10}{
3908}{1.2CA}
\punkta{10/08/91}{22:07}{8479.422}{ -0.090}{0.090}{ 500}{3.22}{
539}{1.2CA}

    \put(7350,0){\setlength{\unitlength}{1cm}\begin{picture}(0,0)(0,0)
        \put(0,-1.3){\makebox(0,0)[tl]{\footnotesize{\bf Fig.~1.} (continued)}}
    \end{picture}}

\end{picture}}

\end{picture}

\vspace*{1.3cm}
\end{figure*}

\begin{figure*}

\vspace*{0.4cm}

\begin{picture}(18 ,2.5 )(0,0)
\put(0,0){\setlength{\unitlength}{0.01059cm}%
\begin{picture}(1700, 236.111)(7350,0)
\put(7350,0){\framebox(1700, 236.111)[tl]{\begin{picture}(0,0)(0,0)
        \put(1700,0){\makebox(0,0)[tr]{\bf{1634+706}\T{0.4}
                                 \hspace*{0.5cm}}}
    \end{picture}}}

\thicklines
\put(7350,0){\setlength{\unitlength}{5cm}\begin{picture}(0,0)(0,-0.25)
   \put(0,0){\setlength{\unitlength}{1cm}\begin{picture}(0,0)(0,0)
        \put(0,0){\line(1,0){0.3}}
        \end{picture}}
   \end{picture}}

\put(9050,0){\setlength{\unitlength}{5cm}\begin{picture}(0,0)(0,-0.25)
   \put(0,0){\setlength{\unitlength}{1cm}\begin{picture}(0,0)(0,0)
        \put(0,0){\line(-1,0){0.3}}
        \end{picture}}
   \end{picture}}

\thinlines
\put(7350,0){\setlength{\unitlength}{5cm}\begin{picture}(0,0)(0,-0.25)
   \multiput(0,0)(0,0.1){3}{\setlength{\unitlength}{1cm}%
\begin{picture}(0,0)(0,0)
        \put(0,0){\line(1,0){0.12}}
        \end{picture}}
   \end{picture}}

\put(7350,0){\setlength{\unitlength}{5cm}\begin{picture}(0,0)(0,-0.25)
   \multiput(0,0)(0,-0.1){3}{\setlength{\unitlength}{1cm}%
\begin{picture}(0,0)(0,0)
        \put(0,0){\line(1,0){0.12}}
        \end{picture}}
   \end{picture}}

\put(9050,0){\setlength{\unitlength}{5cm}\begin{picture}(0,0)(0,-0.25)
   \multiput(0,0)(0,0.1){3}{\setlength{\unitlength}{1cm}%
\begin{picture}(0,0)(0,0)
        \put(0,0){\line(-1,0){0.12}}
        \end{picture}}
   \end{picture}}

\put(9050,0){\setlength{\unitlength}{5cm}\begin{picture}(0,0)(0,-0.25)
   \multiput(0,0)(0,-0.1){3}{\setlength{\unitlength}{1cm}%
\begin{picture}(0,0)(0,0)
        \put(0,0){\line(-1,0){0.12}}
        \end{picture}}
   \end{picture}}

   \put(7527.5, 236.111){\setlength{\unitlength}{1cm}\begin{picture}(0,0)(0,0)
        \put(0,0){\line(0,-1){0.2}}
        \put(0,0.2){\makebox(0,0)[b]{\bf 1989}}
   \end{picture}}
   \put(7892.5, 236.111){\setlength{\unitlength}{1cm}\begin{picture}(0,0)(0,0)
        \put(0,0){\line(0,-1){0.2}}
        \put(0,0.2){\makebox(0,0)[b]{\bf 1990}}
   \end{picture}}
   \put(8257.5, 236.111){\setlength{\unitlength}{1cm}\begin{picture}(0,0)(0,0)
        \put(0,0){\line(0,-1){0.2}}
        \put(0,0.2){\makebox(0,0)[b]{\bf 1991}}
   \end{picture}}
   \put(8622.5, 236.111){\setlength{\unitlength}{1cm}\begin{picture}(0,0)(0,0)
        \put(0,0){\line(0,-1){0.2}}
        \put(0,0.2){\makebox(0,0)[b]{\bf 1992}}
   \end{picture}}
   \put(8987.5, 236.111){\setlength{\unitlength}{1cm}\begin{picture}(0,0)(0,0)
        \put(0,0){\line(0,-1){0.2}}
        \put(0,0.2){\makebox(0,0)[b]{\bf 1993}}
   \end{picture}}
    \multiput(7350,0)(50,0){33}%
        {\setlength{\unitlength}{1cm}\begin{picture}(0,0)(0,0)
        \put(0,0){\line(0,1){0.12}}
    \end{picture}}
    \put(7500,0){\setlength{\unitlength}{1cm}\begin{picture}(0,0)(0,0)
        \put(0,0){\line(0,1){0.2}}
    \end{picture}}
    \put(7750,0){\setlength{\unitlength}{1cm}\begin{picture}(0,0)(0,0)
        \put(0,0){\line(0,1){0.2}}
    \end{picture}}
    \put(8000,0){\setlength{\unitlength}{1cm}\begin{picture}(0,0)(0,0)
        \put(0,0){\line(0,1){0.2}}
    \end{picture}}
    \put(8250,0){\setlength{\unitlength}{1cm}\begin{picture}(0,0)(0,0)
        \put(0,0){\line(0,1){0.2}}
    \end{picture}}
    \put(8500,0){\setlength{\unitlength}{1cm}\begin{picture}(0,0)(0,0)
        \put(0,0){\line(0,1){0.2}}
    \end{picture}}
    \put(8750,0){\setlength{\unitlength}{1cm}\begin{picture}(0,0)(0,0)
        \put(0,0){\line(0,1){0.2}}
    \end{picture}}
    \put(9000,0){\setlength{\unitlength}{1cm}\begin{picture}(0,0)(0,0)
        \put(0,0){\line(0,1){0.2}}
    \end{picture}}

\punkta{12/06/89}{01:43}{7689.572}{ -0.009}{0.006}{1000}{1.39}{
1622}{1.2CA}
\punkt{13/06/89}{01:33}{7690.565}{ -0.005}{0.013}{1000}{1.16}{
1232}{1.2CA}
\punkt{25/06/89}{23:12}{7703.467}{ -0.009}{0.010}{1000}{1.63}{
1080}{1.2CA}
\punkt{28/06/89}{00:39}{7705.527}{ -0.001}{0.015}{1000}{2.15}{
1084}{1.2CA}
\punkt{24/07/89}{00:08}{7731.506}{ -0.020}{0.008}{ 500}{2.33}{
1043}{1.2CA}
\punkt{15/08/89}{21:47}{7754.408}{  0.006}{0.016}{ 500}{1.29}{
1479}{1.2CA}
\punkt{31/08/89}{20:14}{7770.344}{ -0.029}{0.009}{ 500}{3.08}{
1219}{1.2CA}
\punkt{01/10/89}{10:07}{7800.922}{ -0.023}{0.020}{ 500}{2.01}{
566}{1.2CA}
\punkt{31/10/89}{19:24}{7831.309}{ -0.011}{0.020}{ 500}{2.30}{
604}{1.2CA}
\punkta{23/05/90}{02:51}{8034.619}{  0.057}{0.006}{ 400}{2.13}{
1466}{1.2CA}
\punkt{16/10/90}{21:35}{8181.400}{ -0.016}{0.026}{ 500}{2.91}{
870}{1.2CA}
\punkt{17/10/90}{21:35}{8182.400}{ -0.017}{0.011}{ 500}{2.90}{
870}{1.2CA}
\punkt{28/05/91}{00:02}{8404.502}{  0.021}{0.025}{ 300}{2.25}{
1107}{1.2CA}
\punkt{10/08/91}{22:40}{8479.445}{  0.025}{0.010}{ 500}{2.63}{
613}{1.2CA}
\end{picture}}

\end{picture}

\vspace*{-0.02cm}

\begin{picture}(18 ,3.5 )(0,0)
\put(0,0){\setlength{\unitlength}{0.01059cm}%
\begin{picture}(1700, 330.555)(7350,0)
\put(7350,0){\framebox(1700, 330.555)[tl]{\begin{picture}(0,0)(0,0)
        \put(1700,0){\makebox(0,0)[tr]{\bf{1640+396}\T{0.4}
                                 \hspace*{0.5cm}}}
    \end{picture}}}

\thicklines
\put(7350,0){\setlength{\unitlength}{5cm}\begin{picture}(0,0)(0,-0.45)
   \put(0,0){\setlength{\unitlength}{1cm}\begin{picture}(0,0)(0,0)
        \put(0,0){\line(1,0){0.3}}
        \end{picture}}
   \end{picture}}

\put(9050,0){\setlength{\unitlength}{5cm}\begin{picture}(0,0)(0,-0.45)
   \put(0,0){\setlength{\unitlength}{1cm}\begin{picture}(0,0)(0,0)
        \put(0,0){\line(-1,0){0.3}}
        \end{picture}}
   \end{picture}}

\thinlines
\put(7350,0){\setlength{\unitlength}{5cm}\begin{picture}(0,0)(0,-0.45)
   \multiput(0,0)(0,0.1){3}{\setlength{\unitlength}{1cm}%
\begin{picture}(0,0)(0,0)
        \put(0,0){\line(1,0){0.12}}
        \end{picture}}
   \end{picture}}

\put(7350,0){\setlength{\unitlength}{5cm}\begin{picture}(0,0)(0,-0.45)
   \multiput(0,0)(0,-0.1){5}{\setlength{\unitlength}{1cm}%
\begin{picture}(0,0)(0,0)
        \put(0,0){\line(1,0){0.12}}
        \end{picture}}
   \end{picture}}

\put(9050,0){\setlength{\unitlength}{5cm}\begin{picture}(0,0)(0,-0.45)
   \multiput(0,0)(0,0.1){3}{\setlength{\unitlength}{1cm}%
\begin{picture}(0,0)(0,0)
        \put(0,0){\line(-1,0){0.12}}
        \end{picture}}
   \end{picture}}

\put(9050,0){\setlength{\unitlength}{5cm}\begin{picture}(0,0)(0,-0.45)
   \multiput(0,0)(0,-0.1){5}{\setlength{\unitlength}{1cm}%
\begin{picture}(0,0)(0,0)
        \put(0,0){\line(-1,0){0.12}}
        \end{picture}}
   \end{picture}}

   \put(7527.5, 330.555){\setlength{\unitlength}{1cm}\begin{picture}(0,0)(0,0)
        \put(0,0){\line(0,-1){0.2}}
   \end{picture}}
   \put(7892.5, 330.555){\setlength{\unitlength}{1cm}\begin{picture}(0,0)(0,0)
        \put(0,0){\line(0,-1){0.2}}
   \end{picture}}
   \put(8257.5, 330.555){\setlength{\unitlength}{1cm}\begin{picture}(0,0)(0,0)
        \put(0,0){\line(0,-1){0.2}}
   \end{picture}}
   \put(8622.5, 330.555){\setlength{\unitlength}{1cm}\begin{picture}(0,0)(0,0)
        \put(0,0){\line(0,-1){0.2}}
   \end{picture}}
   \put(8987.5, 330.555){\setlength{\unitlength}{1cm}\begin{picture}(0,0)(0,0)
        \put(0,0){\line(0,-1){0.2}}
   \end{picture}}
    \multiput(7350,0)(50,0){33}%
        {\setlength{\unitlength}{1cm}\begin{picture}(0,0)(0,0)
        \put(0,0){\line(0,1){0.12}}
    \end{picture}}
    \put(7500,0){\setlength{\unitlength}{1cm}\begin{picture}(0,0)(0,0)
        \put(0,0){\line(0,1){0.2}}
    \end{picture}}
    \put(7750,0){\setlength{\unitlength}{1cm}\begin{picture}(0,0)(0,0)
        \put(0,0){\line(0,1){0.2}}
    \end{picture}}
    \put(8000,0){\setlength{\unitlength}{1cm}\begin{picture}(0,0)(0,0)
        \put(0,0){\line(0,1){0.2}}
    \end{picture}}
    \put(8250,0){\setlength{\unitlength}{1cm}\begin{picture}(0,0)(0,0)
        \put(0,0){\line(0,1){0.2}}
    \end{picture}}
    \put(8500,0){\setlength{\unitlength}{1cm}\begin{picture}(0,0)(0,0)
        \put(0,0){\line(0,1){0.2}}
    \end{picture}}
    \put(8750,0){\setlength{\unitlength}{1cm}\begin{picture}(0,0)(0,0)
        \put(0,0){\line(0,1){0.2}}
    \end{picture}}
    \put(9000,0){\setlength{\unitlength}{1cm}\begin{picture}(0,0)(0,0)
        \put(0,0){\line(0,1){0.2}}
    \end{picture}}

\put(7000,0){\begin{picture}(0,0)(7000,-94.4287)
\punkt{26/06/89}{01:11}{7703.550}{  0.054}{0.026}{ 500}{1.36}{
997}{1.2CA}
\punkt{24/07/89}{00:46}{7731.532}{  0.052}{0.019}{ 500}{2.24}{
1249}{1.2CA}
\punkta{14/08/89}{22:47}{7753.450}{  0.070}{0.102}{ 500}{2.06}{
1232}{1.2CA}
\punkta{01/08/90}{22:24}{8105.434}{ -0.348}{0.144}{ 500}{1.59}{
1568}{1.2CA}
\punkt{15/02/91}{04:54}{8302.704}{ -0.344}{0.040}{ 500}{2.32}{
512}{1.2CA}
\punkta{}{}{8400.565}{-0.012}{}{}{}{}{}
\punkta{26/05/91}{00:04}{8402.503}{  0.025}{0.325}{ 500}{2.36}{
1617}{1.2CA}
\punkta{29/05/91}{01:38}{8405.568}{  0.017}{0.323}{ 500}{1.47}{
3475}{1.2CA}
\punkt{24/07/91}{22:08}{8462.423}{  0.094}{0.064}{ 500}{1.71}{
1662}{1.2CA}
\punkt{25/07/91}{23:24}{8463.476}{ -0.004}{0.034}{ 500}{1.21}{
1685}{1.2CA}
\punkt{28/07/91}{22:01}{8466.418}{ -0.010}{0.074}{ 500}{1.43}{
1736}{1.2CA}
\punkta{29/07/91}{21:59}{8467.416}{  0.018}{0.500}{ 500}{3.70}{
758}{1.2CA}
\punkta{30/07/91}{21:53}{8468.412}{  0.078}{0.135}{ 100}{2.44}{
392}{1.2CA}
\punkt{31/07/91}{22:18}{8469.430}{  0.029}{0.017}{ 500}{1.05}{
589}{1.2CA}
\punkt{01/08/91}{21:41}{8470.404}{  0.033}{0.013}{ 500}{1.21}{
475}{1.2CA}
\punkt{02/08/91}{21:28}{8471.395}{  0.047}{0.015}{ 500}{1.36}{
507}{1.2CA}
\punkt{03/08/91}{22:33}{8472.440}{  0.045}{0.015}{ 500}{1.30}{
490}{1.2CA}
\punkt{04/08/91}{22:45}{8473.448}{  0.065}{0.012}{ 500}{1.33}{
503}{1.2CA}
\punkt{05/08/91}{22:21}{8474.431}{  0.052}{0.011}{ 500}{1.19}{
462}{1.2CA}
\punkta{11/08/91}{00:46}{8479.532}{  0.012}{0.108}{ 500}{3.81}{
579}{1.2CA}
\punkt{12/08/91}{00:27}{8480.519}{ -0.065}{0.035}{ 500}{1.65}{
634}{1.2CA}
\punkt{12/08/91}{22:16}{8481.428}{  0.045}{0.014}{ 500}{1.16}{
533}{1.2CA}
\punkt{19/08/91}{21:17}{8488.387}{  0.069}{0.041}{ 500}{1.35}{
852}{1.2CA}
\punkt{03/02/92}{05:35}{8655.733}{ -0.261}{0.075}{ 500}{1.48}{
547}{1.2CA}

\end{picture}}

\end{picture}}

\end{picture}

\vspace*{-0.02cm}

\begin{picture}(18 ,2.5 )(0,0)
\put(0,0){\setlength{\unitlength}{0.01059cm}%
\begin{picture}(1700, 236.111)(7350,0)
\put(7350,0){\framebox(1700, 236.111)[tl]{\begin{picture}(0,0)(0,0)
        \put(1700,0){\makebox(0,0)[tr]{\bf{1701+610}\T{0.4}
                                 \hspace*{0.5cm}}}
    \end{picture}}}

\thicklines
\put(7350,0){\setlength{\unitlength}{5cm}\begin{picture}(0,0)(0,-0.25)
   \put(0,0){\setlength{\unitlength}{1cm}\begin{picture}(0,0)(0,0)
        \put(0,0){\line(1,0){0.3}}
        \end{picture}}
   \end{picture}}

\put(9050,0){\setlength{\unitlength}{5cm}\begin{picture}(0,0)(0,-0.25)
   \put(0,0){\setlength{\unitlength}{1cm}\begin{picture}(0,0)(0,0)
        \put(0,0){\line(-1,0){0.3}}
        \end{picture}}
   \end{picture}}

\thinlines
\put(7350,0){\setlength{\unitlength}{5cm}\begin{picture}(0,0)(0,-0.25)
   \multiput(0,0)(0,0.1){3}{\setlength{\unitlength}{1cm}%
\begin{picture}(0,0)(0,0)
        \put(0,0){\line(1,0){0.12}}
        \end{picture}}
   \end{picture}}

\put(7350,0){\setlength{\unitlength}{5cm}\begin{picture}(0,0)(0,-0.25)
   \multiput(0,0)(0,-0.1){3}{\setlength{\unitlength}{1cm}%
\begin{picture}(0,0)(0,0)
        \put(0,0){\line(1,0){0.12}}
        \end{picture}}
   \end{picture}}

\put(9050,0){\setlength{\unitlength}{5cm}\begin{picture}(0,0)(0,-0.25)
   \multiput(0,0)(0,0.1){3}{\setlength{\unitlength}{1cm}%
\begin{picture}(0,0)(0,0)
        \put(0,0){\line(-1,0){0.12}}
        \end{picture}}
   \end{picture}}

\put(9050,0){\setlength{\unitlength}{5cm}\begin{picture}(0,0)(0,-0.25)
   \multiput(0,0)(0,-0.1){3}{\setlength{\unitlength}{1cm}%
\begin{picture}(0,0)(0,0)
        \put(0,0){\line(-1,0){0.12}}
        \end{picture}}
   \end{picture}}

   \put(7527.5, 236.111){\setlength{\unitlength}{1cm}\begin{picture}(0,0)(0,0)
        \put(0,0){\line(0,-1){0.2}}
   \end{picture}}
   \put(7892.5, 236.111){\setlength{\unitlength}{1cm}\begin{picture}(0,0)(0,0)
        \put(0,0){\line(0,-1){0.2}}
   \end{picture}}
   \put(8257.5, 236.111){\setlength{\unitlength}{1cm}\begin{picture}(0,0)(0,0)
        \put(0,0){\line(0,-1){0.2}}
   \end{picture}}
   \put(8622.5, 236.111){\setlength{\unitlength}{1cm}\begin{picture}(0,0)(0,0)
        \put(0,0){\line(0,-1){0.2}}
   \end{picture}}
   \put(8987.5, 236.111){\setlength{\unitlength}{1cm}\begin{picture}(0,0)(0,0)
        \put(0,0){\line(0,-1){0.2}}
   \end{picture}}
    \multiput(7350,0)(50,0){33}%
        {\setlength{\unitlength}{1cm}\begin{picture}(0,0)(0,0)
        \put(0,0){\line(0,1){0.12}}
    \end{picture}}
    \put(7500,0){\setlength{\unitlength}{1cm}\begin{picture}(0,0)(0,0)
        \put(0,0){\line(0,1){0.2}}
    \end{picture}}
    \put(7750,0){\setlength{\unitlength}{1cm}\begin{picture}(0,0)(0,0)
        \put(0,0){\line(0,1){0.2}}
    \end{picture}}
    \put(8000,0){\setlength{\unitlength}{1cm}\begin{picture}(0,0)(0,0)
        \put(0,0){\line(0,1){0.2}}
    \end{picture}}
    \put(8250,0){\setlength{\unitlength}{1cm}\begin{picture}(0,0)(0,0)
        \put(0,0){\line(0,1){0.2}}
    \end{picture}}
    \put(8500,0){\setlength{\unitlength}{1cm}\begin{picture}(0,0)(0,0)
        \put(0,0){\line(0,1){0.2}}
    \end{picture}}
    \put(8750,0){\setlength{\unitlength}{1cm}\begin{picture}(0,0)(0,0)
        \put(0,0){\line(0,1){0.2}}
    \end{picture}}
    \put(9000,0){\setlength{\unitlength}{1cm}\begin{picture}(0,0)(0,0)
        \put(0,0){\line(0,1){0.2}}
    \end{picture}}

%
\punkta{11/06/89}{01:03}{7688.544}{  0.133}{0.063}{1000}{2.54}{
1620}{1.2CA}
\punkta{24/06/89}{00:47}{7701.533}{  0.120}{0.054}{ 500}{1.89}{
998}{1.2CA}
\punkta{25/07/89}{23:39}{7733.485}{ -0.028}{0.065}{1000}{1.72}{
1385}{1.2CA}
\punkta{14/08/89}{23:21}{7753.474}{  0.048}{0.013}{ 500}{2.12}{
1182}{1.2CA}
\punkta{31/07/90}{23:18}{8104.471}{ -0.045}{0.028}{ 500}{1.66}{
954}{1.2CA}
\punkta{24/05/91}{02:14}{8400.593}{ -0.090}{0.057}{ 300}{1.32}{
430}{1.2CA}
\punkta{26/05/91}{01:53}{8402.579}{  0.097}{0.05}{ 250}{2.57}{
676}{1.2CA}
\punkta{29/05/91}{02:36}{8405.609}{  0.129}{0.05}{ 200}{1.69}{
1258}{1.2CA}
\punkta{26/07/91}{22:46}{8464.449}{  0.055}{0.052}{ 301}{1.35}{
4075}{1.2CA}
\punkta{28/07/91}{23:27}{8466.478}{  0.009}{0.125}{ 150}{2.13}{
701}{1.2CA}
\punkta{07/08/91}{22:25}{8476.434}{ -0.008}{0.049}{ 500}{1.80}{
611}{1.2CA}
\punkta{11/08/91}{23:31}{8480.480}{  0.026}{0.032}{ 500}{1.77}{
614}{1.2CA}
\punkta{19/08/91}{21:38}{8488.402}{ -0.097}{0.069}{ 500}{1.36}{
767}{1.2CA}
\punkta{23/08/91}{22:05}{8492.421}{ -0.024}{0.098}{ 500}{1.82}{
1609}{1.2CA}
\punkta{26/08/91}{21:21}{8495.390}{  0.078}{0.05}{ 200}{1.90}{
1038}{1.2CA}
\punkta{28/10/91}{20:25}{8558.351}{ -0.133}{0.050}{ 500}{1.23}{
763}{1.2CA}
\punkta{28/10/91}{20:25}{8558.351}{ -0.133}{0.050}{ 500}{1.23}{
763}{1.2CA}

\end{picture}}

\end{picture}

\vspace*{-0.02cm}

\begin{picture}(18 ,2.5 )(0,0)
\put(0,0){\setlength{\unitlength}{0.01059cm}%
\begin{picture}(1700, 236.111)(7350,0)
\put(7350,0){\framebox(1700, 236.111)[tl]{\begin{picture}(0,0)(0,0)
        \put(1700,0){\makebox(0,0)[tr]{\bf{1704+608}\T{0.4}
                                 \hspace*{0.5cm}}}
    \end{picture}}}

\thicklines
\put(7350,0){\setlength{\unitlength}{5cm}\begin{picture}(0,0)(0,-0.35)
   \put(0,0){\setlength{\unitlength}{1cm}\begin{picture}(0,0)(0,0)
        \put(0,0){\line(1,0){0.3}}
        \end{picture}}
   \end{picture}}

\put(9050,0){\setlength{\unitlength}{5cm}\begin{picture}(0,0)(0,-0.35)
   \put(0,0){\setlength{\unitlength}{1cm}\begin{picture}(0,0)(0,0)
        \put(0,0){\line(-1,0){0.3}}
        \end{picture}}
   \end{picture}}

\thinlines
\put(7350,0){\setlength{\unitlength}{5cm}\begin{picture}(0,0)(0,-0.35)
   \multiput(0,0)(0,0.1){4}{\setlength{\unitlength}{1cm}%
\begin{picture}(0,0)(0,0)
        \put(0,0){\line(1,0){0.12}}
        \end{picture}}
   \end{picture}}

\put(7350,0){\setlength{\unitlength}{5cm}\begin{picture}(0,0)(0,-0.35)
   \multiput(0,0)(0,-0.1){4}{\setlength{\unitlength}{1cm}%
\begin{picture}(0,0)(0,0)
        \put(0,0){\line(1,0){0.12}}
        \end{picture}}
   \end{picture}}

\put(9050,0){\setlength{\unitlength}{5cm}\begin{picture}(0,0)(0,-0.35)
   \multiput(0,0)(0,0.1){4}{\setlength{\unitlength}{1cm}%
\begin{picture}(0,0)(0,0)
        \put(0,0){\line(-1,0){0.12}}
        \end{picture}}
   \end{picture}}

\put(9050,0){\setlength{\unitlength}{5cm}\begin{picture}(0,0)(0,-0.35)
   \multiput(0,0)(0,-0.1){4}{\setlength{\unitlength}{1cm}%
\begin{picture}(0,0)(0,0)
        \put(0,0){\line(-1,0){0.12}}
        \end{picture}}
   \end{picture}}

   \put(7527.5, 236.111){\setlength{\unitlength}{1cm}\begin{picture}(0,0)(0,0)
        \put(0,0){\line(0,-1){0.2}}
   \end{picture}}
   \put(7892.5, 236.111){\setlength{\unitlength}{1cm}\begin{picture}(0,0)(0,0)
        \put(0,0){\line(0,-1){0.2}}
   \end{picture}}
   \put(8257.5, 236.111){\setlength{\unitlength}{1cm}\begin{picture}(0,0)(0,0)
        \put(0,0){\line(0,-1){0.2}}
   \end{picture}}
   \put(8622.5, 236.111){\setlength{\unitlength}{1cm}\begin{picture}(0,0)(0,0)
        \put(0,0){\line(0,-1){0.2}}
   \end{picture}}
   \put(8987.5, 236.111){\setlength{\unitlength}{1cm}\begin{picture}(0,0)(0,0)
        \put(0,0){\line(0,-1){0.2}}
   \end{picture}}
    \multiput(7350,0)(50,0){33}%
        {\setlength{\unitlength}{1cm}\begin{picture}(0,0)(0,0)
        \put(0,0){\line(0,1){0.12}}
    \end{picture}}
    \put(7500,0){\setlength{\unitlength}{1cm}\begin{picture}(0,0)(0,0)
        \put(0,0){\line(0,1){0.2}}
    \end{picture}}
    \put(7750,0){\setlength{\unitlength}{1cm}\begin{picture}(0,0)(0,0)
        \put(0,0){\line(0,1){0.2}}
    \end{picture}}
    \put(8000,0){\setlength{\unitlength}{1cm}\begin{picture}(0,0)(0,0)
        \put(0,0){\line(0,1){0.2}}
    \end{picture}}
    \put(8250,0){\setlength{\unitlength}{1cm}\begin{picture}(0,0)(0,0)
        \put(0,0){\line(0,1){0.2}}
    \end{picture}}
    \put(8500,0){\setlength{\unitlength}{1cm}\begin{picture}(0,0)(0,0)
        \put(0,0){\line(0,1){0.2}}
    \end{picture}}
    \put(8750,0){\setlength{\unitlength}{1cm}\begin{picture}(0,0)(0,0)
        \put(0,0){\line(0,1){0.2}}
    \end{picture}}
    \put(9000,0){\setlength{\unitlength}{1cm}\begin{picture}(0,0)(0,0)
        \put(0,0){\line(0,1){0.2}}
    \end{picture}}

\put(7000,0){\begin{picture}(0,0)(7000,-47.214)
\punkt{16/08/89}{23:09}{7755.465}{ -0.279}{0.011}{1000}{2.21}{
4066}{3.5CA}
\punkt{23/05/90}{02:41}{8034.612}{ -0.126}{0.034}{ 400}{2.43}{
1404}{3.5CA}
\punkt{31/07/90}{00:30}{8103.521}{ -0.166}{0.009}{ 500}{1.52}{
892}{3.5CA}
\punkt{31/07/90}{23:35}{8104.483}{ -0.164}{0.012}{ 500}{1.49}{
954}{3.5CA}
\punkt{29/09/90}{19:53}{8164.329}{ -0.137}{0.020}{ 443}{2.01}{
2464}{3.5CA}
\punkt{23/02/91}{05:01}{8310.709}{ -0.027}{0.006}{ 300}{1.23}{
419}{3.5CA}
\punkt{23/05/91}{03:09}{8399.631}{  0.035}{0.014}{ 500}{1.49}{
516}{3.5CA}
\punkt{}{}{8400.597}{0.027}{0.023}{}{}{}{}
\punkta{26/05/91}{02:06}{8402.588}{  0.048}{0.048}{ 300}{2.84}{
700}{3.5CA}
\punkt{29/05/91}{02:25}{8405.601}{  0.043}{0.025}{ 200}{1.64}{
1025}{3.5CA}
\punkt{26/07/91}{23:02}{8464.460}{  0.025}{0.012}{ 154}{1.28}{
2126}{3.5CA}
\punkta{26/07/91}{23:06}{8464.463}{  0.027}{0.012}{ 150}{1.38}{
2117}{3.5CA}
\punkt{28/07/91}{23:52}{8466.495}{  0.028}{0.015}{ 500}{2.40}{
1655}{3.5CA}
\punkt{04/08/91}{23:36}{8473.484}{  0.035}{0.007}{ 500}{1.28}{
518}{3.5CA}
\punkt{05/08/91}{22:54}{8474.454}{  0.031}{0.006}{ 500}{1.13}{
449}{3.5CA}
\punkt{06/08/91}{23:07}{8475.464}{  0.027}{0.007}{ 500}{1.25}{
490}{3.5CA}
\punkt{07/08/91}{22:42}{8476.446}{  0.029}{0.012}{ 500}{1.87}{
648}{3.5CA}
\punkt{08/08/91}{22:46}{8477.449}{  0.038}{0.011}{ 500}{2.20}{
528}{3.5CA}
\punkt{11/08/91}{23:47}{8480.491}{  0.032}{0.005}{ 500}{1.72}{
612}{3.5CA}
\punkt{19/08/91}{21:54}{8488.413}{  0.031}{0.015}{ 500}{1.29}{
743}{3.5CA}
\punkt{22/08/91}{01:10}{8490.549}{  0.040}{0.032}{ 200}{2.42}{
811}{3.5CA}
\punkt{22/08/91}{20:53}{8491.370}{  0.042}{0.019}{ 500}{2.28}{
1277}{3.5CA}
\punkt{24/08/91}{22:31}{8493.438}{  0.043}{0.019}{ 500}{1.80}{
2205}{3.5CA}
\punkt{25/08/91}{21:12}{8494.384}{  0.040}{0.015}{ 500}{1.43}{
1903}{3.5CA}
\punkt{26/08/91}{21:36}{8495.400}{  0.034}{0.022}{ 200}{2.07}{
955}{3.5CA}
\punkt{20/09/91}{20:52}{8520.370}{  0.024}{0.010}{ 500}{1.64}{
1479}{3.5CA}
\punkt{17/10/91}{19:41}{8547.321}{  0.003}{0.010}{ 500}{2.51}{
1097}{3.5CA}
\punkt{20/10/91}{20:36}{8550.359}{  0.004}{0.016}{ 500}{1.30}{
1505}{3.5CA}
\punkt{20/10/91}{20:36}{8550.359}{  0.004}{0.016}{ 500}{1.30}{
1505}{3.5CA}

\end{picture}}

\end{picture}}

\end{picture}

\vspace*{-0.02cm}

\begin{picture}(18 ,2.5 )(0,0)
\put(0,0){\setlength{\unitlength}{0.01059cm}%
\begin{picture}(1700, 236.111)(7350,0)
\put(7350,0){\framebox(1700, 236.111)[tl]{\begin{picture}(0,0)(0,0)
        \put(1700,0){\makebox(0,0)[tr]{\bf{1718+481}\T{0.4}
                                 \hspace*{0.5cm}}}
        \put(1700,-
236.111){\setlength{\unitlength}{1cm}\begin{picture}(0,0)(0,0)
        \end{picture}}
    \end{picture}}}

\thicklines
\put(7350,0){\setlength{\unitlength}{5cm}\begin{picture}(0,0)(0,-0.25)
   \put(0,0){\setlength{\unitlength}{1cm}\begin{picture}(0,0)(0,0)
        \put(0,0){\line(1,0){0.3}}
        \end{picture}}
   \end{picture}}

\put(9050,0){\setlength{\unitlength}{5cm}\begin{picture}(0,0)(0,-0.25)
   \put(0,0){\setlength{\unitlength}{1cm}\begin{picture}(0,0)(0,0)
        \put(0,0){\line(-1,0){0.3}}
        \end{picture}}
   \end{picture}}

\thinlines
\put(7350,0){\setlength{\unitlength}{5cm}\begin{picture}(0,0)(0,-0.25)
   \multiput(0,0)(0,0.1){3}{\setlength{\unitlength}{1cm}%
\begin{picture}(0,0)(0,0)
        \put(0,0){\line(1,0){0.12}}
        \end{picture}}
   \end{picture}}

\put(7350,0){\setlength{\unitlength}{5cm}\begin{picture}(0,0)(0,-0.25)
   \multiput(0,0)(0,-0.1){3}{\setlength{\unitlength}{1cm}%
\begin{picture}(0,0)(0,0)
        \put(0,0){\line(1,0){0.12}}
        \end{picture}}
   \end{picture}}

\put(9050,0){\setlength{\unitlength}{5cm}\begin{picture}(0,0)(0,-0.25)
   \multiput(0,0)(0,0.1){3}{\setlength{\unitlength}{1cm}%
\begin{picture}(0,0)(0,0)
        \put(0,0){\line(-1,0){0.12}}
        \end{picture}}
   \end{picture}}

\put(9050,0){\setlength{\unitlength}{5cm}\begin{picture}(0,0)(0,-0.25)
   \multiput(0,0)(0,-0.1){3}{\setlength{\unitlength}{1cm}%
\begin{picture}(0,0)(0,0)
        \put(0,0){\line(-1,0){0.12}}
        \end{picture}}
   \end{picture}}

   \put(7527.5, 236.111){\setlength{\unitlength}{1cm}\begin{picture}(0,0)(0,0)
        \put(0,0){\line(0,-1){0.2}}
   \end{picture}}
   \put(7892.5, 236.111){\setlength{\unitlength}{1cm}\begin{picture}(0,0)(0,0)
        \put(0,0){\line(0,-1){0.2}}
   \end{picture}}
   \put(8257.5, 236.111){\setlength{\unitlength}{1cm}\begin{picture}(0,0)(0,0)
        \put(0,0){\line(0,-1){0.2}}
  \end{picture}}
   \put(8622.5, 236.111){\setlength{\unitlength}{1cm}\begin{picture}(0,0)(0,0)
        \put(0,0){\line(0,-1){0.2}}
   \end{picture}}
   \put(8987.5, 236.111){\setlength{\unitlength}{1cm}\begin{picture}(0,0)(0,0)
        \put(0,0){\line(0,-1){0.2}}
   \end{picture}}
    \multiput(7350,0)(50,0){33}%
        {\setlength{\unitlength}{1cm}\begin{picture}(0,0)(0,0)
        \put(0,0){\line(0,1){0.12}}
    \end{picture}}
    \put(7500,0){\setlength{\unitlength}{1cm}\begin{picture}(0,0)(0,0)
        \put(0,0){\line(0,1){0.2}}
    \end{picture}}
    \put(7750,0){\setlength{\unitlength}{1cm}\begin{picture}(0,0)(0,0)
        \put(0,0){\line(0,1){0.2}}
    \end{picture}}
    \put(8000,0){\setlength{\unitlength}{1cm}\begin{picture}(0,0)(0,0)
        \put(0,0){\line(0,1){0.2}}
    \end{picture}}
    \put(8250,0){\setlength{\unitlength}{1cm}\begin{picture}(0,0)(0,0)
        \put(0,0){\line(0,1){0.2}}
   \end{picture}}
    \put(8500,0){\setlength{\unitlength}{1cm}\begin{picture}(0,0)(0,0)
        \put(0,0){\line(0,1){0.2}}
    \end{picture}}
    \put(8750,0){\setlength{\unitlength}{1cm}\begin{picture}(0,0)(0,0)
        \put(0,0){\line(0,1){0.2}}
    \end{picture}}
    \put(9000,0){\setlength{\unitlength}{1cm}\begin{picture}(0,0)(0,0)
        \put(0,0){\line(0,1){0.2}}
    \end{picture}}

\punkt{04/09/88}{21:19}{7409.389}{  0.033}{0.039}{1000}{1.40}{
1053}{1.2CA}
\punkt{10/06/89}{01:30}{7687.563}{  0.022}{0.011}{1000}{1.24}{
1290}{1.2CA}
\punkt{10/06/89}{02:35}{7687.608}{  0.011}{0.009}{1000}{1.35}{
1283}{1.2CA}
\punkt{12/06/89}{02:57}{7689.623}{  0.017}{0.008}{1000}{1.59}{
1532}{1.2CA}
\punkt{25/06/89}{01:41}{7702.571}{ -0.011}{0.021}{ 500}{1.20}{
1014}{1.2CA}
\punkt{27/06/89}{01:50}{7704.577}{ -0.011}{0.018}{ 500}{1.73}{
873}{1.2CA}
\punkt{25/07/89}{01:43}{7732.572}{ -0.007}{0.012}{ 500}{1.51}{
1132}{1.2CA}
\punkt{14/08/89}{23:37}{7753.485}{ -0.008}{0.014}{ 500}{1.59}{
1224}{1.2CA}
\punkt{02/09/89}{21:43}{7772.405}{ -0.007}{0.014}{ 500}{1.61}{
772}{1.2CA}
\punkt{03/09/89}{21:59}{7773.416}{ -0.012}{0.024}{1000}{1.42}{
981}{1.2CA}
\punkt{03/09/89}{23:19}{7773.472}{ -0.013}{0.023}{1000}{1.47}{
926}{1.2CA}
\punkt{01/10/89}{08:07}{7800.838}{  0.004}{0.030}{ 500}{2.82}{
588}{1.2CA}
\punkt{01/11/89}{19:10}{7832.299}{  0.003}{0.020}{ 500}{1.61}{
617}{1.2CA}
\punkt{16/02/91}{05:47}{8303.741}{  0.006}{0.011}{ 100}{1.81}{
323}{1.2CA}
\punkt{16/02/91}{05:53}{8303.745}{ -0.001}{0.015}{ 100}{1.87}{
339}{1.2CA}
\punkt{24/05/91}{04:03}{8400.669}{ -0.009}{0.020}{ 100}{1.47}{
774}{1.2CA}
\punkt{24/05/91}{04:05}{8400.671}{ -0.013}{0.022}{ 100}{1.46}{
983}{1.2CA}
\punkt{29/07/91}{00:06}{8466.505}{  0.000}{0.023}{ 148}{2.40}{
753}{1.2CA}
\punkt{29/07/91}{00:14}{8466.510}{ -0.001}{0.018}{ 300}{2.66}{
1150}{1.2CA}
\punkt{12/08/91}{22:46}{8481.449}{ -0.002}{0.014}{ 300}{1.18}{
512}{1.2CA}
\punkt{12/08/91}{22:46}{8481.449}{ -0.002}{0.014}{ 300}{1.18}{
512}{1.2CA}

\end{picture}}

\end{picture}

\vspace*{-0.02cm}

\begin{picture}(18 ,2.5 )(0,0)
\put(0,0){\setlength{\unitlength}{0.01059cm}%
\begin{picture}(1700,  236.111)(7350,0)
\put(7350,0){\framebox(1700,  236.111)[tl]{\begin{picture}(0,0)(0,0)
        \put(1700,0){\makebox(0,0)[tr]{\bf{1821+643}\T{0.4}
                                 \hspace*{0.5cm}}}
    \end{picture}}}

\thicklines
\put(7350,0){\setlength{\unitlength}{5cm}\begin{picture}(0,0)(0,-0.35)
   \put(0,0){\setlength{\unitlength}{1cm}\begin{picture}(0,0)(0,0)
        \put(0,0){\line(1,0){0.3}}
        \end{picture}}
   \end{picture}}

\put(9050,0){\setlength{\unitlength}{5cm}\begin{picture}(0,0)(0,-0.35)
   \put(0,0){\setlength{\unitlength}{1cm}\begin{picture}(0,0)(0,0)
        \put(0,0){\line(-1,0){0.3}}
        \end{picture}}
   \end{picture}}

\thinlines
\put(7350,0){\setlength{\unitlength}{5cm}\begin{picture}(0,0)(0,-0.35)
   \multiput(0,0)(0,0.1){2}{\setlength{\unitlength}{1cm}%
\begin{picture}(0,0)(0,0)
        \put(0,0){\line(1,0){0.12}}
        \end{picture}}
   \end{picture}}

\put(7350,0){\setlength{\unitlength}{5cm}\begin{picture}(0,0)(0,-0.35)
   \multiput(0,0)(0,-0.1){4}{\setlength{\unitlength}{1cm}%
\begin{picture}(0,0)(0,0)
        \put(0,0){\line(1,0){0.12}}
        \end{picture}}
   \end{picture}}

\put(9050,0){\setlength{\unitlength}{5cm}\begin{picture}(0,0)(0,-0.35)
   \multiput(0,0)(0,0.1){2}{\setlength{\unitlength}{1cm}%
\begin{picture}(0,0)(0,0)
        \put(0,0){\line(-1,0){0.12}}
        \end{picture}}
   \end{picture}}

\put(9050,0){\setlength{\unitlength}{5cm}\begin{picture}(0,0)(0,-0.35)
   \multiput(0,0)(0,-0.1){4}{\setlength{\unitlength}{1cm}%
\begin{picture}(0,0)(0,0)
        \put(0,0){\line(-1,0){0.12}}
        \end{picture}}
   \end{picture}}

   \put(7527.5,  236.111){\setlength{\unitlength}{1cm}\begin{picture}(0,0)(0,0)
        \put(0,0){\line(0,-1){0.2}}
   \end{picture}}
   \put(7892.5,  236.111){\setlength{\unitlength}{1cm}\begin{picture}(0,0)(0,0)
        \put(0,0){\line(0,-1){0.2}}
   \end{picture}}
   \put(8257.5,  236.111){\setlength{\unitlength}{1cm}\begin{picture}(0,0)(0,0)
        \put(0,0){\line(0,-1){0.2}}
   \end{picture}}
   \put(8622.5,  236.111){\setlength{\unitlength}{1cm}\begin{picture}(0,0)(0,0)
        \put(0,0){\line(0,-1){0.2}}
   \end{picture}}
   \put(8987.5,  236.111){\setlength{\unitlength}{1cm}\begin{picture}(0,0)(0,0)
        \put(0,0){\line(0,-1){0.2}}
   \end{picture}}
    \multiput(7350,0)(50,0){33}%
        {\setlength{\unitlength}{1cm}\begin{picture}(0,0)(0,0)
        \put(0,0){\line(0,1){0.12}}
    \end{picture}}
    \put(7500,0){\setlength{\unitlength}{1cm}\begin{picture}(0,0)(0,0)
        \put(0,0){\line(0,1){0.2}}
    \end{picture}}
   \put(7750,0){\setlength{\unitlength}{1cm}\begin{picture}(0,0)(0,0)
        \put(0,0){\line(0,1){0.2}}
    \end{picture}}
   \put(8000,0){\setlength{\unitlength}{1cm}\begin{picture}(0,0)(0,0)
        \put(0,0){\line(0,1){0.2}}
    \end{picture}}
   \put(8250,0){\setlength{\unitlength}{1cm}\begin{picture}(0,0)(0,0)
        \put(0,0){\line(0,1){0.2}}
    \end{picture}}
    \put(8500,0){\setlength{\unitlength}{1cm}\begin{picture}(0,0)(0,0)
        \put(0,0){\line(0,1){0.2}}
    \end{picture}}
    \put(8750,0){\setlength{\unitlength}{1cm}\begin{picture}(0,0)(0,0)
       \put(0,0){\line(0,1){0.2}}
    \end{picture}}
    \put(9000,0){\setlength{\unitlength}{1cm}\begin{picture}(0,0)(0,0)
        \put(0,0){\line(0,1){0.2}}
    \end{picture}}

\put(7000,0){\begin{picture}(0,0)(7000,-47.214)
\punkt{24/06/89}{01:04}{7701.544}{  0.038}{0.021}{ 500}{1.74}{
1047}{1.2CA}
\punkt{24/07/89}{02:03}{7731.586}{  0.027}{0.004}{ 500}{2.04}{
1269}{1.2CA}
\punkt{10/08/89}{22:08}{7749.422}{  0.005}{0.009}{ 500}{1.63}{
1517}{1.2CA}
\punkt{16/08/89}{23:49}{7755.493}{  0.012}{0.013}{ 500}{2.05}{
2314}{1.2CA}
\punkt{31/08/89}{12:06}{7770.004}{ -0.002}{0.009}{ 500}{2.59}{
956}{1.2CA}
\punkt{01/09/89}{00:06}{7770.504}{  0.000}{0.006}{ 500}{2.76}{
918}{1.2CA}
\punkt{01/11/89}{20:30}{7832.355}{ -0.021}{0.007}{ 500}{1.85}{
606}{1.2CA}
\punkta{27/01/90}{05:33}{7918.731}{ -0.232}{0.029}{ 300}{2.63}{
1331}{1.2CA}
\punkt{27/01/90}{05:40}{7918.736}{ -0.031}{0.010}{ 300}{2.64}{
1338}{1.2CA}
\punkt{24/09/90}{20:52}{8159.370}{ -0.007}{0.011}{ 500}{2.36}{
1124}{1.2CA}
\punkt{25/09/90}{20:52}{8160.370}{  0.004}{0.013}{ 500}{2.34}{
1117}{1.2CA}
\punkt{25/05/91}{01:24}{8401.559}{  0.060}{0.021}{ 500}{2.42}{
1118}{1.2CA}
\punkt{27/07/91}{00:09}{8464.507}{  0.064}{0.016}{ 300}{2.00}{
4434}{1.2CA}
%

\end{picture}}

\end{picture}}

\end{picture}

\vspace*{-0.02cm}

\begin{picture}(18 ,2.5 )(0,0)
\put(0,0){\setlength{\unitlength}{0.01059cm}%
\begin{picture}(1700, 236.111)(7350,0)
\put(7350,0){\framebox(1700, 236.111)[tl]{\begin{picture}(0,0)(0,0)
        \put(1700,0){\makebox(0,0)[tr]{\bf{2126-$\!$-158}\T{0.4}
                                 \hspace*{0.5cm}}}
        \put(1700,-
236.111){\setlength{\unitlength}{1cm}\begin{picture}(0,0)(0,0)
        \end{picture}}
    \end{picture}}}

\thicklines
\put(7350,0){\setlength{\unitlength}{5cm}\begin{picture}(0,0)(0,-0.25)
   \put(0,0){\setlength{\unitlength}{1cm}\begin{picture}(0,0)(0,0)
        \put(0,0){\line(1,0){0.3}}
        \end{picture}}
   \end{picture}}

\put(9050,0){\setlength{\unitlength}{5cm}\begin{picture}(0,0)(0,-0.25)
   \put(0,0){\setlength{\unitlength}{1cm}\begin{picture}(0,0)(0,0)
        \put(0,0){\line(-1,0){0.3}}
        \end{picture}}
   \end{picture}}

\thinlines
\put(7350,0){\setlength{\unitlength}{5cm}\begin{picture}(0,0)(0,-0.25)
   \multiput(0,0)(0,0.1){3}{\setlength{\unitlength}{1cm}%
\begin{picture}(0,0)(0,0)
        \put(0,0){\line(1,0){0.12}}
        \end{picture}}
   \end{picture}}

\put(7350,0){\setlength{\unitlength}{5cm}\begin{picture}(0,0)(0,-0.25)
   \multiput(0,0)(0,-0.1){3}{\setlength{\unitlength}{1cm}%
\begin{picture}(0,0)(0,0)
        \put(0,0){\line(1,0){0.12}}
        \end{picture}}
   \end{picture}}

\put(9050,0){\setlength{\unitlength}{5cm}\begin{picture}(0,0)(0,-0.25)
   \multiput(0,0)(0,0.1){3}{\setlength{\unitlength}{1cm}%
\begin{picture}(0,0)(0,0)
        \put(0,0){\line(-1,0){0.12}}
        \end{picture}}
   \end{picture}}

\put(9050,0){\setlength{\unitlength}{5cm}\begin{picture}(0,0)(0,-0.25)
   \multiput(0,0)(0,-0.1){3}{\setlength{\unitlength}{1cm}%
\begin{picture}(0,0)(0,0)
        \put(0,0){\line(-1,0){0.12}}
        \end{picture}}
   \end{picture}}

   \put(7527.5, 236.111){\setlength{\unitlength}{1cm}\begin{picture}(0,0)(0,0)
        \put(0,0){\line(0,-1){0.2}}
   \end{picture}}
   \put(7892.5, 236.111){\setlength{\unitlength}{1cm}\begin{picture}(0,0)(0,0)
        \put(0,0){\line(0,-1){0.2}}
   \end{picture}}
   \put(8257.5, 236.111){\setlength{\unitlength}{1cm}\begin{picture}(0,0)(0,0)
        \put(0,0){\line(0,-1){0.2}}
   \end{picture}}
   \put(8622.5, 236.111){\setlength{\unitlength}{1cm}\begin{picture}(0,0)(0,0)
        \put(0,0){\line(0,-1){0.2}}
   \end{picture}}
   \put(8987.5, 236.111){\setlength{\unitlength}{1cm}\begin{picture}(0,0)(0,0)
        \put(0,0){\line(0,-1){0.2}}
   \end{picture}}
    \multiput(7350,0)(50,0){33}%
        {\setlength{\unitlength}{1cm}\begin{picture}(0,0)(0,0)
        \put(0,0){\line(0,1){0.12}}
    \end{picture}}
    \put(7500,0){\setlength{\unitlength}{1cm}\begin{picture}(0,0)(0,0)
        \put(0,0){\line(0,1){0.2}}
    \end{picture}}
    \put(7750,0){\setlength{\unitlength}{1cm}\begin{picture}(0,0)(0,0)
        \put(0,0){\line(0,1){0.2}}
    \end{picture}}
    \put(8000,0){\setlength{\unitlength}{1cm}\begin{picture}(0,0)(0,0)
        \put(0,0){\line(0,1){0.2}}
    \end{picture}}
    \put(8250,0){\setlength{\unitlength}{1cm}\begin{picture}(0,0)(0,0)
        \put(0,0){\line(0,1){0.2}}
    \end{picture}}
    \put(8500,0){\setlength{\unitlength}{1cm}\begin{picture}(0,0)(0,0)
        \put(0,0){\line(0,1){0.2}}
    \end{picture}}
    \put(8750,0){\setlength{\unitlength}{1cm}\begin{picture}(0,0)(0,0)
        \put(0,0){\line(0,1){0.2}}
    \end{picture}}
    \put(9000,0){\setlength{\unitlength}{1cm}\begin{picture}(0,0)(0,0)
        \put(0,0){\line(0,1){0.2}}
    \end{picture}}

%
\punkt{02/09/88}{22:51}{7407.453}{ -0.040}{0.023}{ 600}{1.92}{
664}{1.2CA}
\punkt{03/09/88}{23:03}{7408.461}{ -0.001}{0.031}{ 600}{1.59}{
684}{1.2CA}
\punkt{08/10/88}{19:04}{7443.295}{  0.013}{0.025}{ 500}{1.51}{
592}{1.2CA}
\punkt{26/06/89}{01:41}{7703.571}{ -0.019}{0.012}{ 500}{1.64}{
1203}{1.2CA}
\punkt{28/06/89}{02:01}{7705.585}{ -0.020}{0.046}{ 500}{2.48}{
809}{1.2CA}
\punkt{13/08/89}{00:15}{7751.511}{ -0.015}{0.020}{ 500}{1.79}{
1170}{1.2CA}
\punkt{02/09/89}{22:42}{7772.446}{  0.024}{0.015}{ 500}{2.04}{
724}{1.2CA}
\punkt{01/08/90}{01:25}{8104.559}{  0.019}{0.015}{ 500}{1.52}{
739}{1.2CA}
\punkt{04/08/91}{00:47}{8472.533}{  0.015}{0.011}{ 500}{1.57}{
620}{1.2CA}
\punkt{06/08/91}{01:23}{8474.558}{  0.017}{0.012}{ 500}{1.51}{
559}{1.2CA}
\punkt{06/08/91}{01:23}{8474.558}{  0.017}{0.012}{ 500}{1.51}{
559}{1.2CA}

\end{picture}}

\end{picture}

\vspace*{-0.02cm}

\begin{picture}(18 ,2.5 )(0,0)
\put(0,0){\setlength{\unitlength}{0.01059cm}%
\begin{picture}(1700, 236.111)(7350,0)
\put(7350,0){\framebox(1700, 236.111)[tl]{\begin{picture}(0,0)(0,0)
        \put(1700,0){\makebox(0,0)[tr]{\bf{2134+004}\T{0.4}
                                 \hspace*{0.5cm}}}
        \put(1700,-
236.111){\setlength{\unitlength}{1cm}\begin{picture}(0,0)(0,0)
            \put(0,-1){\makebox(0,0)[br]{\bf J.D.\,2,440,000\,+}}
        \end{picture}}
    \end{picture}}}

\thicklines
\put(7350,0){\setlength{\unitlength}{5cm}\begin{picture}(0,0)(0,-0.25)
   \put(0,0){\setlength{\unitlength}{1cm}\begin{picture}(0,0)(0,0)
        \put(0,0){\line(1,0){0.3}}
        \end{picture}}
   \end{picture}}

\put(9050,0){\setlength{\unitlength}{5cm}\begin{picture}(0,0)(0,-0.25)
   \put(0,0){\setlength{\unitlength}{1cm}\begin{picture}(0,0)(0,0)
        \put(0,0){\line(-1,0){0.3}}
        \end{picture}}
   \end{picture}}

\thinlines
\put(7350,0){\setlength{\unitlength}{5cm}\begin{picture}(0,0)(0,-0.25)
   \multiput(0,0)(0,0.1){3}{\setlength{\unitlength}{1cm}%
\begin{picture}(0,0)(0,0)
        \put(0,0){\line(1,0){0.12}}
        \end{picture}}
   \end{picture}}

\put(7350,0){\setlength{\unitlength}{5cm}\begin{picture}(0,0)(0,-0.25)
   \multiput(0,0)(0,-0.1){3}{\setlength{\unitlength}{1cm}%
\begin{picture}(0,0)(0,0)
        \put(0,0){\line(1,0){0.12}}
        \end{picture}}
   \end{picture}}

\put(9050,0){\setlength{\unitlength}{5cm}\begin{picture}(0,0)(0,-0.25)
   \multiput(0,0)(0,0.1){3}{\setlength{\unitlength}{1cm}%
\begin{picture}(0,0)(0,0)
        \put(0,0){\line(-1,0){0.12}}
        \end{picture}}
   \end{picture}}

\put(9050,0){\setlength{\unitlength}{5cm}\begin{picture}(0,0)(0,-0.25)
   \multiput(0,0)(0,-0.1){3}{\setlength{\unitlength}{1cm}%
\begin{picture}(0,0)(0,0)
        \put(0,0){\line(-1,0){0.12}}
        \end{picture}}
   \end{picture}}

   \put(7527.5, 236.111){\setlength{\unitlength}{1cm}\begin{picture}(0,0)(0,0)
        \put(0,0){\line(0,-1){0.2}}
   \end{picture}}
   \put(7892.5, 236.111){\setlength{\unitlength}{1cm}\begin{picture}(0,0)(0,0)
        \put(0,0){\line(0,-1){0.2}}
   \end{picture}}
   \put(8257.5, 236.111){\setlength{\unitlength}{1cm}\begin{picture}(0,0)(0,0)
        \put(0,0){\line(0,-1){0.2}}
   \end{picture}}
   \put(8622.5, 236.111){\setlength{\unitlength}{1cm}\begin{picture}(0,0)(0,0)
        \put(0,0){\line(0,-1){0.2}}
   \end{picture}}
   \put(8987.5, 236.111){\setlength{\unitlength}{1cm}\begin{picture}(0,0)(0,0)
        \put(0,0){\line(0,-1){0.2}}
   \end{picture}}
    \multiput(7350,0)(50,0){33}%
        {\setlength{\unitlength}{1cm}\begin{picture}(0,0)(0,0)
        \put(0,0){\line(0,1){0.12}}
    \end{picture}}
    \put(7500,0){\setlength{\unitlength}{1cm}\begin{picture}(0,0)(0,0)
        \put(0,0){\line(0,1){0.2}}
        \put(0,-0.2){\makebox(0,0)[t]{\bf 7500}}
    \end{picture}}
    \put(7750,0){\setlength{\unitlength}{1cm}\begin{picture}(0,0)(0,0)
        \put(0,0){\line(0,1){0.2}}
        \put(0,-0.2){\makebox(0,0)[t]{\bf 7750}}
    \end{picture}}
    \put(8000,0){\setlength{\unitlength}{1cm}\begin{picture}(0,0)(0,0)
        \put(0,0){\line(0,1){0.2}}
        \put(0,-0.2){\makebox(0,0)[t]{\bf 8000}}
    \end{picture}}
    \put(8250,0){\setlength{\unitlength}{1cm}\begin{picture}(0,0)(0,0)
        \put(0,0){\line(0,1){0.2}}
        \put(0,-0.2){\makebox(0,0)[t]{\bf 8250}}
    \end{picture}}
    \put(8500,0){\setlength{\unitlength}{1cm}\begin{picture}(0,0)(0,0)
        \put(0,0){\line(0,1){0.2}}
        \put(0,-0.2){\makebox(0,0)[t]{\bf 8500}}
    \end{picture}}
    \put(8750,0){\setlength{\unitlength}{1cm}\begin{picture}(0,0)(0,0)
        \put(0,0){\line(0,1){0.2}}
        \put(0,-0.2){\makebox(0,0)[t]{\bf 8750}}
    \end{picture}}
    \put(9000,0){\setlength{\unitlength}{1cm}\begin{picture}(0,0)(0,0)
        \put(0,0){\line(0,1){0.2}}
        \put(0,-0.2){\makebox(0,0)[t]{\bf 9000}}
    \end{picture}}

\punkt{05/09/88}{22:46}{7410.449}{  0.048}{0.021}{ 500}{1.66}{
668}{1.2CA}
\punkt{04/10/88}{20:03}{7439.336}{  0.060}{0.022}{ 500}{1.74}{
566}{1.2CA}
\punkt{07/10/88}{19:57}{7442.331}{  0.046}{0.020}{ 500}{1.71}{
676}{1.2CA}
\punkt{10/10/88}{19:01}{7445.292}{  0.003}{0.024}{ 500}{1.99}{
630}{1.2CA}
\punkt{16/10/88}{21:10}{7451.382}{  0.031}{0.029}{ 500}{1.95}{
630}{1.2CA}
\punkt{14/06/89}{01:52}{7691.578}{  0.007}{0.042}{ 500}{2.32}{
765}{1.2CA}
\punkt{27/06/89}{15:05}{7705.129}{  0.009}{0.031}{ 500}{2.80}{
856}{1.2CA}
\punkt{10/08/89}{23:10}{7749.466}{ -0.013}{0.014}{ 500}{1.58}{
994}{1.2CA}
\punkt{02/09/89}{22:59}{7772.458}{ -0.012}{0.012}{ 500}{1.97}{
689}{1.2CA}
\punkt{15/12/89}{18:17}{7876.262}{ -0.034}{0.029}{ 500}{2.87}{
769}{1.2CA}
\punkt{02/08/90}{02:55}{8105.622}{ -0.012}{0.013}{ 500}{1.43}{
798}{1.2CA}
\punkt{24/09/90}{22:23}{8159.433}{  0.020}{0.017}{ 500}{2.65}{
875}{1.2CA}
\punkt{25/09/90}{22:23}{8160.433}{  0.001}{0.016}{ 500}{2.63}{
878}{1.2CA}
\punkt{24/05/91}{03:21}{8400.640}{ -0.033}{0.022}{ 300}{1.44}{
448}{1.2CA}
\punkt{01/08/91}{23:18}{8470.471}{ -0.038}{0.010}{ 300}{1.30}{
532}{1.2CA}
\punkt{01/08/91}{23:24}{8470.476}{ -0.036}{0.012}{ 301}{1.29}{
468}{1.2CA}
\punkt{06/08/91}{01:40}{8474.570}{ -0.016}{0.012}{ 500}{1.21}{
546}{1.2CA}
\punkt{06/08/91}{01:40}{8474.570}{ -0.016}{0.012}{ 500}{1.21}{
546}{1.2CA}

    \put(7350,0){\setlength{\unitlength}{1cm}\begin{picture}(0,0)(0,0)
        \put(0,-1.3){\makebox(0,0)[tl]{\footnotesize{\bf Fig.~1.} (continued)}}
    \end{picture}}

\end{picture}}

\end{picture}

\vspace*{1.3cm}
\end{figure*}

\begin{figure*}

\vspace*{0.4cm}

\begin{picture}(18 ,3.5 )(0,0)
\put(0,0){\setlength{\unitlength}{0.01059cm}%
\begin{picture}(1700, 330.555)(7350,0)
\put(7350,0){\framebox(1700, 330.555)[tl]{\begin{picture}(0,0)(0,0)
        \put(1700,0){\makebox(0,0)[tr]{\bf{2251+244}\T{0.4}
                                 \hspace*{0.5cm}}}
%
    \end{picture}}}

\thicklines
\put(7350,0){\setlength{\unitlength}{5cm}\begin{picture}(0,0)(0,-0.45)
   \put(0,0){\setlength{\unitlength}{1cm}\begin{picture}(0,0)(0,0)
        \put(0,0){\line(1,0){0.3}}
        \end{picture}}
   \end{picture}}

\put(9050,0){\setlength{\unitlength}{5cm}\begin{picture}(0,0)(0,-0.45)
   \put(0,0){\setlength{\unitlength}{1cm}\begin{picture}(0,0)(0,0)
        \put(0,0){\line(-1,0){0.3}}
        \end{picture}}
   \end{picture}}

\thinlines
\put(7350,0){\setlength{\unitlength}{5cm}\begin{picture}(0,0)(0,-0.45)
   \multiput(0,0)(0,0.1){3}{\setlength{\unitlength}{1cm}%
\begin{picture}(0,0)(0,0)
        \put(0,0){\line(1,0){0.12}}
        \end{picture}}
   \end{picture}}

\put(7350,0){\setlength{\unitlength}{5cm}\begin{picture}(0,0)(0,-0.45)
   \multiput(0,0)(0,-0.1){5}{\setlength{\unitlength}{1cm}%
\begin{picture}(0,0)(0,0)
        \put(0,0){\line(1,0){0.12}}
        \end{picture}}
   \end{picture}}

\put(9050,0){\setlength{\unitlength}{5cm}\begin{picture}(0,0)(0,-0.45)
   \multiput(0,0)(0,0.1){3}{\setlength{\unitlength}{1cm}%
\begin{picture}(0,0)(0,0)
        \put(0,0){\line(-1,0){0.12}}
        \end{picture}}
   \end{picture}}

\put(9050,0){\setlength{\unitlength}{5cm}\begin{picture}(0,0)(0,-0.45)
   \multiput(0,0)(0,-0.1){5}{\setlength{\unitlength}{1cm}%
\begin{picture}(0,0)(0,0)
        \put(0,0){\line(-1,0){0.12}}
        \end{picture}}
   \end{picture}}

   \put(7527.5, 330.555){\setlength{\unitlength}{1cm}\begin{picture}(0,0)(0,0)
        \put(0,0){\line(0,-1){0.2}}
        \put(0,0.2){\makebox(0,0)[b]{\bf 1989}}
   \end{picture}}
   \put(7892.5, 330.555){\setlength{\unitlength}{1cm}\begin{picture}(0,0)(0,0)
        \put(0,0){\line(0,-1){0.2}}
        \put(0,0.2){\makebox(0,0)[b]{\bf 1990}}
   \end{picture}}
   \put(8257.5, 330.555){\setlength{\unitlength}{1cm}\begin{picture}(0,0)(0,0)
        \put(0,0){\line(0,-1){0.2}}
        \put(0,0.2){\makebox(0,0)[b]{\bf 1991}}
   \end{picture}}
   \put(8622.5, 330.555){\setlength{\unitlength}{1cm}\begin{picture}(0,0)(0,0)
        \put(0,0){\line(0,-1){0.2}}
        \put(0,0.2){\makebox(0,0)[b]{\bf 1992}}
   \end{picture}}
   \put(8987.5, 330.555){\setlength{\unitlength}{1cm}\begin{picture}(0,0)(0,0)
        \put(0,0){\line(0,-1){0.2}}
        \put(0,0.2){\makebox(0,0)[b]{\bf 1993}}
  \end{picture}}
    \multiput(7350,0)(50,0){33}%
        {\setlength{\unitlength}{1cm}\begin{picture}(0,0)(0,0)
        \put(0,0){\line(0,1){0.12}}
    \end{picture}}
    \put(7500,0){\setlength{\unitlength}{1cm}\begin{picture}(0,0)(0,0)
        \put(0,0){\line(0,1){0.2}}
    \end{picture}}
    \put(7750,0){\setlength{\unitlength}{1cm}\begin{picture}(0,0)(0,0)
        \put(0,0){\line(0,1){0.2}}
    \end{picture}}
    \put(8000,0){\setlength{\unitlength}{1cm}\begin{picture}(0,0)(0,0)
        \put(0,0){\line(0,1){0.2}}
    \end{picture}}
    \put(8250,0){\setlength{\unitlength}{1cm}\begin{picture}(0,0)(0,0)
        \put(0,0){\line(0,1){0.2}}
    \end{picture}}
    \put(8500,0){\setlength{\unitlength}{1cm}\begin{picture}(0,0)(0,0)
        \put(0,0){\line(0,1){0.2}}
    \end{picture}}
    \put(8750,0){\setlength{\unitlength}{1cm}\begin{picture}(0,0)(0,0)
        \put(0,0){\line(0,1){0.2}}
    \end{picture}}
    \put(9000,0){\setlength{\unitlength}{1cm}\begin{picture}(0,0)(0,0)
        \put(0,0){\line(0,1){0.2}}
    \end{picture}}

\put(7000,0){\begin{picture}(0,0)(7000,-94.4287)
\punkt{01/09/89}{01:20}{7770.556}{  0.041}{0.017}{ 500}{2.33}{
984}{1.2CA}
\punkta{02/09/89}{00:54}{7771.538}{  0.088}{0.014}{ 500}{2.21}{
1000}{1.2CA}
\punkta{}{}{7798.938}{0.008}{0.05}{}{}{}{}
\punkta{19/12/89}{18:13}{7880.260}{  0.137}{0.102}{ 100}{1.46}{
728}{1.2CA}
\punkt{01/08/90}{02:16}{8104.595}{  0.045}{0.015}{ 500}{1.15}{
713}{1.2CA}
\punkt{24/09/90}{22:56}{8159.456}{  0.049}{0.022}{ 500}{2.32}{
841}{1.2CA}
\punkt{25/09/90}{22:56}{8160.456}{  0.035}{0.021}{ 500}{2.33}{
841}{1.2CA}
\punkt{28/09/90}{01:02}{8162.544}{  0.038}{0.062}{ 500}{1.29}{
826}{1.2CA}
\punkt{16/10/90}{23:47}{8181.492}{  0.065}{0.055}{ 500}{2.28}{
763}{1.2CA}
\punkt{17/10/90}{23:47}{8182.492}{  0.074}{0.017}{ 500}{2.27}{
763}{1.2CA}
\punkta{20/10/90}{00:19}{8184.514}{  0.096}{0.055}{ 500}{3.21}{
824}{1.2CA}
\punkta{21/10/90}{00:19}{8185.514}{  0.051}{0.068}{ 500}{3.05}{
823}{1.2CA}
\punkta{25/05/91}{02:37}{8401.609}{ -0.059}{0.05}{ 300}{2.70}{
509}{1.2CA}
\punkt{25/07/91}{01:57}{8462.582}{ -0.038}{0.016}{ 500}{1.24}{
1402}{1.2CA}
\punkt{26/07/91}{02:38}{8463.610}{ -0.039}{0.028}{ 500}{1.56}{
1676}{1.2CA}
\punkta{27/07/91}{02:08}{8464.589}{ -0.224}{0.05}{ 150}{1.84}{
2780}{1.2CA}
\punkt{01/08/91}{03:10}{8469.632}{ -0.057}{0.016}{ 300}{1.15}{
899}{1.2CA}
\punkt{02/08/91}{01:56}{8470.581}{ -0.042}{0.012}{ 500}{1.28}{
939}{1.2CA}
\punkt{03/08/91}{01:28}{8471.561}{ -0.029}{0.058}{ 500}{1.30}{
725}{1.2CA}
\punkt{04/08/91}{03:14}{8472.635}{ -0.040}{0.060}{ 500}{1.32}{
660}{1.2CA}
\punkt{05/08/91}{03:21}{8473.640}{ -0.051}{0.010}{ 500}{1.21}{
589}{1.2CA}
\punkt{06/08/91}{02:14}{8474.593}{ -0.031}{0.069}{ 500}{1.20}{
469}{1.2CA}
\punkt{11/08/91}{02:01}{8479.585}{ -0.072}{0.035}{ 500}{2.93}{
483}{1.2CA}
\punkt{12/08/91}{02:41}{8480.612}{ -0.026}{0.062}{ 500}{1.50}{
467}{1.2CA}
\punkta{22/08/91}{22:19}{8491.430}{ -0.089}{0.05}{ 500}{2.32}{
1987}{1.2CA}

\end{picture}}

\end{picture}}

\end{picture}

\vspace*{-0.02cm}

\begin{picture}(18 ,2.5 )(0,0)
\put(0,0){\setlength{\unitlength}{0.01059cm}%
\begin{picture}(1700, 236.111)(7350,0)
\put(7350,0){\framebox(1700, 236.111)[tl]{\begin{picture}(0,0)(0,0)
        \put(1700,0){\makebox(0,0)[tr]{\bf{2308+098}\T{0.4}
                                 \hspace*{0.5cm}}}
    \end{picture}}}

\thicklines
\put(7350,0){\setlength{\unitlength}{5cm}\begin{picture}(0,0)(0,-0.25)
   \put(0,0){\setlength{\unitlength}{1cm}\begin{picture}(0,0)(0,0)
        \put(0,0){\line(1,0){0.3}}
        \end{picture}}
   \end{picture}}

\put(9050,0){\setlength{\unitlength}{5cm}\begin{picture}(0,0)(0,-0.25)
   \put(0,0){\setlength{\unitlength}{1cm}\begin{picture}(0,0)(0,0)
        \put(0,0){\line(-1,0){0.3}}
        \end{picture}}
   \end{picture}}

\thinlines
\put(7350,0){\setlength{\unitlength}{5cm}\begin{picture}(0,0)(0,-0.25)
   \multiput(0,0)(0,0.1){3}{\setlength{\unitlength}{1cm}%
\begin{picture}(0,0)(0,0)
        \put(0,0){\line(1,0){0.12}}
        \end{picture}}
   \end{picture}}

\put(7350,0){\setlength{\unitlength}{5cm}\begin{picture}(0,0)(0,-0.25)
   \multiput(0,0)(0,-0.1){3}{\setlength{\unitlength}{1cm}%
\begin{picture}(0,0)(0,0)
        \put(0,0){\line(1,0){0.12}}
        \end{picture}}
   \end{picture}}

\put(9050,0){\setlength{\unitlength}{5cm}\begin{picture}(0,0)(0,-0.25)
   \multiput(0,0)(0,0.1){3}{\setlength{\unitlength}{1cm}%
\begin{picture}(0,0)(0,0)
        \put(0,0){\line(-1,0){0.12}}
        \end{picture}}
   \end{picture}}

\put(9050,0){\setlength{\unitlength}{5cm}\begin{picture}(0,0)(0,-0.25)
   \multiput(0,0)(0,-0.1){3}{\setlength{\unitlength}{1cm}%
\begin{picture}(0,0)(0,0)
        \put(0,0){\line(-1,0){0.12}}
        \end{picture}}
   \end{picture}}

   \put(7527.5, 236.111){\setlength{\unitlength}{1cm}\begin{picture}(0,0)(0,0)
        \put(0,0){\line(0,-1){0.2}}
   \end{picture}}
   \put(7892.5, 236.111){\setlength{\unitlength}{1cm}\begin{picture}(0,0)(0,0)
        \put(0,0){\line(0,-1){0.2}}
   \end{picture}}
   \put(8257.5, 236.111){\setlength{\unitlength}{1cm}\begin{picture}(0,0)(0,0)
        \put(0,0){\line(0,-1){0.2}}
   \end{picture}}
   \put(8622.5, 236.111){\setlength{\unitlength}{1cm}\begin{picture}(0,0)(0,0)
        \put(0,0){\line(0,-1){0.2}}
   \end{picture}}
   \put(8987.5, 236.111){\setlength{\unitlength}{1cm}\begin{picture}(0,0)(0,0)
        \put(0,0){\line(0,-1){0.2}}
   \end{picture}}
    \multiput(7350,0)(50,0){33}%
        {\setlength{\unitlength}{1cm}\begin{picture}(0,0)(0,0)
        \put(0,0){\line(0,1){0.12}}
    \end{picture}}
    \put(7500,0){\setlength{\unitlength}{1cm}\begin{picture}(0,0)(0,0)
        \put(0,0){\line(0,1){0.2}}
    \end{picture}}
    \put(7750,0){\setlength{\unitlength}{1cm}\begin{picture}(0,0)(0,0)
        \put(0,0){\line(0,1){0.2}}
    \end{picture}}
    \put(8000,0){\setlength{\unitlength}{1cm}\begin{picture}(0,0)(0,0)
        \put(0,0){\line(0,1){0.2}}
    \end{picture}}
    \put(8250,0){\setlength{\unitlength}{1cm}\begin{picture}(0,0)(0,0)
        \put(0,0){\line(0,1){0.2}}
    \end{picture}}
    \put(8500,0){\setlength{\unitlength}{1cm}\begin{picture}(0,0)(0,0)
        \put(0,0){\line(0,1){0.2}}
    \end{picture}}
    \put(8750,0){\setlength{\unitlength}{1cm}\begin{picture}(0,0)(0,0)
        \put(0,0){\line(0,1){0.2}}
    \end{picture}}
    \put(9000,0){\setlength{\unitlength}{1cm}\begin{picture}(0,0)(0,0)
        \put(0,0){\line(0,1){0.2}}
    \end{picture}}

\punkt{04/10/88}{22:57}{7439.457}{ -0.072}{0.017}{ 500}{1.57}{
643}{1.2CA}
\punkt{07/10/88}{22:19}{7442.430}{ -0.067}{0.012}{ 500}{1.56}{
664}{1.2CA}
\punkt{10/10/88}{21:16}{7445.386}{ -0.053}{0.028}{ 500}{2.93}{
629}{1.2CA}
\punkta{13/10/88}{22:12}{7448.425}{ -0.077}{0.066}{ 500}{3.63}{
996}{1.2CA}
\punkt{16/10/88}{23:47}{7451.491}{ -0.048}{0.030}{ 500}{1.75}{
725}{1.2CA}
\punkt{27/06/89}{03:06}{7704.630}{  0.028}{0.012}{ 500}{2.07}{
1036}{1.2CA}
\punkt{02/08/90}{03:26}{8105.643}{  0.022}{0.011}{ 500}{1.46}{
773}{1.2CA}
\punkt{03/08/91}{02:03}{8471.586}{  0.060}{0.010}{ 500}{1.35}{
802}{1.2CA}
\punkt{08/08/91}{02:28}{8476.603}{  0.078}{0.010}{ 500}{1.32}{
514}{1.2CA}
\punkt{08/08/91}{02:28}{8476.603}{  0.078}{0.010}{ 500}{1.32}{
514}{1.2CA}

\end{picture}}

\end{picture}

\vspace*{-0.02cm}

\begin{picture}(18 ,2.5 )(0,0)
\put(0,0){\setlength{\unitlength}{0.01059cm}%
\begin{picture}(1700, 236.111)(7350,0)
\put(7350,0){\framebox(1700, 236.111)[tl]{\begin{picture}(0,0)(0,0)
        \put(1700,0){\makebox(0,0)[tr]{\bf{2354+144}\T{0.4}
                                 \hspace*{0.5cm}}}
        \put(1700,-
236.111){\setlength{\unitlength}{1cm}\begin{picture}(0,0)(0,0)
            \put(0,-1){\makebox(0,0)[br]{\bf J.D.\,2,440,000\,+}}
        \end{picture}}
    \end{picture}}}

\thicklines
\put(7350,0){\setlength{\unitlength}{5cm}\begin{picture}(0,0)(0,-0.25)
   \put(0,0){\setlength{\unitlength}{1cm}\begin{picture}(0,0)(0,0)
        \put(0,0){\line(1,0){0.3}}
        \end{picture}}
   \end{picture}}

\put(9050,0){\setlength{\unitlength}{5cm}\begin{picture}(0,0)(0,-0.25)
   \put(0,0){\setlength{\unitlength}{1cm}\begin{picture}(0,0)(0,0)
        \put(0,0){\line(-1,0){0.3}}
        \end{picture}}
   \end{picture}}

\thinlines
\put(7350,0){\setlength{\unitlength}{5cm}\begin{picture}(0,0)(0,-0.25)
   \multiput(0,0)(0,0.1){3}{\setlength{\unitlength}{1cm}%
\begin{picture}(0,0)(0,0)
        \put(0,0){\line(1,0){0.12}}
        \end{picture}}
   \end{picture}}

\put(7350,0){\setlength{\unitlength}{5cm}\begin{picture}(0,0)(0,-0.25)
   \multiput(0,0)(0,-0.1){3}{\setlength{\unitlength}{1cm}%
\begin{picture}(0,0)(0,0)
        \put(0,0){\line(1,0){0.12}}
        \end{picture}}
   \end{picture}}

\put(9050,0){\setlength{\unitlength}{5cm}\begin{picture}(0,0)(0,-0.25)
   \multiput(0,0)(0,0.1){3}{\setlength{\unitlength}{1cm}%
\begin{picture}(0,0)(0,0)
        \put(0,0){\line(-1,0){0.12}}
        \end{picture}}
   \end{picture}}

\put(9050,0){\setlength{\unitlength}{5cm}\begin{picture}(0,0)(0,-0.25)
   \multiput(0,0)(0,-0.1){3}{\setlength{\unitlength}{1cm}%
\begin{picture}(0,0)(0,0)
        \put(0,0){\line(-1,0){0.12}}
        \end{picture}}
   \end{picture}}

   \put(7527.5, 236.111){\setlength{\unitlength}{1cm}\begin{picture}(0,0)(0,0)
        \put(0,0){\line(0,-1){0.2}}
   \end{picture}}
   \put(7892.5, 236.111){\setlength{\unitlength}{1cm}\begin{picture}(0,0)(0,0)
        \put(0,0){\line(0,-1){0.2}}
   \end{picture}}
   \put(8257.5, 236.111){\setlength{\unitlength}{1cm}\begin{picture}(0,0)(0,0)
        \put(0,0){\line(0,-1){0.2}}
   \end{picture}}
   \put(8622.5, 236.111){\setlength{\unitlength}{1cm}\begin{picture}(0,0)(0,0)
        \put(0,0){\line(0,-1){0.2}}
   \end{picture}}
   \put(8987.5, 236.111){\setlength{\unitlength}{1cm}\begin{picture}(0,0)(0,0)
        \put(0,0){\line(0,-1){0.2}}
   \end{picture}}
    \multiput(7350,0)(50,0){33}%
        {\setlength{\unitlength}{1cm}\begin{picture}(0,0)(0,0)
        \put(0,0){\line(0,1){0.12}}
    \end{picture}}
    \put(7500,0){\setlength{\unitlength}{1cm}\begin{picture}(0,0)(0,0)
        \put(0,0){\line(0,1){0.2}}
        \put(0,-0.2){\makebox(0,0)[t]{\bf 7500}}
    \end{picture}}
    \put(7750,0){\setlength{\unitlength}{1cm}\begin{picture}(0,0)(0,0)
        \put(0,0){\line(0,1){0.2}}
        \put(0,-0.2){\makebox(0,0)[t]{\bf 7750}}
    \end{picture}}
    \put(8000,0){\setlength{\unitlength}{1cm}\begin{picture}(0,0)(0,0)
        \put(0,0){\line(0,1){0.2}}
        \put(0,-0.2){\makebox(0,0)[t]{\bf 8000}}
    \end{picture}}
    \put(8250,0){\setlength{\unitlength}{1cm}\begin{picture}(0,0)(0,0)
        \put(0,0){\line(0,1){0.2}}
        \put(0,-0.2){\makebox(0,0)[t]{\bf 8250}}
    \end{picture}}
    \put(8500,0){\setlength{\unitlength}{1cm}\begin{picture}(0,0)(0,0)
        \put(0,0){\line(0,1){0.2}}
        \put(0,-0.2){\makebox(0,0)[t]{\bf 8500}}
    \end{picture}}
    \put(8750,0){\setlength{\unitlength}{1cm}\begin{picture}(0,0)(0,0)
        \put(0,0){\line(0,1){0.2}}
        \put(0,-0.2){\makebox(0,0)[t]{\bf 8750}}
    \end{picture}}
    \put(9000,0){\setlength{\unitlength}{1cm}\begin{picture}(0,0)(0,0)
        \put(0,0){\line(0,1){0.2}}
        \put(0,-0.2){\makebox(0,0)[t]{\bf 9000}}
    \end{picture}}

\punkta{10/06/89}{15:22}{7688.141}{  0.018}{0.110}{ 500}{1.87}{
804}{1.2CA}
\punkta{21/07/89}{01:57}{7728.581}{ -0.092}{0.050}{1000}{1.79}{
5948}{1.2CA}
\punkt{19/12/89}{19:33}{7880.315}{ -0.015}{0.057}{ 500}{1.40}{
647}{1.2CA}
\punkt{31/07/90}{02:57}{8103.623}{  0.177}{0.016}{ 500}{1.52}{
591}{1.2CA}
\punkt{18/10/90}{23:34}{8183.482}{  0.131}{0.015}{ 500}{1.45}{
826}{1.2CA}
\punkt{19/10/90}{23:34}{8184.482}{  0.129}{0.016}{ 500}{1.45}{
825}{1.2CA}
\punkt{03/08/91}{02:19}{8471.597}{ -0.064}{0.018}{ 500}{1.12}{
948}{1.2CA}
\punkt{08/08/91}{03:02}{8476.626}{ -0.093}{0.020}{ 500}{1.41}{
653}{1.2CA}
\punkt{10/08/91}{02:50}{8478.619}{ -0.035}{0.052}{1000}{3.06}{
758}{1.2CA}
\punkta{22/08/91}{04:40}{8490.695}{  0.033}{0.05}{ 500}{2.02}{
2369}{1.2CA}
\punkt{23/08/91}{00:09}{8491.506}{ -0.039}{0.006}{ 500}{1.93}{
1707}{1.2CA}
\punkt{19/09/91}{01:39}{8518.569}{ -0.087}{0.019}{ 500}{1.31}{
488}{1.2CA}
\punkt{19/09/91}{23:45}{8519.490}{ -0.104}{0.016}{1000}{1.04}{
3072}{1.2CA}
\punkt{19/09/91}{23:45}{8519.490}{ -0.104}{0.016}{1000}{1.04}{
3072}{1.2CA}

    \put(7350,0){\setlength{\unitlength}{1cm}\begin{picture}(0,0)(0,0)
        \put(0,-1.3){\makebox(0,0)[tl]{\footnotesize{\bf Fig.~1.} (continued)}}
    \end{picture}}

\end{picture}}

\end{picture}

\vspace*{1.3cm}
\end{figure*}

\section{Discussion of the lightcurves}

\noindent
{\bf PHL\,658 (0003+158)} has also been found in large area optical and
X-ray surveys. The historical lightcurve shows no features (the standard
deviation of the data is $\sigma=0.16$\,mag, A73). Between 1966 and 73, a
(monotonic but not linear) 0.5\,mag increase has been measured (TS71, STW,
BRZ); in addition, BRZ report the detection of three flares, not confirmed
by others. L84 found $\sigma=0.062$\,mag (1966-80), PS83 found
$\sigma=0.12$\,mag (1970-78). SNLC report a long-term brightening ($\Delta
B \simeq 0.4$) between 1982 and 1991 as well as short-term variations
around this trend. Our data have $\sigma=0.081$\,mag (1988-92);
there are no signs of short-term variations. Our
POSS photometry suggests variability of some 0.1\,mag.

\noindent
{\bf UM\,208 (0007$-$000)}. NS83 found no significant long-term variation.
In our survey, the object shows a relatively large amplitude
($\Delta R=0.4$\,mag, $R'\simeq0.1$\,mag\,yr$^{-1}$).

\noindent
{\bf UM\,224 (0013$-$004)} also showed no long-term variation
(NS83). Our observations yield weak variability of marginal significance.
Pica et al.\ (\cite{PPSL80}) report on a faint red companion of the
quasar.

\noindent
{\bf PHL\,938 (0058+019)} was almost constant ($\sigma=0.13$\,mag, PS83)
during the '70s, compatible with our data.

\noindent
{\bf 1E\,0104$+$318}. For this BAL quasar, Gioia et al.\ (\cite{GMSG86})
reported a variability of $\Delta R \simeq 0.5$\,mag during the period
1982--84. The HQM lightcurve for this faint quasar is not well sampled;
indications for weak variability are only tentative. Our POSS photometry
indicates moderate variations on very long timescales.

\noindent
{\bf PHL\,1226 (0151+045)} showed the largest variation
($\Delta V=-1.22\pm0.2$\,mag) in the survey of NS83; PS83 found no
significant variability ($\sigma=0.13$\,mag). Our lightcurve shows
relatively strong variations.

\noindent
{\bf 0731$+$653 (W1)}. The HQM lightcurve is dominated by data points
obtained under poor conditions; however, an overall increase in brightness
is obvious from the few good exposures. On our best CCD frames, three faint
galaxies are seen 3\arcsec, 10\arcsec\ and 18\arcsec\ from the quasar.

\noindent
{\bf 1E\,0745+557}. There are no variability data in the literature. The
HQM measurements indicate a steady fading; only data obtained under poor
conditions deviate slightly from this trend.

\noindent
{\bf 4C\,05.34 (0805$+$046)}. During the Rosemary Hill program (PS83), the
object was only marginally variable with $\sigma=0.114$\,mag (for more
recent data see SNLC). The HQM
lightcurve is possibly undersampled so that only an overall brightening
can be stated.

\noindent
{\bf 3C\,196 (0809$+$483)}. All literature data indicate only marginal
variability (GT74). The HQM lightcurve is well described by a second
order polynomial.

\noindent
{\bf 0903$+$175}. There are no variability data in the literature. Our
measurements indicate only variations near the detection limit; the
percentage of observations carried out under poor conditions is relatively
high.

\noindent
{\bf Ton\,490 (1011$+$250)}. MS84 report, without more details, a 1.3\,mag
brightness change. The HQM data are consistent with a constant flux.

\noindent
{\bf 1E\,1109+357}. There are no variability data in the literature. The
HQM data are sparse and partly obtained under poor conditions; only an
overall fading is obvious.

\noindent
{\bf 1209$+$107 (KP\,9)}. There are no variability data in the literature.
The HQM data show a steady decline of $\simeq 0.1$\,mag\,yr$^{-1}$.

\noindent
{\bf Ton\,1530 (1222$+$228)}. There are no variability data in the
literature. Our lightcurve seems to show very weak long-term variations at
the detection limit. On our deepest frames, there is an excess of
faint galaxies inside 30\arcsec.

\noindent
{\bf 4C\,55.27 (1332$+$552)}. MS84 report, without more details, a 1.9\,mag
brightness change. From the HQM data, there are no indications for
variability; whereas our POSS photometry shows a significant change on
the very long scale. On deep images, a number excess of faint galaxies inside
30\arcsec\ around the quasar is seen; additionally, a probably interacting
pair of galaxies (tidal arm) of totally $R\simeq15.5$ lies 68\arcsec\ SW
and a single, relatively bright ($R\simeq14.0$) galaxy 114\arcsec\ SW.

\noindent
{\bf Mkn\,679 (1421$+$330)}. There are no variability data in the literature.
The HQM data seem to indicate very weak long-term variations at
the detection limit.

\noindent
{\bf S4\,1435$+$638}. There are no variability data in the literature.
The HQM lightcurve is not well sampled. Obvious is only a slight
brightening. Our POSS photometry indicates a large variation
($\Delta R_{40}\simeq1.6$).

\noindent
{\bf 1520$+$413 (SP\,43)}. There are no variability data in the literature.
Our measurements indicate weak variations, determined however by only few
data points.

\noindent
{\bf 1604$+$290 (KP\,63)}. There are no variability data in the literature.
Our measurements show that the quasar is relatively strongly variable; the
lightcurve is well fitted by a second order polynomial. On our best frames,
a number excess of faint galaxies inside 30\arcsec\ around the quasar is
seen; probably, a compact cluster of galaxies lies about 1\arcmin\ NE.

\noindent
{\bf PG\,1630$+$377}. There are no variability data in the literature.
Our lightcurve shows very weak variations at the detection limit.

\noindent
{\bf 1633+267 (KP\,83)}. There are no variability data in the literature.
There is an overall decrease in the HQM fluxes; the most recent data are
however obtained under poor conditions.

\noindent
{\bf PG\,1634$+$706}. There are no variability data in the literature.
Our lightcurve is consistent with a constant flux.

\noindent
{\bf 1E\,1640$+$396 }. There are no variability data in the literature.
Our measurements show that the object varies on a timescale comparable
with the timelags between the observing campaigns.

\noindent
{\bf 1701$+$610}. There are no variability data in the literature. There is
only one useful reference star in the CCD frame so that no error
calculation could be made for the HQM data. The scatter in the flux values
is therefore possibly not real.

\noindent
{\bf 3C\,351 (1704$+$608)}. The historical lightcurve between 1895 and
1965 measured by A73 shows variations with $\sigma=0.25$\,mag; a 0.9\,mag
change between two nights was reported. Later brightness values (L84,
Angione et al.\ \cite{AMRS81}, BRZ, Kinman \cite{Kin68}, L\"u \cite{Lu72})
indicate only low amplitude variations. The most recent photometry
shows variability of some 0.1\,mag between 1983 and 1986
(Corso et al.\ \cite{CSPD85}, \cite{CSD86}, \cite{CRSH87}). The Rosemary
Hill data (SNLC) suggest a brightness increase since 1988. The HQM
lightcurve is well fitted by a third order polynomial; a maximum occurring
in summer 1991 is covered by a large number of measurements. On our deepest
exposures, an enhancement of faint galaxies is seen within a few arcsec
around the quasar.

\noindent
{\bf PG\,1718$+$481}. There are no variability data in the literature. The
HQM lightcurve is perfectly flat.

\noindent
{\bf E\,1821+643}. There are no variability data in the literature. The
HQM data for this bright quasar have very low errors. The lightcurve is
well described by a second order polynomial; only one point, obtained under
poor conditions, deviates from the best fit. POSS photometry indicates
variations on the very long scale, too. Its a funny
mistake that Monk et al.\ (\cite{MPPB86}) quote the existence of a
foreground galaxy for E\,1821+643 (cf.\ Hewitt \& Burbidge \cite{HB87});
actually, one has here the very rare case of an angular coincidence between
a quasar and a planetary nebula (Kohoutek, private communication).

\noindent
{\bf PKS\,2126$-$158}. There are no variability data in the literature.
Our photometric measurements are consistent with a constant flux although
there is slight evidence for a weak brightening.

\noindent
{\bf PHL\,61 (2134$+$004)}. L84 found the quasar almost constant between
1977 and 1979. The historical lightcurve of Gottlieb \& Liller
(\cite{GL78}) between 1922 and 1954 shows two very bright outbursts
($B\simeq14.8$\,mag), compared to a usual level of $B\simeq17.5$\,mag.
The mm-radio lightcurve between 1982 and 1986 measured by Fiedler et al.\
(\cite{Fie87}) shows significant variability in the percent range. At
lower radio frequencies the object was also found to be variable during
the period 1972--74 (Wardle et al.\ \cite{WBK81}). The only significant
feature in our optical data is a linear fading by
$\sim$\,0.015\,mag\,yr$^{-1}$.

\noindent
{\bf 4C\,24.61 (2251$+$244)}. There are no variability data in the
literature. A large fraction of the HQM data was obtained under poor
conditions. The better data are well fitted by a second order polynomial.

\noindent
{\bf 4C\,09.72 (2308$+$098)}. BRZ searched for variability during 1973/74,
with a negative result. The HQM lightcurve is not well sampled so that
only an overall brightening can be stated.

\noindent
{\bf PKS\,2354$+$144}. This object was included in the early Rosemary Hill
sample; Folsom et al.\ (\cite{FSHH71}) reported only moderate variability
with $\Delta B \la 0.5$\,mag. The HQM data are in agreement with this;
whereas our POSS photometry indicates a larger variation on the very long
scale.

\bigskip

\acknowledgements{We thank
          the MPIA and the Calar Alto staff members for excellent support
          and D.~Mehlert, L.~Nieser, M.~Schaaf, T.~Schramm
          and H.J.~Witt for their help during observations.
          This work has
          been supported by the Deutsche Forschungsgemeinschaft
          under Bo$\,$904/1, Re$\,$439/5 and Schr\,292/6.}

\end{document}